\documentclass[12pt]{article}

\pdfoutput=1
\usepackage{jheppub}
\usepackage{subcaption}
\graphicspath{{feyndiags/}}

\newcommand{\ab}{\ensuremath{\alpha_0}}
\newcommand{\als}{\ensuremath{\alpha_s}}

\newcommand{\mub}{\ensuremath{\mu_0}}
\newcommand{\mur}{\ensuremath{\mu_R}}
\newcommand{\amp}{\ensuremath{\mathcal{A}}}
\newcommand{\se}{\ensuremath{S_\epsilon}}
\newcommand{\msbar}{\ensuremath{\overline{\mathrm{MS}}}}

\def\eps{\epsilon}
\def\shat{\hat{s}}
\def\that{\hat{t}}
\def\be{\begin{equation}}
\def\ee{\end{equation}}
\def\bea{\begin{eqnarray}}
\def\eea{\end{eqnarray}}
\def\nn{\nonumber}

\def\ket#1{|{#1}\rangle}
\def\bra#1{\langle{#1}|}

\def\bom#1{{\mbox{\boldmath $\mathrm{#1}$}}}

\newcommand{\dd}[1]{\mathrm{d}#1\,}

\title{Full top quark mass dependence in Higgs boson pair production at NLO}

\author[a]{S.~Borowka,}
\author[a]{N.~Greiner,}
\author[b]{G.~Heinrich,}
\author[b]{S.~P.~Jones,}
\author[b]{M.~Kerner,}
\author[b]{J.~Schlenk,}
\author[b]{T.~Zirke}

\affiliation[a]{Institute for Physics, Universit{\"a}t Z{\"u}rich, Winterthurerstr.190, 8057 Z\"urich,Switzerland}
\affiliation[b]{Max-Planck-Institute for Physics, F\"ohringer Ring 6, 80805 M\"unchen, Germany}

\preprint{{\small  MPP-2016-261, ZU-TH-31/16}}

\abstract{
 We study the effects of the exact top quark mass-dependent two-loop corrections 
 to  Higgs boson pair production by gluon
 fusion at the LHC and at a 100 TeV hadron collider. 
 We perform a detailed comparison of the full next-to-leading order result to various 
 approximations at the level of differential distributions
 and also analyse non-standard Higgs self-coupling scenarios.
 We find that the different next-to-leading order approximations differ from the full 
 result by up to 50 percent in relevant differential distributions. This clearly stresses
 the importance of the full NLO result.
}


\begin{document}

\maketitle

\section{Introduction}


After the discovery of a boson~\cite{Aad:2012tfa,Chatrchyan:2012ufa} whose
characteristics have so far been consistent with 
the Standard Model Higgs boson, it is a primary
goal of the LHC and future colliders
to further scrutinize its properties. 
In particular, the form of the Higgs potential needs to be reconstructed
by experimental measurements, in order to  confirm the 
mechanism of electroweak symmetry breaking  postulated by
the Standard Model. One of the parameters entering the Higgs
potential, the mass of the Higgs boson, already has been measured to
an impressive precision~\cite{Aad:2015zhl}. The other parameter, the Higgs boson
self-coupling, is more difficult to constrain, as it requires the
production of at least two Higgs bosons.
The cross sections
for Higgs boson pair production at the LHC are about three orders of magnitude smaller
than the ones for single Higgs production.
The dominant production channel is  the gluon fusion channel, as for
single Higgs boson production at the LHC.

In the gluon fusion channel, there are two categories of contributions to di-Higgs production: either
 a virtual Higgs boson, produced by the same mechanism as in 
 single Higgs production, is decaying into a Higgs boson pair, 
involving the self-coupling $\lambda_{hhh}$,
or the two Higgs bosons are both directly radiated
from a heavy quark. At leading order (LO), these two mechanisms can be
 attributed to ``triangle'' and ``box'' contributions, respectively.
However, at NLO, i.e. at the level of two-loop diagrams, the
diagram topologies are more complicated, such that the association of
 ``triangle diagrams'' to diagrams containing the self-coupling
 $\lambda_{hhh}$ becomes invalid.

The Higgs boson pair production cross section is additionally
suppressed by the fact that there is destructive interference
between contributions containing the Higgs boson self-coupling and 
the ones containing only Yukawa couplings to heavy quarks, and that for larger values of $\sqrt{\hat s}$, the
 contributions with an s-channel virtual Higgs boson propagator are strongly suppressed.

Therefore, narrowing the window of possible values for the
triple-Higgs coupling experimentally 
will have to wait until the high-luminosity run 
of the LHC~\cite{ATL-PHYS-PUB-2015-046,ATL-PHYS-PUB-2014-019,Contardo:2020886}, if Standard Model rates are assumed. 
However, the Higgs boson pair production rate could be modified by
physics beyond the Standard Model (BSM), and hence 
it is important to be able to distinguish BSM effects from
Standard Model higher order corrections.
In this paper we will study the effects of a modified  Higgs boson
self-coupling and show that the Higgs boson invariant mass
distribution is quite sensitive to changes in $\lambda_{hhh}$,
as such changes modify the interference pattern.

Both ATLAS and CMS have published measurements of Higgs boson pair production in the
decay channels $\gamma\gamma
b\bar{b}$~\cite{ATLAS-CONF-2016-004,Khachatryan:2016sey,Aad:2015xja,Aad:2014yja}, 
$b\bar{b}b\bar{b}$~\cite{Aaboud:2016xco,CMS:2016tlj,Aad:2015xja,Khachatryan:2015yea,Aad:2015uka}, 
$\gamma\gamma W W^*$, $b\bar{b}W W^*$,
$\tau^+\tau^-b\bar{b}$~\cite{ATLAS:2016qmt,CMS:2016cdj,CMS:2016ymn,CMS:2016rec,CMS-PAS-HIG-16-013,CMS:2016ugf,CMS:2016zxv,Aad:2015xja}.

Phenomenological studies about Higgs boson pair production and the
feasibility of Higgs boson self-coupling measurements can be found
e.g. in Refs.~\cite{Baglio:2012np,Goertz:2013kp,Barr:2013tda,Gouzevitch:2013qca,Barger:2013jfa,Dolan:2013rja,Maierhofer:2013sha,Li:2013flc,Slawinska:2014vpa,deLima:2014dta,Frederix:2014hta,Buschmann:2014sia,Dolan:2012rv,Dolan:2015zja,Dall'Osso:2015aia,Azatov:2015oxa,Grober:2015cwa,Ghezzi:2015vva,Papaefstathiou:2015iba,Dicus:2015yva,Dawson:2015oha,Dawson:2015haa,Behr:2015oqq,Degrassi:2016wml,Li:2016nrr,Kanemura:2016lkz}.

The leading order (one-loop) calculation of Higgs boson pair production in gluon fusion 
has been performed in Refs.~\cite{Eboli:1987dy,Glover:1987nx,Plehn:1996wb}.
NLO corrections were calculated 
in the $m_t\to\infty$ limit, where the top quark degrees of freedom are
integrated out, leading to point-like effective couplings of gluons
to  Higgs bosons (``Higgs Effective Field Theory'', HEFT). 

Top quark mass effects have been included in various approximations. 
Calculating the NLO corrections within the heavy top limit and then
rescaling the result differentially by a
factor $B_{FT}/B_{HEFT}$, where $B_{FT}$ denotes the leading order
matrix element squared in the full theory, is denoted ``Born-improved HEFT"
approximation.
This calculation~\cite{Dawson:1998py}, implemented in the program {\sc
Hpair}, led to a K-factor of about two.
In Ref.~\cite{Maltoni:2014eza}, 
another approximation, called ``FT$_{approx}$", was introduced, which contains the full top quark mass dependence in the real radiation, 
while the virtual part is calculated in the HEFT approximation and rescaled by the re-weighting factor $B_{FT}/B_{HEFT}$.
The ``FT$_{approx}^{\prime}$" result~\cite{Maltoni:2014eza} in addition
uses partial NLO results for the virtual part, i.e. it employs the exact results where they are known 
from  single Higgs production.  The ``FT$_{approx}$" calculation leads to a
cross section which is about 10\% smaller than the Born-improved NLO
HEFT cross section. Using the ``FT$_{approx}^{\prime}$" procedure, the
reduction is about 9\% with respect to the Born-improved NLO
HEFT result.
It was also found that top width effects can reach up to $-4\%$ above
the $t\bar{t}$ threshold~\cite{Maltoni:2014eza}.
At LO, a finite top width reduces the total cross section at $\sqrt{s}=14$\,TeV by
about 2\%. In our calculation we do not include a finite top width.

In addition, the HEFT results at NLO and NNLO have been improved by an expansion in
$1/m_t^{2\rho}$ in Refs.~\cite{Grigo:2013rya,Grigo:2014jma,Grigo:2015dia,Degrassi:2016vss},
with $\rho^{\rm{max}}=6$ at NLO, and $\rho^{\rm{max}}=2$ for the
soft-virtual part at NNLO ~\cite{Grigo:2015dia}. 
In the latter reference it is also demonstrated that the sign of the
finite top mass corrections, amounting to about $\pm 10$\%, depends on whether the re-weighting factor
is applied at differential level, i.e. before the integration over the
partonic centre of mass energy, or at total cross section level. 

The NNLO QCD corrections in the heavy top limit have been 
performed in Refs.~\cite{deFlorian:2013uza,deFlorian:2013jea,Grigo:2014jma}, 
and they have been supplemented by an expansion in $1/m_t^2$ in Ref.~\cite{Grigo:2015dia}
and by resummation at NLO+NNLL in Ref.~\cite{Shao:2013bz}.
The most precise results within the infinite top mass approximation
are  NNLO+NNLL resummed results, calculated in
Ref.~\cite{deFlorian:2015moa}, leading to K-factors of about 1.2 relative to the Born-improved HEFT result.
 Very recently, fully differential NNLO
results in the HEFT approximation have become
available~\cite{deFlorian:2016uhr}.

As the different approximations partly led to corrections with
opposite sign, there was a rather large uncertainty associated with
the unknown effect of the exact top quark mass dependence at NLO,  
which was estimated to be of the order of 10\% at $\sqrt{s}=14$\,TeV.

The full NLO calculation which became available
recently~\cite{Borowka:2016ehy}, revealed a 14\% reduction 
of the total cross section compared to the Born improved HEFT at
$\sqrt{s}=14$\,TeV and a 24\% reduction at $\sqrt{s}=100$\,TeV.

At differential distribution level, we found that the deviation 
from the Born-improved HEFT approximation can be as large as 50\% in the
tails of distributions like the Higgs boson pair invariant mass or
Higgs boson transverse momentum distributions.

This paper is structured as follows. 
In Section 2  we give details of the calculation, in particular about
the calculation of the two-loop amplitude and about the $1/m_t$
expansion which we also performed. In Section 3 we discuss our 
phenomenological results.
We study various distributions at $\sqrt{s}=14$\,TeV and $\sqrt{s}=100$\,TeV, comparing the full NLO result to
different approximations. We also analyze the effects of non-Standard Model values of
the triple Higgs coupling.

\section{Details of the calculation}
\label{sec:calculation}

\subsection{Amplitude structure}
\label{sec:amp}


\begin{figure}
\centering
\includegraphics[width=12cm]{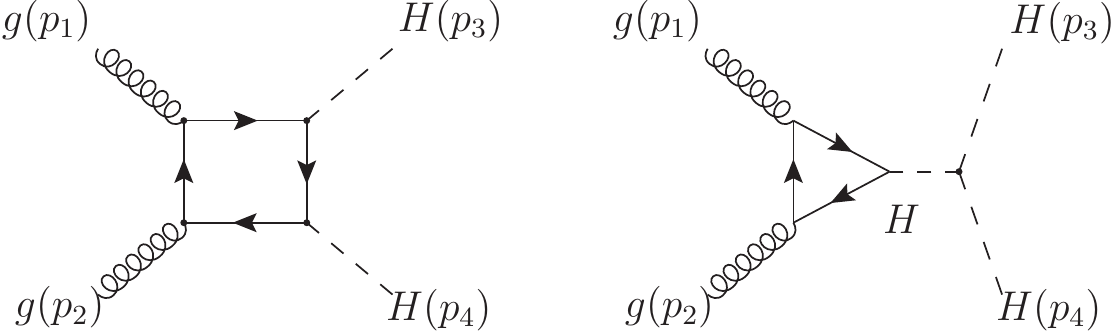}
\caption{Diagrams contributing to the process $gg\to hh$ at leading order.\label{fig:LOdiagrams}}
\end{figure}

The leading order diagrams contributing to the process $gg\to hh$ are shown in Fig.~\ref{fig:LOdiagrams}.
As the  cross section does not have a tree level contribution, the 
virtual contribution at next-to-leading order
involves two-loop diagrams, and the NLO real radiation part involves 
one-loop diagrams up to pentagons.

The amplitude for the process $g(p_1,\mu)+g(p_2,\nu)\to h(p_3)+h(p_4)$
can be decomposed into form factors as
\begin{align}
&{\cal M}_{ab}=\delta_{ab}\,\eps^\mu (p_1,n_1)\eps^\nu (p_2,n_2)\,{\cal
  M}_{\mu\nu}\label{eq:FFdeco}\\
&{\cal M}^{\mu\nu}=\frac{\alpha_s}{8\pi v^2}\left\{F_1(\hat{s},\hat{t},m_h^2,m_t^2,D)\;T_1 ^{\mu\nu}+F_2(\hat{s},\hat{t},m_h^2,m_t^2,D)\;T_2 ^{\mu\nu}\right\}\;,\nn
\end{align}
where $n_1,n_2$ are arbitrary reference momenta for the two gluon
polarization vectors $\eps^{\mu},\eps^{\nu}$. Colour indices are
denoted by $a,b$ and
\begin{align}
  \hat{s}=(p_1+p_2)^2,\quad \hat{t}=(p_1-p_3)^2,\quad \hat{u}=(p_2-p_3)^2\;.
\end{align}
The decomposition into tensors carrying the Lorentz structure is not
unique. 
It is however convenient to define the form factors such that~\cite{Glover:1987nx}
\bea
{\cal M}^{++}&=&{\cal M}^{--}=-\frac{\alpha_s}{8\pi v^2} \,F_1\\
{\cal M}^{+-}&=&{\cal M}^{-+}=-\frac{\alpha_s}{8\pi v^2} \,F_2\;,\nn
\eea
which is fulfilled with the following definitions 
\begin{eqnarray}
T_1 ^{\mu\nu}&=& g^{\mu\nu}-\frac{p_1^{\nu}\,p_2^{\mu}}{p_1\cdot  p_2} \;,\label{eq:Ttensors} \\
T_2 ^{\mu\nu}&=& g^{\mu\nu}+\frac{1}{p_T^2\,(p_1\cdot p_2)}\,\left\{
  m_h^2 \,p_1^{\nu}\,p_2^{\mu} - 2\,(p_1\cdot p_3) \,p_3^{\nu}\,p_2^{\mu}- 2\,(p_2\cdot p_3) \,p_3^{\mu}\,p_1^{\nu}+ 2\,(p_1\cdot p_2) \,p_3^{\nu}\,p_3^{\mu}\right\}\nn\\
\mbox{ where } && p_T^2=(\hat{u}\,\hat{t}-m_h^4)/\hat{s}\;,\; T_1\cdot T_2=D-4\;, \;T_1\cdot T_1= T_2\cdot T_2=D-2\;.\nn
\end{eqnarray}   
At leading order, we can further split $F_1$ into a ``triangle'' and a ``box'' contribution 
\be
F_1(\hat{s},\hat{t},m_h^2,m_t^2,D)=F_\triangle(\hat{s},\hat{t},m_h^2,m_t^2,D)+F_\Box(\hat{s},\hat{t},m_h^2,m_t^2,D)\;.
\ee
As the LO form factor $F_\triangle$  
only contains the triangle diagrams, which  
have no angular momentum dependence, it can be attributed entirely to an s-wave contribution.
The form factors $F_\Box$ and $F_2$ can be attributed to the spin-0 and spin-2 states 
of the scattering amplitude, respectively.

We can get an idea about the angular dependence of $F_1$ and $F_2$ by
 considering the partial wave decomposition of the scattering
 amplitude, which is independent of the loop order.
It should be noted however that this analysis is valid for $2\to 2$ scattering. 
At NLO, the cross section for the process $gg\to HH$ also contains a
$2\to 3$ scattering contribution from the real radiation. 
Therefore the analysis of the angular dependence below does not apply
to the full NLO cross section. 
 
 In general, for a scattering process $a+b\to c+d$ with the corresponding
helicities $\lambda_{a},...,\lambda_{d}$, 
the partial wave decomposition reads~\cite{Jacob:1959at,Degrande:2012wf,Dawson:2012mk}
\begin{equation}
\label{eq:partial}
 \bra{\theta\phi\lambda_c\lambda_d}T(E)\ket{00\lambda_a\lambda_b} \;=\; 16\pi\sum_J(2J+1)\bra{\lambda_c,\lambda_d}T^J(E)
 \ket{\lambda_a,\lambda_b}e^{i(s_i-s_f)\phi}d^J_{s_i,s_f}(\theta)\;,
\end{equation}
with $s_i=\lambda_a-\lambda_b$ and $s_f=\lambda_c-\lambda_d$, and where $\bra{\theta\phi\lambda_c\lambda_d}T(E)\ket{00\lambda_a\lambda_b}$
denotes the transition matrix element. Unitarity must hold for each partial wave independently, {\it i.e.}
$|T^J|\;\le\; 1\;.$
Thus the amplitude is decomposed into (orthogonal)  Wigner $d$-functions $d^J_{s_i,s_f}(\theta)$, where $J$ denotes 
the total angular momentum and $s_i,s_f$ the total spin of the initial
and final state, respectively.
The structure of the amplitude is such that 
$F_1$ only contributes to $s_i=0$, while $F_2$ only contributes to $s_i=2$.
$F_1$ can have both a component proportional to 
$d^0_{0,0}(\theta)$ as well as one proportional to
$d^2_{0,0}(\theta)$, while $F_2$ is proportional to
$d^2_{2,0}(\theta)$.
The $d$-functions $d^J_{0,0}(\theta)$ are proportional to the
Legendre-Polynomials $P_J(\cos\theta)$.
As $P_0(x)=1, P_2(x)=\frac{1}{2}\,(3x^2-1)$ and $d^2_{2,0}(\theta)\sim
\sin^2\theta$, we can conclude that the angular dependence of $F_2$
should be $\sim \sin^2\theta$. 
From the analytic expression for $F_2$ at leading
order~\cite{Glover:1987nx}, 
we can verify that indeed $F_2\sim
p_T^2=(\hat{u}\,\hat{t}-m_h^4)/\hat{s}=\frac{\hat{s}}{4}\beta_h^2\,\sin^2\theta$
where $\beta_h^2=1-4m_h^2/\hat{s}$.

Further, it is known that the leading contributions to the amplitude
come from the lower partial waves in Eq.~(\ref{eq:partial}). Therefore we
also conclude that the contribution from $F_2$ should be subleading
with respect to $F_1$ in most of the kinematic regions.
Indeed we observe that the contribution of the form factor $F_2$ 
to the virtual two-loop amplitude is suppressed as compared to $F_1$.

\subsection{Leading Order cross section}

The functions $F_i$ at leading order with full mass dependence can be found e.g. in 
Refs.~\cite{Glover:1987nx,Plehn:1996wb}. 
At LO, the ``triangle'' form factor has the simple form
\begin{eqnarray}
F_\triangle & = & C_\triangle \bar{F}_\triangle\; , \; C_\triangle = \frac{\lambda_{hhh}}{\hat s - m_h^2}\;,\;
\lambda_{hhh} = 3 m_h^2\lambda\;,\label{eq:lambda} \\
\bar{F}_\triangle & = & 4m_q^2\left\{ 2+(4\,m_q^2 - \shat)  C_{0} \right\}
= 2\hat{s}\,\tau_q \left[ 1+(1-\tau_q) f(\tau_q)\right]\;,\nn
\end{eqnarray}
where $\lambda=1$ in the Standard Model, $\tau_q=4\,m_q^2/\shat$ and
\be
f(\tau_q)=\left\{
\begin{array}{ll}  \displaystyle
\arcsin^2\frac{1}{\sqrt{\tau_q}} &\mbox{for } \tau_q \geq 1 \\
 -\frac{1}{4}\left[ \log\frac{1+\sqrt{1-\tau_q}}
{1-\sqrt{1-\tau_q}}-i\pi \right]^2 \hspace{0.5cm} &\mbox{for } \tau_q<1
\end{array} \right.
\ee
$$C_{0}=\int \frac{d^4q}{i\pi^2}~\frac{1}
{(q^2-m_q^2)\left[ (q+p_1)^2-m_q^2\right]
\left[ (q+p_1+p_2)^2-m_q^2\right]}\;. $$


The partonic leading order cross section for $gg\to hh$  can be written as
\begin{equation}
\hat \sigma^{\mathrm{LO}}(gg\to hh)
 =\frac{\alpha_s^2(\mu_R)}{2^{12}v^4 (2\pi)^3\hat{s}^2} \int_{\hat t_-}^{\hat t_+} d\hat t \,
 \left\{ \left|
F_1 \right|^2 + \left| F_2
\right|^2 \right\}.
\label{eq:sigmahatLO}
\end{equation}
The integration limits $\that^\pm$ are derived from a 
momentum parametrisation in the centre-of-mass frame,  leading to 
$\hat{t}_{\pm}=m_h^2-\frac{\shat}{2}\,(1\mp\beta_h)$, where $\beta_h^2=1-4\frac{m_h^2}{\shat}$.

\vspace*{3mm}

To obtain the hadronic cross section, we also have to integrate over the PDFs. 
Defining the luminosity function as
\begin{equation}
{d{\cal L}_{ij}
\over d\tau}=\sum_{ij} \int_{\tau}^1{dx\over x} f_i(x,\mu_F) f_j\biggl({\tau\over x},\mu_F\biggr)\, ,
\end{equation}
the total cross section reads
\begin{equation}
\sigma^{\mathrm{LO}}  =  \int_{\tau_0}^1 d\tau~\frac{d{\cal L}_{gg}}{d\tau}~
\hat\sigma^{\mathrm{LO}}(\hat{s} = \tau s)\, ,
\label{eq:sigmalo}
\end{equation}
where $s$ is the square of the hadronic centre of mass energy, $\tau_0 = 4m_h^2/s$,
and $\mu_F$ is the factorization scale.

\subsubsection{Heavy top limit}

In the $m_t\to \infty$ approximation
the  LO form factors are given by
\begin{eqnarray} \label{eq:FHTL}
\bar{F}_\triangle & \to & \frac{4}{3}\hat{s}\;, \;
F_\Box \to -\frac{4}{3}\hat{s}\;, \;
F_2  \to  0\;,
\end{eqnarray}
which implies for the
the effective $ggH$ and $ggHH$ couplings $c_h$ and $c_{hh}$\footnote{Higher order corrections to these effective couplings, and to couplings
involving more than two Higgs bosons, can be found in
Ref.~\cite{Spira:2016zna} and references therein.}
\begin{align}
 c_h= -c_{hh} =
 -\frac{\alpha_s}{4 \pi}\,\frac{i}{3}  + {\cal O} \left(\frac{m_h^2}{4m_t^2}\right) \; .
\end{align}
From the expressions above we can derive the following expression for the squared amplitude in
the heavy top limit :
\begin{align}
|{\cal M}|^2&\sim  \frac{2}{9} -  \frac{4}{3}\,m_h^2\,\frac{\lambda}{\hat{s}-m_h^2} +
  2\,m_h^4\,\frac{\lambda^2}{(\hat{s}-m_h^2)^2}\;.\label{eq:loheft}
\end{align}
For $\lambda=1$, this expression vanishes at the Higgs boson pair production threshold 
$\shat \sim 4 m_h^2$.
This explains why near  the threshold 
the contributions containing the triple Higgs boson coupling
and the  ones which do not contain  an s-channel Higgs boson exchange  almost cancel.
On the other hand, if the triple Higgs boson coupling was different from the 
Standard Model value, for example equal to zero, 
this should be clearly seen from the behaviour of the 
$m_{hh}$ distribution.
We investigate the effects of non-standard values for the triple
Higgs boson coupling in Section \ref{sec:tripleH}.

\subsection{NLO cross section}

The NLO cross section is composed of various parts, which we discuss separately in the following.
\begin{equation}
\sigma^{\mathrm{NLO}}(pp \rightarrow hh) = 
\sigma^{\mathrm{LO}} + 
\sigma^{\mathrm{virt}} + \sigma_{gg}^{\mathrm{r}}
 + \sigma_{gq}^{\mathrm{r}}+ \sigma_{g\bar{q}}^{\mathrm{r}}  + \sigma_{q\bar{q}}^{\mathrm{r}}\;.
\label{eq:sigmanlo}
\end{equation}
The contributions from the real radiation, $\sigma^{\mathrm{r}}$,  can be divided 
into four channels, according to the partons in the initial state. The $q\bar{q}$ 
 channel is infrared finite.
Details are given in Section \ref{sec:real}.

\subsubsection{Calculation of the virtual two-loop amplitude}
\label{sec:2loop}


\vspace*{3mm}


{\bf Amplitude generation}

\vspace*{3mm}

For the virtual two-loop amplitude, we use projectors $P_j^{\mu\nu}$ to achieve a separation into objects 
carrying the Lorentz structure $T_i^{\mu\nu}$ and the form factors $F_1$ and $F_2$,
\begin{eqnarray*}
P_1^{\mu\nu} {\cal M}_{\mu\nu}&=&\frac{\alpha_s}{8\pi v^2}\,F_1(\hat{s},\hat{t},m_h^2,m_t^2,D)\;, \\
P_2^{\mu\nu} {\cal M}_{\mu\nu}&=&\frac{\alpha_s}{8\pi v^2}\,F_2(\hat{s},\hat{t},m_h^2,m_t^2,D)\;. 
\end{eqnarray*}
In $D$ dimensions we can use the tensors $T_i^{\mu\nu}$,
defined in Eqs.~(\ref{eq:Ttensors}), to build the projectors
\begin{eqnarray}
P_1^{\mu\nu} &=&\quad\frac{1}{4}\,\frac{D-2}{D-3} \,T_1^{\mu\nu}
-\frac{1}{4}\,\frac{D-4}{D-3} \,T_2^{\mu\nu}\;,\label{eq:proj1}\\
P_2^{\mu\nu}&=& -\frac{1}{4}\,\frac{D-4}{D-3} \,T_1^{\mu\nu}
+\frac{1}{4}\,\frac{D-2}{D-3} \,T_2^{\mu\nu}\;.\label{eq:proj2}
\end{eqnarray}

\begin{figure}
\centering
\begin{subfigure}{0.3\textwidth}
\includegraphics[width=\textwidth,angle=0]{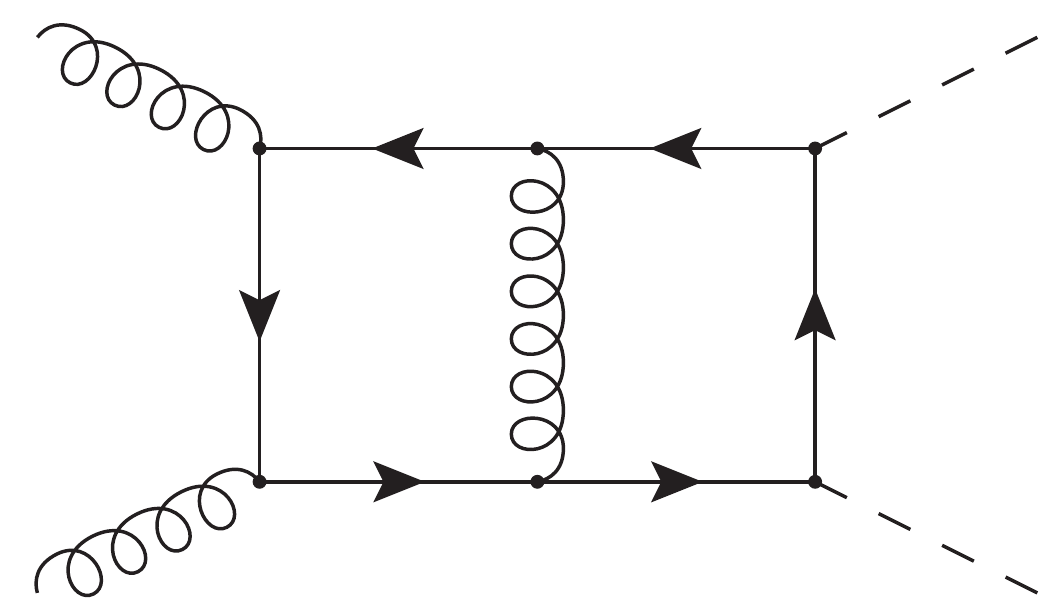}
\caption{}
\end{subfigure}
\begin{subfigure}{0.3\textwidth}
\includegraphics[width=\textwidth,angle=0]{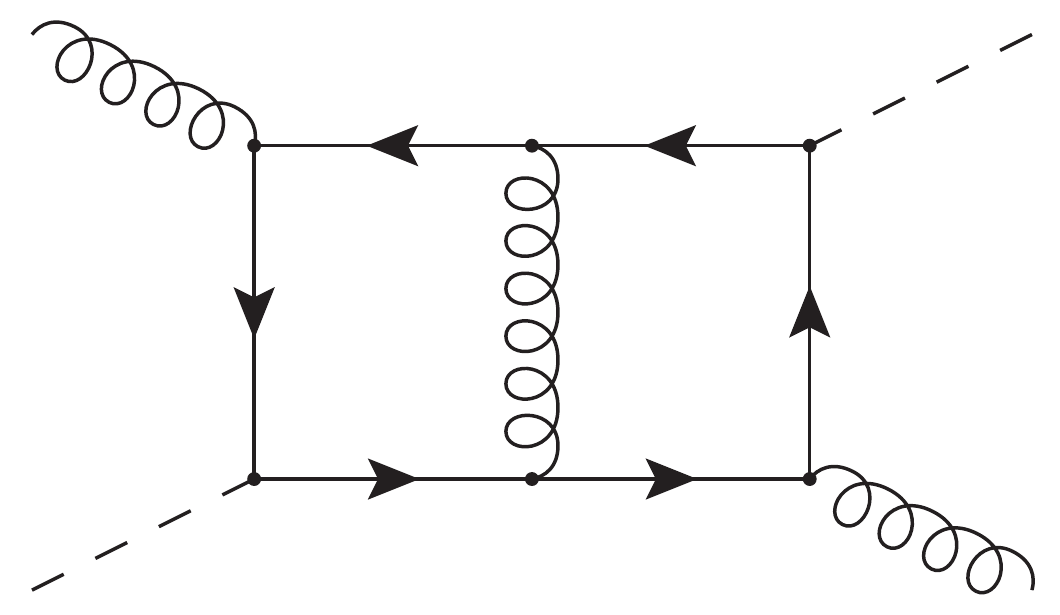}
\caption{}
\end{subfigure}
\begin{subfigure}{0.3\textwidth}
\includegraphics[width=\textwidth,angle=0]{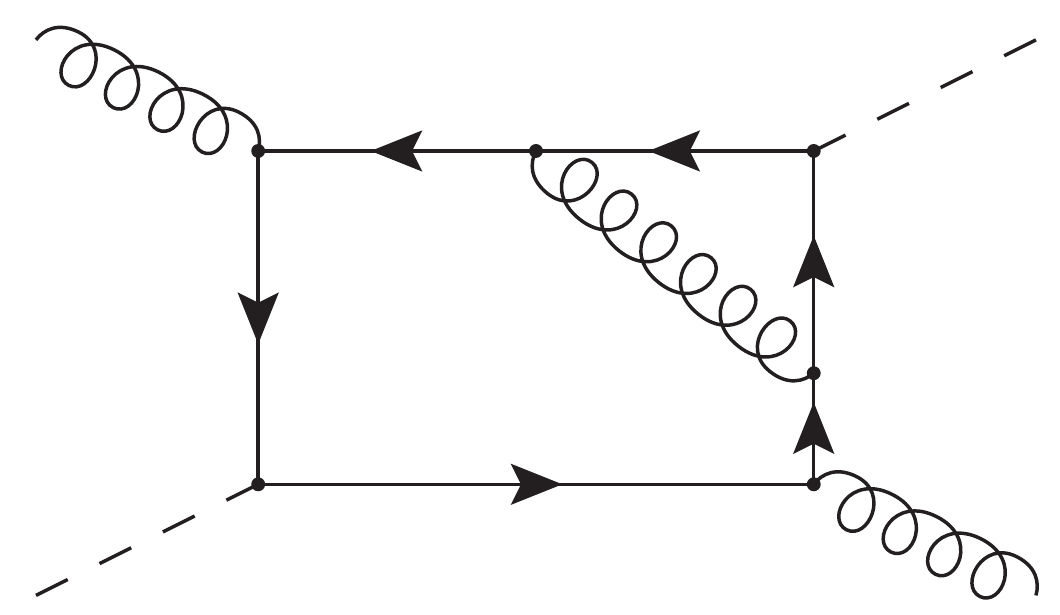}
\caption{}
\end{subfigure}
\begin{subfigure}{0.3\textwidth}
\includegraphics[width=\textwidth,angle=0]{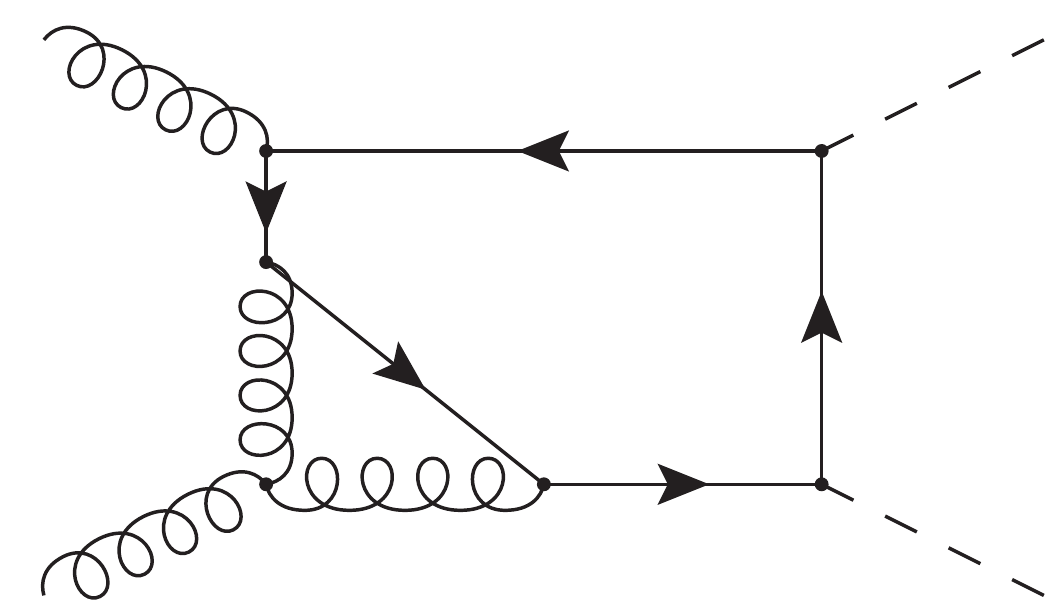}
\caption{}
\end{subfigure}
\begin{subfigure}{0.3\textwidth}
\includegraphics[width=\textwidth,angle=0]{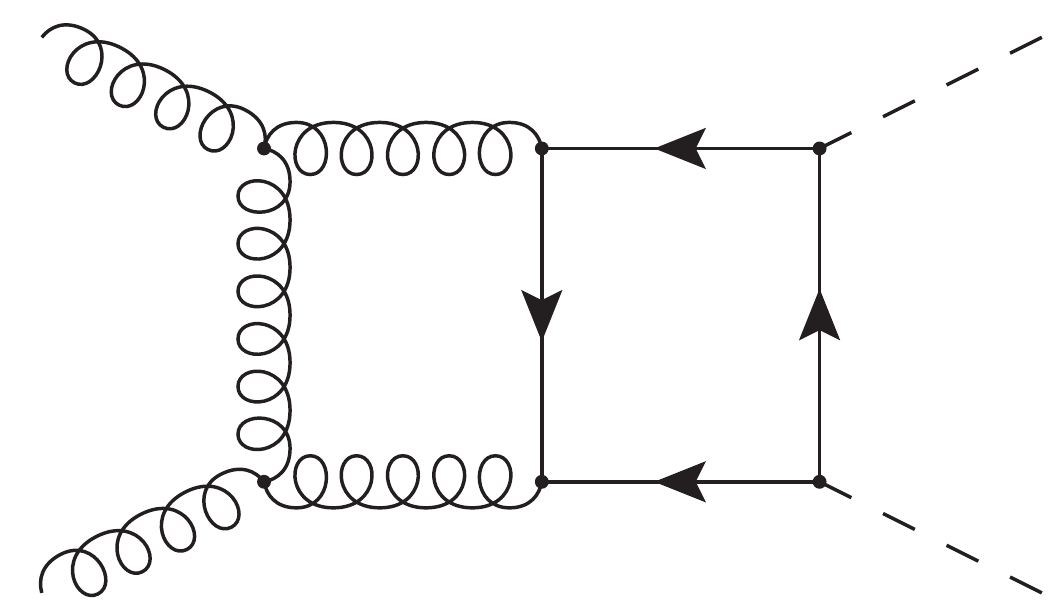}
\caption{}
\end{subfigure}
\begin{subfigure}{0.3\textwidth}
\includegraphics[width=\textwidth,angle=0]{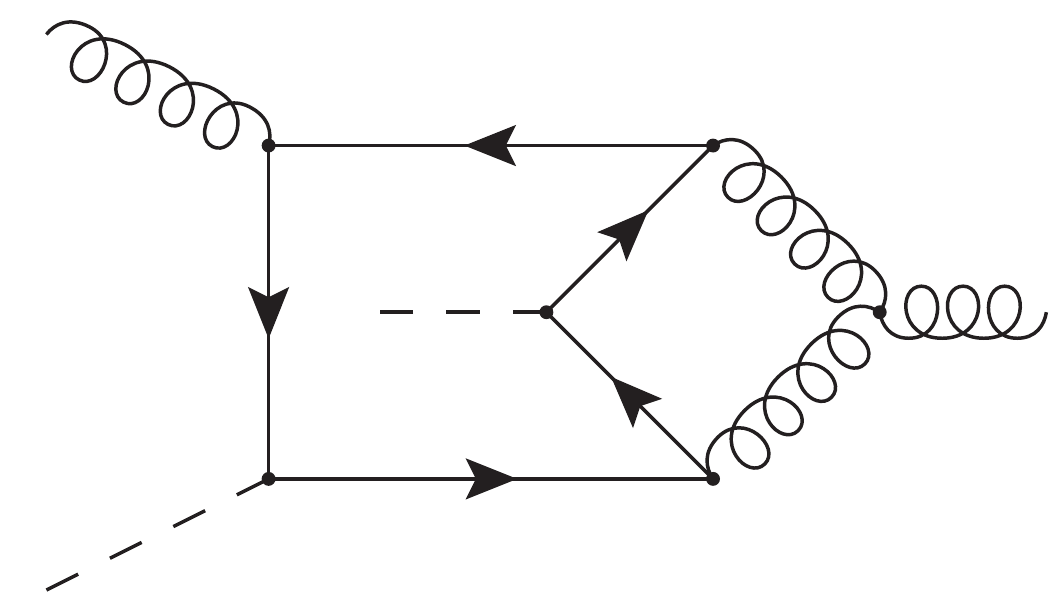}
\caption{}
\end{subfigure}
\begin{subfigure}{0.3\textwidth}
\includegraphics[width=\textwidth,angle=0]{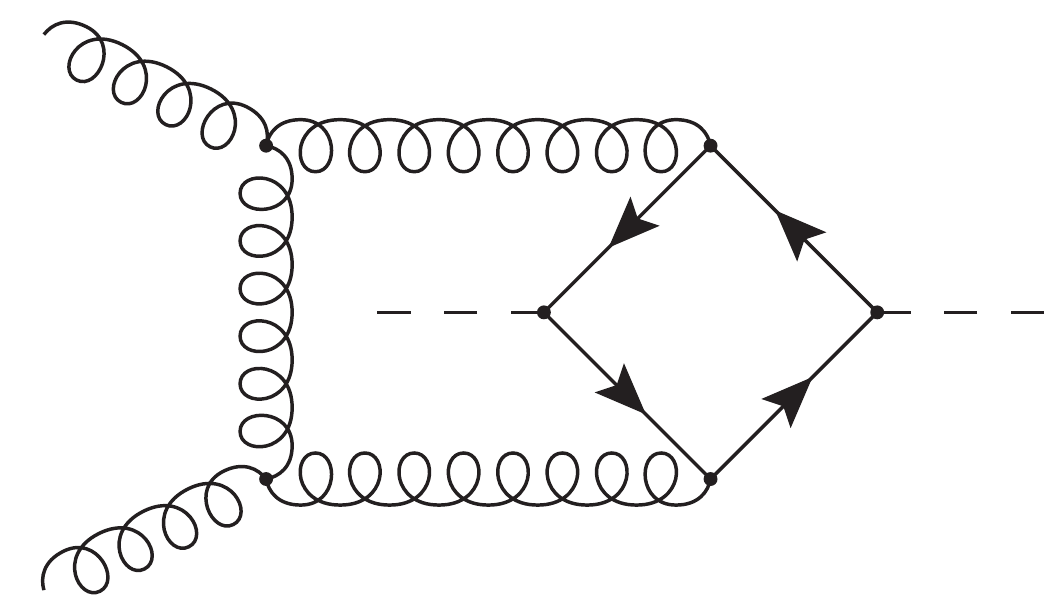}
\caption{}
\end{subfigure}
\begin{subfigure}{0.3\textwidth}
\includegraphics[width=\textwidth,angle=0]{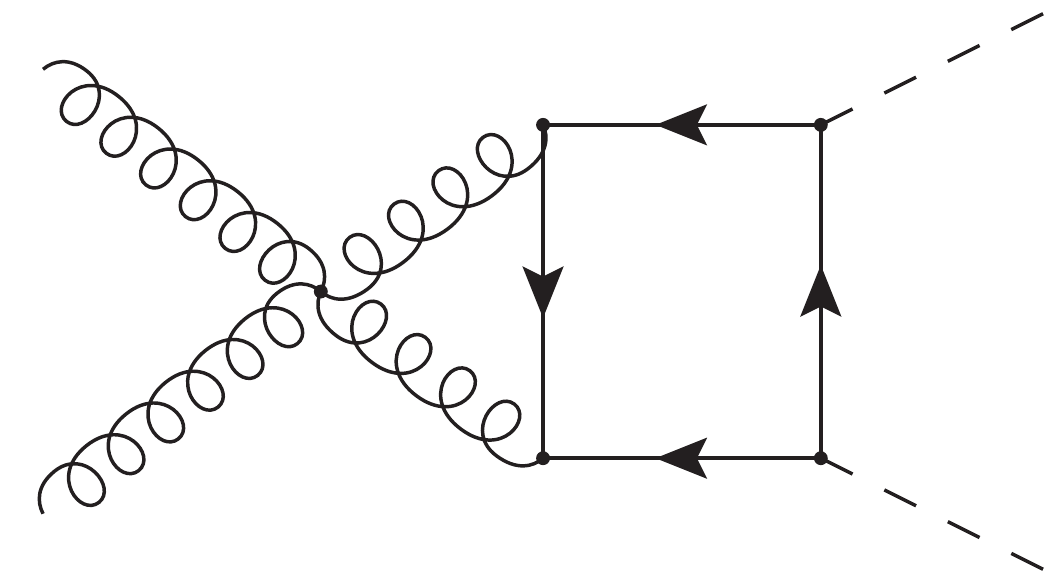}
\caption{}
\end{subfigure}
\begin{subfigure}{0.3\textwidth}
\includegraphics[width=\textwidth,angle=0]{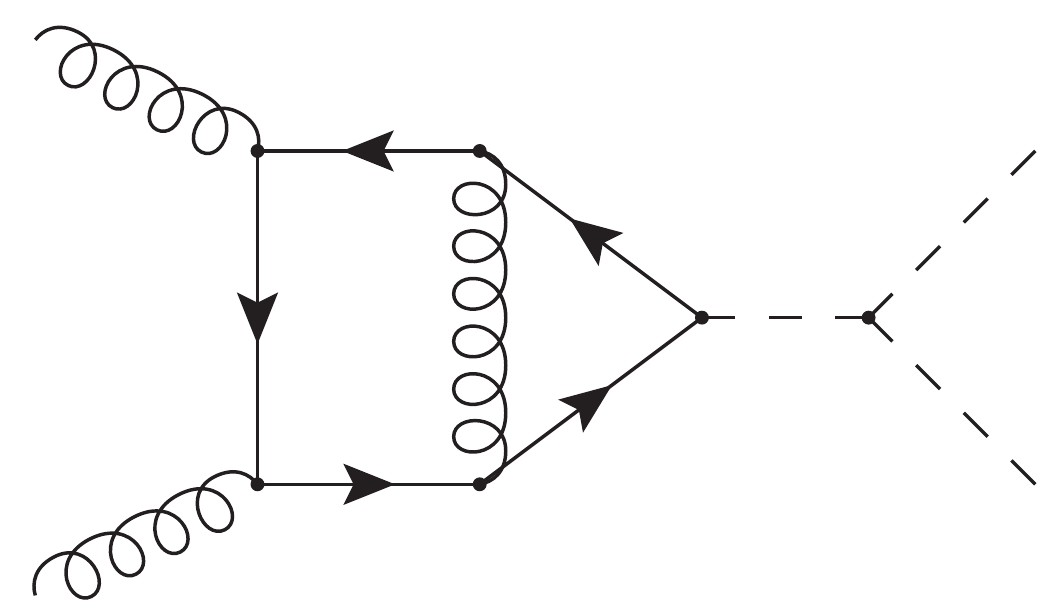}
\caption{}
\end{subfigure}
\begin{subfigure}{0.3\textwidth}
\includegraphics[width=\textwidth,angle=0]{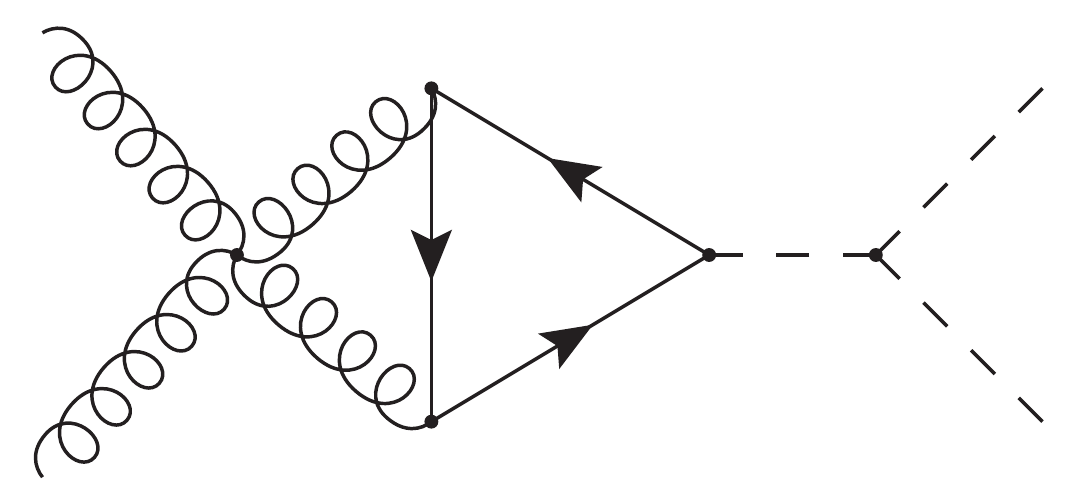}
\caption{}
\end{subfigure}
\qquad
\begin{subfigure}{0.2\textwidth}
\includegraphics[width=\textwidth,angle=0]{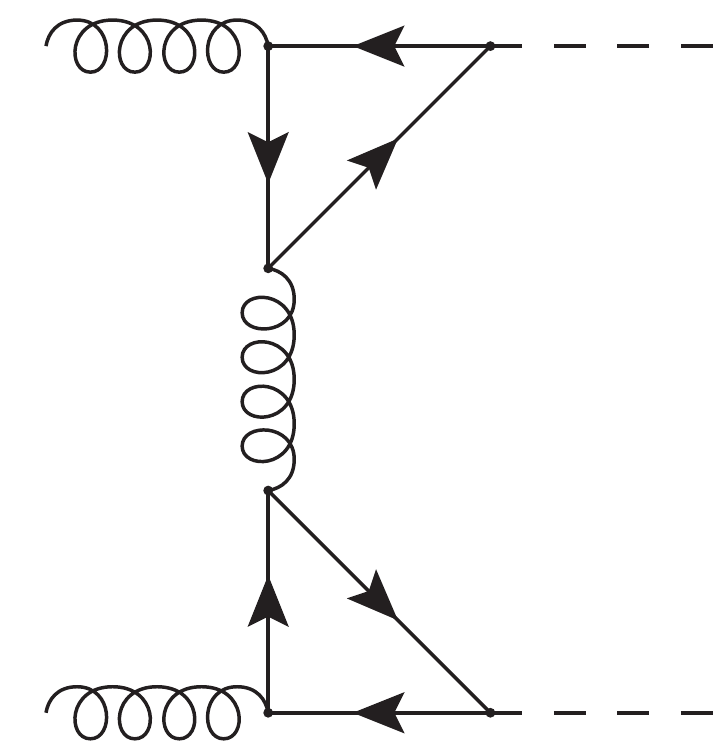}
\caption{\label{subfig:degrassi}}
\end{subfigure}
 \caption{Examples of two-loop diagrams entering the virtual amplitude.\label{fig:2loopdiags}}
 \end{figure}

The virtual amplitude has been generated with an extension of the program {\sc GoSam}~\cite{Cullen:2011ac,Cullen:2014yla}, 
where the diagrams are generated using {\sc Qgraf}~\cite{Nogueira:1991ex} and then further processed 
using {\sc Form}~\cite{Vermaseren:2000nd,Kuipers:2012rf}. 
The two-loop extension of {\sc GoSam}  contains 
an automated python interface to {\sc Reduze}~\cite{vonManteuffel:2012np}, which implies that the user has to provide the 
integral families when running {\sc GoSam}-2loop. 
The other input files needed by {\sc Reduze} are generated automatically by {\sc GoSam}-2loop, 
based on the kinematics of the given process. 
The reduction of the integrals occurring in the amplitude to master integrals should be performed separately, 
where in principle either of the codes  {\sc
  Reduze}~\cite{vonManteuffel:2012np}, {\sc
  Fire5}~\cite{Smirnov:2014hma} or {\sc LiteRed}~\cite{Lee:2013mka}
can be used.
Examples of two-loop diagrams contributing to Higgs boson pair production are shown in Fig.~\ref{fig:2loopdiags}.

 We would like to point out again that the distinction between ``triangle diagrams" and ``box diagrams"  
 becomes ambiguous beyond the leading order. At two-loop and beyond there are diagrams which contain triangle sub-diagrams 
 but which do not contain the Higgs boson self coupling, see Fig.~\ref{subfig:degrassi}.

\vspace*{3mm}

{\bf Integral families and reduction}

\vspace*{3mm}

For the reduction of planar diagrams we have defined five integral families. 
Each family contains nine propagators which allows irreducible
scalar products in the numerator to be written in terms of inverse propagators prior to reduction. We chose a
non-minimal set of integral families in favour of preserving
symmetries  as much as possible. 
We find that integrals with up to four inverse propagators appear in the amplitude and must be reduced.
The families are listed in Table~\ref{tab:families}.
\begin{table}
\centering
\begin{tabular}{|c|c|c|}
\hline
$F_1$ & $F_2$ & $F_3$ \\
\hline
$k_1^2-m_t^2$ & $k_1^2-m_t^2$ & $k_1^2$ \\
$k_2^2-m_t^2$ & $k_2^2-m_t^2$ & $(k_1-k_2)^2-m_t^2$ \\
$(k_1-k_2)^2$ & $(k_1-k_2)^2$ & $(k_1+p_1)^2$ \\
$(k_1+p_1)^2-m_t^2$ & $(k_1+p_1)^2-m_t^2$ & $(k_2+p_1)^2-m_t^2$ \\
$(k_2+p_1)^2-m_t^2$ & $(k_2+p_1)^2-m_t^2$ & $(k_1-p_2)^2$ \\
$(k_1-p_2)^2-m_t^2$ & $(k_1-p_3)^2-m_t^2$ & $(k_2-p_2)^2-m_t^2$ \\
$(k_2-p_2)^2-m_t^2$ & $(k_2-p_3)^2-m_t^2$ & $(k_2-p_2-p_3)^2-m_t^2$ \\
$(k_1-p_2-p_3)^2-m_t^2$ & $(k_1-p_2-p_3)^2-m_t^2$ & $(k_1+p_1+p_3)^2$ \\
$(k_2-p_2-p_3)^2-m_t^2$ & $(k_2-p_2-p_3)^2-m_t^2$  & $(k_2+p_1-p_2)^2$ \\
\hline
\end{tabular}
\begin{tabular}{|c|c|}
$F_4$ & $F_5$ \\
\hline
$k_1^2-m_t^2$ & $k_1^2$ \\
$k_2^2$ &  $k_2^2-m_t^2$ \\
$(k_1-k_2)^2-m_t^2$ & $(k_1-k_2)^2-m_t^2$ \\
$(k_1+p_1)^2-m_t^2$ &  $(k_1+p_1)^2$ \\
$(k_2+p_1)^2$ &  $(k_2+p_1)^2-m_t^2$ \\
$(k_1-p_2)^2-m_t^2$ & $(k_1-p_3)^2$ \\
$(k_2-p_2)^2$ & $(k_2-p_3)^2-m_t^2$ \\
$(k_1-p_2-p_3)^2-m_t^2$ & $(k_1-p_2-p_3)^2$ \\
$(k_2-p_2-p_3)^2$ & $(k_2-p_2-p_3)^2-m_t^2$ \\
\hline
\end{tabular}
\caption{Integral families for the reduction of the planar
  diagrams. The non-planar integrals were computed as tensor
  integrals, see text.\label{tab:families}}
\end{table}

The amplitude generation leads to about 10000 integrals before any symmetries are taken into account. 
After accounting for symmetries and after reduction (complete
reduction of the planar sectors and partial reduction of the
non-planar ones),
we end up with  145 planar master
integrals plus 70 non-planar integrals, and a further 112 integrals
that differ by a crossing. 
As these integrals contain four independent mass scales, 
$\hat{s},\hat{t},m_t^2, m_h^2$, only a small subset is known analytically. 
Besides the diagrams which are factorizing into two one-loop diagrams~\cite{Degrassi:2016vss},  
the known integrals are the two-loop diagrams with two light-like legs and one massive leg, 
which enter single Higgs boson production,
calculated e.g. in Refs.~\cite{Spira:1995rr,Bonciani:2003te,Bonciani:2003hc,Harlander:2005rq,Anastasiou:2006hc}, 
and the triangles with one light-like and two off-shell legs occurring in 
the two-loop calculation of $H\to Z\gamma$~\cite{Gehrmann:2015dua,Bonciani:2015eua}.
However, we calculate all integrals numerically using the program {\sc SecDec}~\cite{Carter:2010hi,Borowka:2012yc,Borowka:2015mxa}.

As the integral basis is not unique, we choose to have two set-ups, relying on
different sets of basis integrals. This serves as a strong check
of the calculation of the virtual amplitude. It has previously been
noted that using a finite basis~\cite{vonManteuffel:2014qoa} along with sector decomposition can
increase the precision obtained by numerical integration for a given
number of sampling points~\cite{vonManteuffel:2015gxa}. 
We also observed that switching to a finite
basis in some of the planar sectors
turned out to be beneficial for the numerical evaluation of the master integrals.

A complete reduction could not be obtained for the non-planar 4-point integrals. 
The inverse propagators appearing in unreduced integrals were
rewritten in terms of scalar products such that the resulting
integrals had the lowest possible tensor rank. The tensor integrals (up to rank 4) were then directly computed with {\sc SecDec}.

We would like to mention that non-planar diagrams also contribute to the leading colour coefficient.
Therefore we could not identify a contribution which is 
both dominant and gauge invariant where only planar integrals
contribute.

\vspace*{3mm}

{\bf Renormalization}

\vspace*{3mm}

We expand the amplitude in $a_0 = \ab/(4 \pi)$, where $\ab$ is the bare QCD coupling. The bare amplitude can be written as
\begin{equation}
\amp_B = a_0 \amp^{(1)}_B + a_0^2 \amp^{(2)}_B + \mathcal{O}(a_0^3),
\end{equation}
where the one- and two-loop coefficients are given by
\begin{align}
\amp^{(1)}_B &= \se \mub^{2\epsilon} \left[ \tilde{b}_0^{(1)} + \tilde{b}_1^{(1)} \epsilon + \tilde{b}_2^{(1)} \epsilon^2 + \mathcal{O}(\epsilon^3) \right], \label{eq:coeff1_bare} \\
\amp^{(2)}_B &= \se^2 \mub^{4\epsilon} \left[ \frac{\tilde{b}_{-2}^{(2)}}{\epsilon^2} + \frac{\tilde{b}_{-1}^{(2)}}{ \epsilon} + \tilde{b}_0^{(2)} + \mathcal{O}(\epsilon) \right]. \label{eq:coeff2_bare} 
\end{align}
Here $\mub^2$ is a parameter introduced in dimensional regularisation to maintain a dimensionless bare coupling 
and $\se = e^{-\gamma_E \epsilon} (4 \pi)^\epsilon,$ with $\gamma_E$ the Euler constant. 
The one-loop amplitude is expanded to $\mathcal{O}(\epsilon^2)$ as it appears multiplied by the Catani-Seymour insertion operator stemming from the integrated dipoles, 
$\mathbf{I}$, which has poles of $\mathcal{O}(\epsilon^{-2})$.

To renormalize the gluon wave function we must multiply the amplitude by $(Z_A)^\frac{1}{2}$ for each external gluon leg, where $Z_A$ is the gluon field renormalization constant. 
We renormalize the QCD coupling  using the relation 
\begin{align}
& a_0 =  a\, Z_{a} \left( \frac{\mur^2}{\mub^2} \right)^\epsilon \;,\;
a = \frac{\als}{4\pi}
\end{align}
where $\als$ is the renormalized coupling  and $Z_{a}$ is the associated renormalization constant. Here $\mur$ is the renormalization scale 
and the dependence of $\als$ on $\mur$ is implicit. 
The top mass is renormalized by relating the bare top mass $m_{t_0}^2$ to the renormalized top mass $m_t^2$ via 
\begin{equation}
m_{t_0}^2 = m_t^2 + a \,\delta m_t^2.
\end{equation} 
In practice, we compute top mass counter-term diagrams, treating $a\,\delta m_t^2$ as a counter-term insertion in top quark lines and renormalize the top Yukawa coupling using
\begin{equation}
y_{t_0} = \left(1 + a \frac{\delta m_t^2}{m_t^2} \right) y_t.
\end{equation}
No Higgs wave function or mass renormalization is required as we compute only QCD corrections.

In our calculation we use conventional dimensional regularization (CDR) with $D=4-2\epsilon$. 
We renormalize the
top mass in the on-shell scheme and the QCD coupling in the $\msbar$ five-flavour scheme ($N_f = 5$) 
with the top quark loops in the gluon self-energy subtracted at zero
momentum.

The one-loop renormalization constants are given to first order in $a$ by\footnote{Note that $Z_{a}$ corresponds to the 
renormalization factor of the coupling $g_s$ {\it squared}, therefore it is twice the expression 
for $Z_{g_s}$ found in the literature, see e.g. Eq.~(3.4) of Ref.~\cite{Dittmaier:2008uj}.}
\begin{align}
Z_A & = 1 + a\, \delta Z_A + \mathcal{O}(a^2), \\
Z_{a} &=  S_\epsilon^{-1} \left[ 1 + a \,\delta Z_{a} + \mathcal{O}(a^2) \right],
\end{align}
where
\begin{align}
\delta Z_A &= \left(\frac{m_t^2}{\mur^2}\right)^{-\eps}\,\left(-\frac{4}{3\eps}\,T_R\right)\;,\nn\\
\delta Z_{a} &=  -\frac{1}{\eps}\,\beta_0+\delta Z_a^{\mathrm{hq}}\;,\;\beta_0= \frac{11}{3}C_A-\frac{4}{3}\,T_R N_f^{\mathrm{light}} \;,\nn\\
\delta Z_a^{\mathrm{hq}}&=\left(\frac{m_t^2}{\mur^2}\right)^{-\eps}\frac{4}{3\eps}\,T_R\;,
\end{align}
and the mass counter-term in the on-shell scheme  is given by
\begin{equation}
\delta m_t^2 = \left(\frac{m_t^2}{\mur^2}\right)^{-\eps}\,2\,m_t^2\,C_F\,\left(-\frac{3}{\eps}-4\right)+ \mathcal{O}(\epsilon)\;.
\end{equation}

The coefficients $\tilde{b}_i$ in \eqref{eq:coeff1_bare},
\eqref{eq:coeff2_bare} contain integrals $I_{r,s}(\hat{s},\hat{t},m_h^2,m_t^2)$,
where $r$ denotes the number of propagators in the denominator and $s$
denotes the number of propagators in the numerator and therefore defines the
tensor rank of the integral.
The integrals have mass dimension 
$[I_{r,s}] = D\,L-2r+2s$, with $L$ the number of loops. We may therefore factor a dimensionful parameter $M$ out of each integral such that they depend only on dimensionless ratios
\begin{equation}
I_{r,s}(\hat{s},\hat{t},m_h^2,m_t^2) = (M^2)^{-L\epsilon} (M^2)^{2L - r+ s} I_{r,s}\left(\frac{\hat{s}}{M^2},\frac{\hat{t}}{M^2},\frac{m_h^2}{M^2},\frac{m_t^2}{M^2}\right).
\end{equation}
The renormalized amplitude may then be written as
\begin{align}
\amp^{\rm{virt}} &=\prod_{n_g} Z_A^\frac12\,\amp_B\left(a_0\to a\, Z_{a} \, \left(\mur^2/\mub^2 \right)^\epsilon,
 m_{t_0}^2 \to m_t^2 + a \,\delta m_t^2\right)\nn\\
 &= a \amp^{(1)} + a^2 ( \frac{n_g}{2}\,\delta Z_A + \delta Z_a) \amp^{(1)} + a^2 \delta m_t^2 \amp^{ct,(1)} + a^2 \amp^{(2)} + O(a^3), \label{eq:amp_ren} \\
\amp^{(1)} &= \left( \frac{\mur^2}{M^2} \right)^{\epsilon} \left[ b_0^{(1)} + b_1^{(1)} \epsilon + b_2^{(1)} \epsilon^2 + \mathcal{O}(\epsilon^3) \right], 
\label{eq:A1ren} \\
\amp^{ct,(1)} &= \left( \frac{\mur^2}{M^2} \right)^{\epsilon} \left[ c_0^{(1)} + c_1^{(1)} \epsilon + \mathcal{O}(\epsilon^2) \right], \\
\amp^{(2)} &= \left( \frac{\mur^2}{M^2} \right)^{2\epsilon} \left[ \frac{b_{-2}^{(2)}}{\epsilon^2} + \frac{b_{-1}^{(2)}}{ \epsilon} + b_0^{(2)} + \mathcal{O}(\epsilon) \right],
\label{eq:A2ren}
\end{align}
where
\begin{align}
\tilde{b}^{(L)} &= (M^2)^{-L\epsilon} b^{(L)} \; , \;
\tilde{c}^{(L)} = (M^2)^{-L\epsilon} c^{(L)}.
\end{align}
Since $\delta m_t^2$ contains poles of $\mathcal{O}(\epsilon^{-1})$ the coefficient $c$ of the top mass counter-term must be expanded to $\mathcal{O}(\epsilon)$. It is obtained by the insertion of a mass counter-term into the heavy quark propagators,
\be
\Pi_{ab}^{\delta m}(p) = \frac{i\delta_{ac}}{\not p-m}\,\left(-i\delta m\right) \,\frac{i\delta_{cb}}{\not p-m}\;,
\ee
where $a,b,c$ are colour indices in the fundamental representation.
Alternatively, the mass counter-term can be obtained by taking the derivative of the one-loop amplitude with respect to $m$.

The coefficients $b$ and $c$ in \eqref{eq:amp_ren} are calculated
numerically. We have extracted the
dependence of the coefficients on the renormalization scale and
introduced a dependence on a new scale, $M$, which we keep fixed in
our numerics. 

For the infrared singularities stemming from the unresolved real
radiation, we use the Catani-Seymour subtraction
scheme~\cite{Catani:1996vz}.
The infrared poles of the virtual amplitude are cancelled after
combination with the $\mathbf{I}$-operator, which is given by 
\bea 
{\bom I}_{gg}(\eps) &=& \frac{\als}{2\pi}\,\frac{(4\pi)^\eps}{\Gamma(1-\eps)}\left(\frac{\mur^2}{\shat}\right)^\eps\cdot 2\cdot \left\{\frac{C_A}{\eps^2}+\frac{\beta_0}{2\eps} -C_A\frac{\pi^2}{3}+\frac{\beta_0}{2}+K_g\right\}\;,
\eea
where $K_g$ is also defined by the Catani-Seymour subtraction scheme~\cite{Catani:1996vz}.
Inserting the $\mathbf{I}$-operator into the Born amplitude leads to\footnote{
  The factor of $\frac{1}{2}$ is necessary to cancel the factor of $2$ obtained from squaring
  $\amp^{\rm{virt}}+\amp^{\rm{IR\,ct}}$ to get the cross section.}
\begin{align}\label{eq:Iop}
  \amp^{\rm{IR\,ct}}&= \frac{1}{2} \cdot {\bom I}_{gg}(\eps) \otimes \amp^{(1)}\nn\\
  &=a^2\left(\frac{\mur^2}{\shat}\right)^\eps\left(\frac{\mur^2}{M^2}\right)^{\eps}
\left(1-\eps^2\frac{\pi^2}{12}\right)\left\{\frac{2\,C_A}{\eps^2}+\frac{\beta_0}{\eps} -C_A\frac{2\pi^2}{3}+\beta_0+2K_g\right\}\nn\\
&\times\left[ b_0^{(1)} + b_1^{(1)} \epsilon + b_2^{(1)} \epsilon^2 \right]\;,
\end{align}
where we again have extracted a factor $(M^2)^{-\epsilon}$ from the integrals 
contained in the one-loop amplitude.
Using \eqref{eq:amp_ren} and \eqref{eq:Iop} we therefore have
\begin{align}
\amp^{\rm{virt}}+\amp^{\rm{IR\,ct}}&= a \amp^{(1)} + 
a^2 \,\left( \frac{\mur^2}{M^2} \right)^{\epsilon}\Bigg\{ 
\delta m_t^2 \left[ c_0^{(1)} + c_1^{(1)} \eps \right] 
+\left( \frac{\mur^2}{M^2} \right)^{\epsilon}
\left[ \frac{b_{-2}^{(2)}}{\epsilon^2} + \frac{b_{-1}^{(2)}}{ \epsilon} + b_0^{(2)} \right]\nn\\
&+
\left[ b_0^{(1)} + b_1^{(1)} \epsilon + b_2^{(1)} \epsilon^2 \right]\,
\left[
\left(\frac{\mur^2}{\shat}\right)^\eps\left\{\frac{2\,C_A}{\eps^2}+\frac{\beta_0}{\eps}+\mbox{fin.}\right\}\, -\frac{\beta_0}{\eps}\right]  \Bigg\} \nn\\
&\nn\\
&= a \amp^{(1)} + 
a^2 \,\left( \frac{\mur^2}{M^2} \right)^{\epsilon}\Bigg\{ 
\frac{1}{\eps^2}\left[2\,C_A \,b_0^{(1)}  +b_{-2}^{(2)}\right]\nn\\
&+\frac{1}{\eps}\left[2\,C_A \,b_0^{(1)}\ln\left(\frac{\mur^2}{\shat}\right)+b_{-2}^{(2)}\ln\left(\frac{\mur^2}{M^2}\right)+
b_{-1}^{(2)}-6m_t^2C_F\,c_0^{(1)}+2\,C_A \,b_1^{(1)}\right]\nn\\
&+b_0^{(1)}\,\beta_0\ln\left(\frac{\mur^2}{\shat}\right)+
\ln\left(\frac{\mur^2}{M^2}\right)\,b_{-1}^{(2)}-
\ln\left(\frac{\mur^2}{m_t^2}\right)\,6m_t^2C_F\,c_0^{(1)}+2\,C_A \,b_1^{(1)}\ln\left(\frac{\mur^2}{\shat}\right)\nn\\
&+C_A \,b_0^{(1)}\ln^2\left(\frac{\mur^2}{\shat}\right)+\frac{b_{-2}^{(2)}}{2}\ln^2\left(\frac{\mur^2}{M^2}\right)\nn\\
&+\mbox{finite  non-logarithmic terms}\Bigg\}\label{eq:renlogarithms}\,.
\end{align}

By construction the double pole in $\epsilon$ must vanish, thus \eqref{eq:renlogarithms} implies
\be
b_{-2}^{(2)}=-2\,C_A \,b_0^{(1)}\;.
\ee
Substituting the above relation back into \eqref{eq:renlogarithms} we 
see that the dependence on the renormalization scale $\mur$ cancels in
the single pole term. 
The dependence of the cross section on the factorization scale is encoded in the $\mathbf{P}$ and $\mathbf{K}$ terms of 
the Catani-Seymour framework~\cite{Catani:1996vz}.

\vspace*{3mm}
{\bf Integration of the two-loop amplitude}

\vspace*{3mm}

To evaluate the two-loop integrals appearing in the amplitude we first apply sector decomposition as implemented in {\sc SecDec}. 
In the Euclidean region sector decomposition resolves singularities in the regulator $\epsilon$, 
leaving only finite integrals over the Feynman parameters which can be evaluated numerically. 
In the physical region we treat the integrable singularities by 
contour deformation~\cite{Soper:1999xk,Binoth:2005ff,Nagy:2006xy,Borowka:2012yc}. 
To obtain the differential cross section we have to evaluate integrals at phase space points  very close to  threshold, 
where no special treatment was necessary but numerical convergence was considerably harder to achieve.

After sector decomposition each loop-integral $I_j$ can be written as a sum over sectors~$s$ which have a Laurent series starting at some $\eps$-order  $e_{s}^{\mathrm{min}}$
  \begin{align}
    I_j(\epsilon) = \sum_s\sum_{e>e_s^{\mathrm{min}}} \epsilon^e I_{j,s,e}.
    \label{<+label+>}
  \end{align}
  For the numerical evaluation of the amplitude we structured the code such that the integrand of each sector-decomposed loop integral $I_{j,s,e}$ is stored along with the Laurent series of their coefficients $a_j$ appearing in the expressions for the amplitudes~\eqref{eq:A1ren}-\eqref{eq:A2ren}. 
  E.g. at two-loop we write the amplitude as
\begin{align}
  \amp^{(2)} &= \left( \frac{\mur^2}{M^2} \right)^{2\epsilon} \,\sum_{j,s,e} I_{j,s,e} \cdot a_j(\epsilon)
  \label{eq:amp2}
\end{align}
and store $a_j$ as a vector containing the coefficients of $I_j$ in the expressions for $b_k^{(2)}$, leading to the amplitude structure given in Eq.~\eqref{eq:A2ren}.

Structuring the code this way allows us to
dynamically set the number of sampling points used for each integral according to its contribution to the amplitude. After calculating each integral with a fixed number of sampling points, we 
assume that the integration error $\Delta_j$ of the integrals scales as $\Delta_j\propto t_j^{-\alpha}$ with the integration time $t_j$. To efficiently calculate the results $b_{k}^{(i)}$ with a given relative accuracy $\varepsilon_k^{(i)} = \Delta_{k}^{(i)}/b_k^{(i)}$, we estimate the required number of sampling points for each integral such that the total time 
\begin{align}
  T_k^{(i)}=\sum_j t_j + \bar\lambda\left( (\Delta_k^{(i)})^2-\sum_j (\Delta_{j,k}^{(i)})^2 \right)
  \label{eq:timek}
\end{align}
is minimal. $\Delta_{j,k}^{(i)}$ is the error estimate of integral $I_j$ including its coefficients in $b_k^{(i)}$ and $\bar\lambda$ is a Lagrange multiplier.  
Since the loop integrals can contribute to several results~$b_k^{(i)}$, we apply the above optimization formula for each required order in $\epsilon$ and for both form factors. For each integral, we then use the maximum of the estimated number of required sampling points.
Instead of directly evaluating each integral with the calculated number of sampling points, we limit the number of new sampling points and iterate this procedure to reach the desired accuracy, updating the estimated number of sampling points after each iteration. The desired accuracy for the finite part of the two-loop amplitude ($\varepsilon_0^{(2)}$) is set to 3\% for form factor $F_1$ and (depending on the ratio $F_2/F_1$) to a value of 5-20\% for form factor $F_2$.

For the integration we use a quasi-Monte Carlo method based on a rank-one lattice rule~\cite{Li:2015foa,QMCActaNumerica,nuyens2006fast}. 
For suitable integrands, this rule provides a convergence rate of $\mathcal{O}(1/n)$ as opposed to Monte Carlo or adaptive 
Monte Carlo techniques, such as {\sc Vegas}~\cite{Lepage:1980dq}, which converge $\mathcal{O}(1/\sqrt{n})$, where $n$ is the number 
of sampling points. 
While we observe a convergence rate of $\mathcal O(1/n)$ for most of the integrals, the convergence of some integrals is worse and we therefore assume a scaling of $\Delta_j(t_j)$ with exponent $\alpha=0.7$  when estimating the number of required sampling points.

The integration rule is implemented in {\sc OpenCL\,1.1} and a further ({\sc OpenMP} threaded) {\sc C++} 
implementation is used as a partial cross-check. The 913 
phase-space points at 14 TeV (1029 phase-space points at 100 TeV) used for the current publication
were computed with $\sim$16 dual {\sc Nvidia Tesla K20X} GPGPU nodes. 
More details on the numeric evaluation of the amplitudes can be found in Refs.~\cite{Stephen:LL2016,Matthias:LL2016}.

\subsubsection{Real radiation}
\label{sec:real}


\begin{figure}
\includegraphics[width=1\textwidth,angle=0]{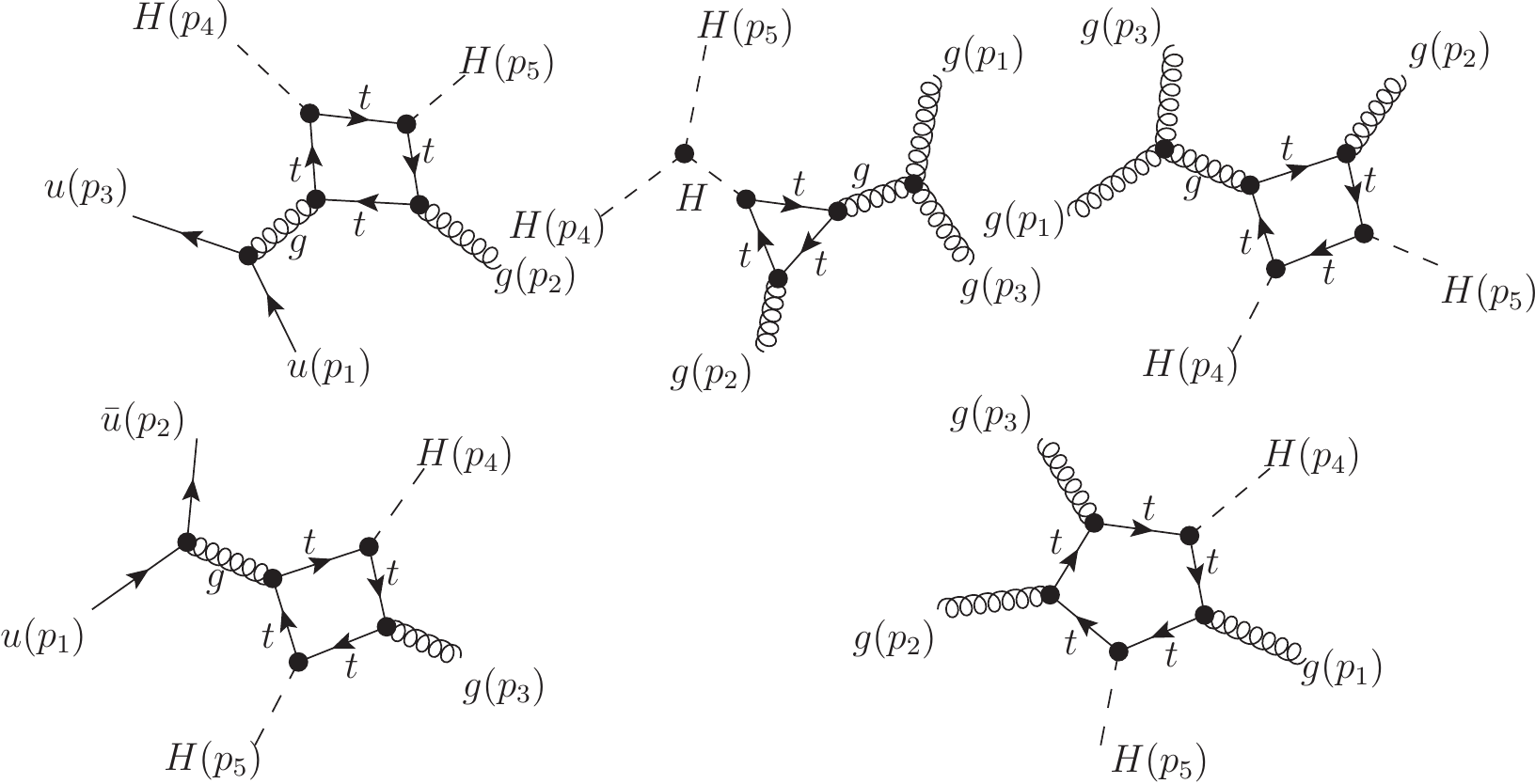}
\caption{Examples of diagrams contributing to the real radiation part at NLO.
The diagrams in the second row do not lead to infrared singularities.\label{fig:NLOrealdiagrams}}
\end{figure}

As we calculate a process which is loop-induced, 
the NLO corrections involve two-loop
integrals. But, for the real part
only single-unresolved  radiation can occur. This means that 
a standard NLO infrared subtraction scheme can be used.
We use the Catani-Seymour dipole formalism~\cite{Catani:1996vz}, 
combined with a phase space restriction parameter $\alpha$ to restrict the dipole subtraction 
to a limited region, as suggested in Ref.~\cite{Nagy:2003tz}.

There are four partonic channels for the real radiation contribution to the cross section:
\be
\sigma^{\mathrm{r}}(gg\to hh+g), \sigma^{\mathrm{r}}(gq\to hh+q), \sigma^{\mathrm{r}}(g\bar{q}\to hh+\bar{q}), \sigma^{\mathrm{r}}(q\bar{q}\to hh+g)\;.
\ee 
Including all crossings, there are 78 real radiation
diagrams. Infrared singularities only originate from initial state
radiation, diagrams with extra gluons radiated from a heavy quark line
are infrared finite, which  implies that the  $q\bar{q}$ channel is  finite.
Example diagrams are depicted in Fig.~\ref{fig:NLOrealdiagrams}.

\subsection{Validation of the calculation and expansion in $1/m_t$}
\label{sec:checks}

\subsubsection{Expansion in $1/m_t^{2}$}

We have  calculated top mass corrections as an expansion in $1/m_t^{2}$
in the following way:
we write the partonic differential cross section as
\begin{align}\label{eq:mtexp}
  d\hat{\sigma}_{\text{exp},N} = \sum_{\rho=0}^N d\hat{\sigma}^{(\rho)} \left(\frac{\Lambda}{m_t}\right)^{2\rho},
\end{align}
where $\Lambda\in\left\{\sqrt{\hat{s}}, \sqrt{\hat{t}}, \sqrt{\hat{u}}, m_h\right\}$,
and determine the first few terms (up to $N=3$) of this asymptotic series.
The case $N=0$ reproduces to the usual effective theory approach, without the need to
calculate Wilson coefficients separately, however.

To generate the diagrams we again use {\sc qgraf}~\cite{Nogueira:1991ex}.
The generation and expansion of the amplitude in small external momenta is then performed using
{\sc q2e}/{\sc exp}~\cite{Harlander:1997zb,Seidensticker:1999bb} and leads to
two-loop vacuum integrals inserted into tree-level diagrams as well as one-loop vacuum integrals 
inserted into massless one-loop triangles. 
Whereas the vacuum integrals are evaluated with {\sc Matad}~\cite{Steinhauser:2000ry},
the massless integrals can be expressed in terms of a single one-loop bubble, 
which we achieve with the help of {\sc Reduze}~\cite{vonManteuffel:2012np}.
Again, the algebraic processing of the amplitude is done with {\sc Form}~\cite{Vermaseren:2000nd,Kuipers:2012rf}.

The exact and expanded matrix elements were combined in the following way:
a series expansion  for the virtual corrections was performed then rescaled with the exact born,
  \begin{align}\label{eq:V+I}
    d\sigma^V + d\sigma^{LO}(\eps) \otimes {\bom I}
    &\approx d\sigma_{\text{exp},N}^V \frac{d\sigma^{LO}(\eps)}{d\sigma_{\text{exp},N}^{LO}(\eps)}
    + d\sigma^{LO}(\eps) \otimes {\bom I} \nonumber\\
    &= \left(d\sigma_{\text{exp},N}^V + d\sigma^{LO}_{\text{exp},N}(\eps) \otimes {\bom I}\right) 
    \frac{d\sigma^{LO}(\eps)}{d\sigma_{\text{exp},N}^{LO}(\eps)} \nonumber\\
    &= \underbrace{\left(d\sigma_{\text{exp},N}^V + d\sigma^{LO}_{\text{exp},N}(\eps) \otimes {\bom I}\right) }_{\equiv V_N}
    \frac{d\sigma^{LO}(\eps=0)}{d\sigma_{\text{exp},N}^{LO}(\eps=0)} + \mathcal{O}\left(\eps\right).
  \end{align}
  The first identity is valid because the colour structure of the exact and the expanded LO cross
  section are identical, and the second because the sum in the bracket is finite.
  Thus one needs to know only the $\eps$ dependence of the expanded LO cross section in this
  approximation.

  There is some ambiguity when to do the rescaling, i.e. before or after the phase-space integration, 
  and convolution with the PDFs. We opt to do it on a fully differential level, 
  i.e. the rescaling is done for each phase-space point individually.


\subsubsection{Checks of the calculation}

We have verified  for all calculated phase space points that the 
coefficients of the poles in  $\epsilon$ are zero within the numerical uncertainties.
For a  randomly chosen sample of phase-space points we have calculated the pole coefficients 
with higher accuracy and obtained a median cancellation of five digits.

Our implementation of the virtual two-loop amplitude has been checked to be invariant under the interchange of $\hat{t}$ and $\hat{u}$ 
at various randomly selected phase-space points. 
Single Higgs boson production has been re-calculated with the same setup for the virtual corrections and 
compared to the results obtained with the program {\sc Sushi}~\cite{Harlander:2012pb}. 
Further, the one-loop amplitude has been computed 
using an identical framework to the two-loop amplitude and has been checked against the result of Ref.~\cite{Glover:1987nx}.

\begin{figure}
\centering
\includegraphics[width=0.75\textwidth]{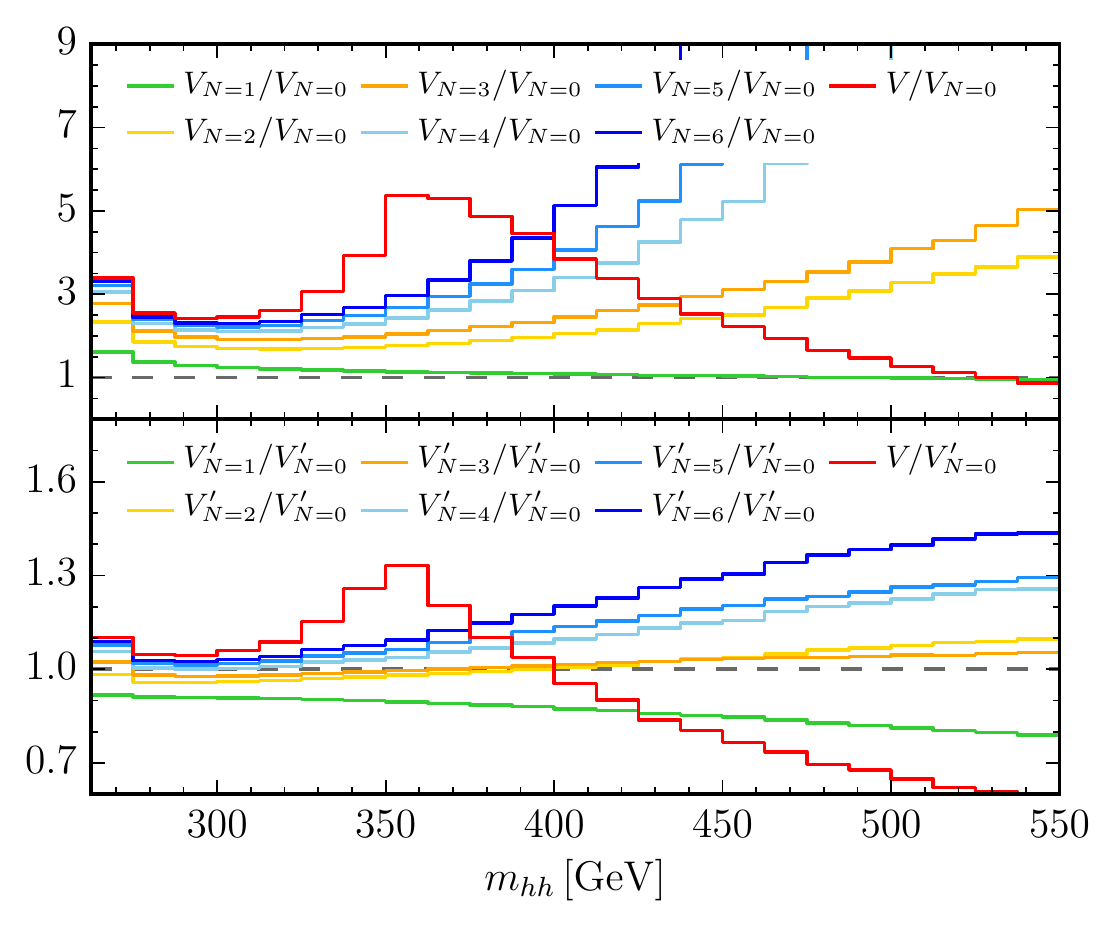}
\caption{Comparison of the virtual part as defined in Eq.~(\ref{eq:V+I}) with full top-quark mass dependence to
  various orders in a $1/m_t^{2}$ expansion. $V^\prime_N$ denotes the
  Born-improved HEFT result to order $N$ in the $1/m_t^{2}$ expansion,
  i.e. $V^\prime_N=V_N\,B_{FT}/B_N$.
The results for the orders $N=4,5,6$ have been provided to us by Jens Hoff~\cite{Grigo:2015dia}.\label{fig:ampexpand_TZJH}}
\end{figure}

As a further cross-check we 
have also calculated top mass corrections as an expansion in $1/m_t^{2}$ as explained above.
We have also compared to results provided to us by Jens Hoff for the orders $N=4,5,6$ in the expansion above, 
worked out in ~\cite{Grigo:2015dia}. The result of the comparison is
shown in Fig.~\ref{fig:ampexpand_TZJH}.
One can see that below the $2m_t$ threshold, where agreement is to be
expected, the expansion converges towards the full result.

The computation of the mass counter-term diagrams has been cross-checked by expanding the one-loop amplitude about the bare top mass
\begin{align}
\amp^{(1)}_B(m_t^2) &= \amp^{(1)}_B(m_{t_0}^2) - a \delta m_t^2 \left. \left( \frac{\partial}{\partial m_t^2} \amp^{(1)}_B(m_t^2) \right) \right|_{m_{t_0}^2} \nonumber \\
&= \amp^{(1)}_B(m_{t_0}^2) - a \delta m_t^2 \amp^{ct,(1)}_B(m_{t_0}^2),
\end{align}
where $\amp^{ct,(1)}$ is the one-loop top quark mass counter-term. 

On the real radiation side, we have verified the independence of the amplitude from the phase space restriction parameter $\alpha$.
We have also varied the technical cut 
$p_{T}^{\rm{min}}$ in the range $10^{-2} \leq p_{T}^{\rm{min}}/\sqrt{\hat{s}} \leq 10^{-6}$ to verify that the
contribution to the total cross section is stable and independent of
the cut within the numerical accuracy.

Further, we have compared to the results of Ref.~\cite{Maltoni:2014eza} 
for the Born-improved HEFT and FT$_{approx}$ approximations  
and found agreement within the numerical uncertainties~\cite{YR4}.

\section{Phenomenological results}
\label{sec:results}


\subsection{Setup and total cross sections}

We use the PDF4LHC15\_nlo\_100\_pdfas~\cite{Butterworth:2015oua,CT14,MMHT14,NNPDF} parton distribution functions,
along with the corresponding value for $\alpha_s$ for both the NLO and
the LO calculation.
The masses have been set to $m_h=125$\,GeV, $m_t=173$\,GeV,
and the top quark width has been set to zero.
We use no cuts except a technical cut in the real radiation of
$p_{T}^{\rm{min}}=10^{-4}\,\sqrt{\hat{s}}$.
The scale variation bands are the result of a 7-point scale variation~\cite{YR4} around the central scale $\mu_0 = m_{hh}/2$,
with $\mu_{R,F}=c_{R,F}\,\mu_0$, where 
$c_R,c_F\in \{2,1,0.5\}$, except that the extreme variations $(c_R,c_F)=(2,0.5)$ and $(c_R,c_F)=(0.5,2)$
are omitted.
The values we obtain for the total cross sections are shown in Table~\ref{tab:sigtot}.
The full NLO result has a statistical uncertainty of 0.3\% at 14 TeV
(0.16\% at 100 TeV) stemming from the phase space integration and an
additional uncertainty stemming from the numerical integration of the
virtual amplitude
of 0.04\% at 14 TeV and 0.2\% at 100 TeV.
These uncertainties are not included in Table~\ref{tab:sigtot}, where only scale variation uncertainties are shown.

\begin{table}
\begin{center}
\begingroup
\renewcommand\tabcolsep{8pt}
\begin{tabular}{|c|c|c|c|c|}
\hline
$\sqrt{s}$& LO& B-i.\,NLO HEFT& NLO FT$_{approx}$& NLO \\
\hline
14 TeV& 19.85$^{+27.6\%}_{-20.5\%}$& 38.32$^{+18.1\%}_{-14.9\%}$&34.26$^{+14.7\%}_{-13.2\%}$&32.91$^{+13.6\%}_{-12.6\%}$ \\
100 TeV& 731.3$^{+20.9\%}_{-15.9\%}$& 1511$^{+16.0\%}_{-13.0\%}$& 1220$^{+11.9\%}_{-10.7\%}$& 1149$^{+10.8\%}_{-10.0\%}$ \\
\hline
\end{tabular}
\endgroup
\caption{Total cross sections at various centre of mass energies (in femtobarns). The uncertainty in percent is from 7-point scale variations as explained in the text.
The central scale is $m_{hh}/2$. We used $m_t=173$\,GeV, $m_h=125$\,GeV.
The PDF set is {\sc PDF4LHC15}\_nlo\_100\_pdfas.\label{tab:sigtot}
}
\end{center}
\end{table}

\subsection{NLO distributions}

In this section we show differential distributions at $\sqrt{s}=14$\,TeV and
$\sqrt{s}=100$\,TeV for various observables and compare to the
approximate results in order to assess the effect of the full top
quark mass dependence at NLO. Results which are obtained within the effective field theory approach 
without reweighting by the leading order results in the full theory are always denoted by ``basic HEFT", 
while ``B-i.\,NLO HEFT" stands for the Born-improved NLO HEFT result, where the NLO corrections have been calculated in the 
$m_t\to\infty$ limit and then a reweighting factor $B_{FT}/B_{HEFT}$ is applied (on differential level, 
$B_{FT}$ stands for the Born amplitude squared in the full theory).\\
We decided to take the same bin sizes as in
Ref.~\cite{deFlorian:2016uhr}, such that the differences to the
effective theory results can be exhibited most clearly.
\begin{figure}
\centering
\begin{subfigure}{0.49\textwidth}
\includegraphics[width=\textwidth]{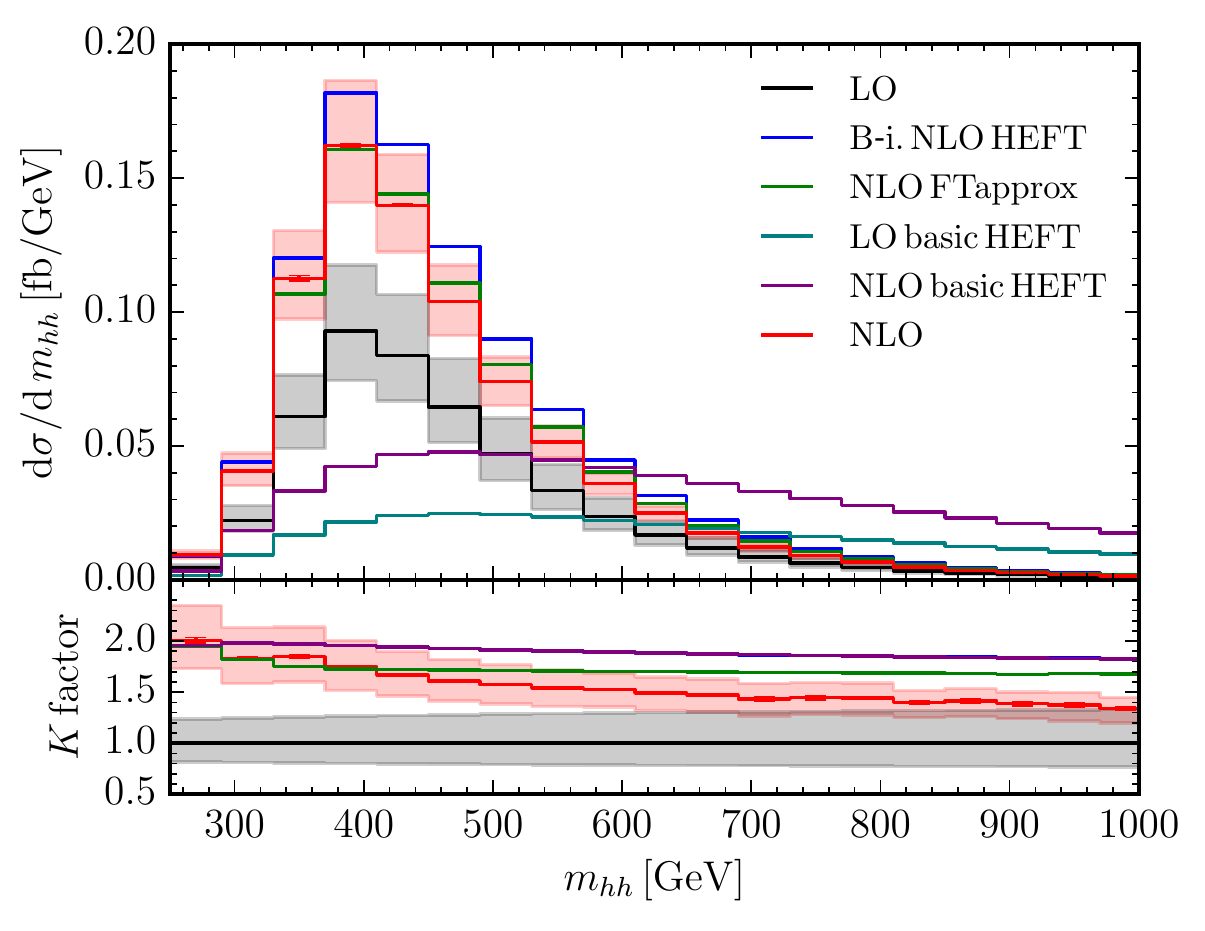}
\caption{14 TeV\label{subfig:mhh}}
\end{subfigure}
\begin{subfigure}{0.49\textwidth}
\includegraphics[width=\textwidth]{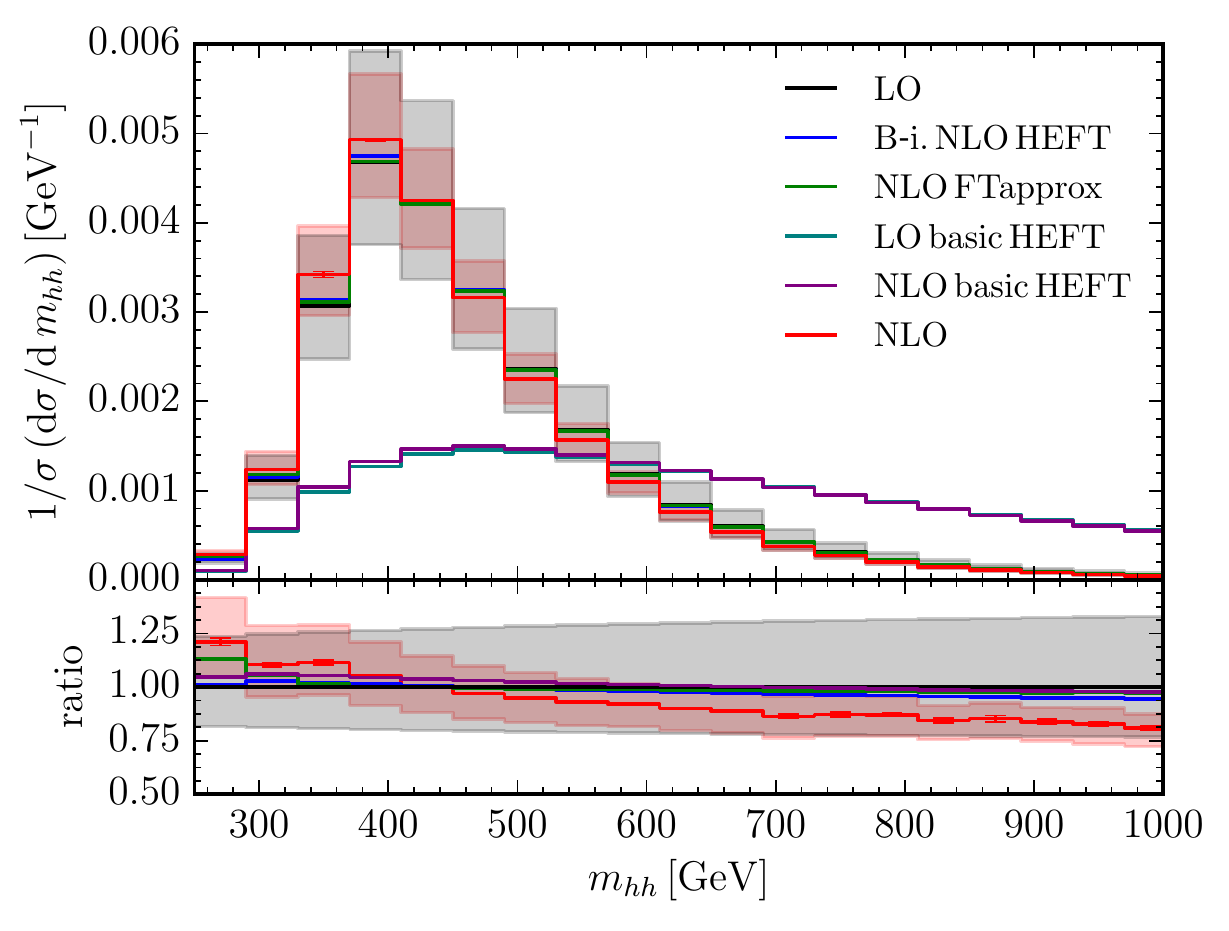}
\caption{14 TeV, normalised}
\end{subfigure}
\begin{subfigure}{0.49\textwidth}
\includegraphics[width=\textwidth]{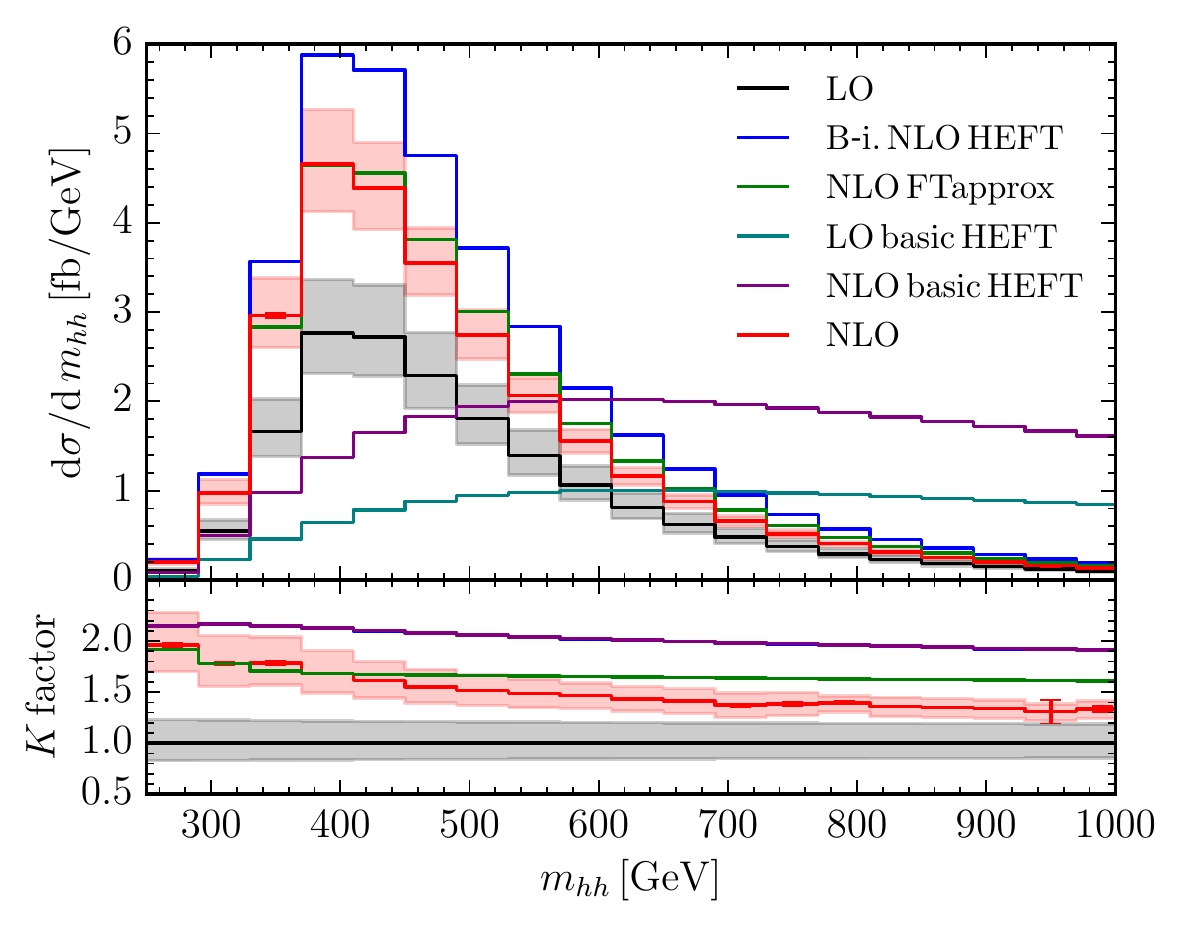}
\caption{100 TeV}
\end{subfigure}
\begin{subfigure}{0.49\textwidth}
\includegraphics[width=\textwidth]{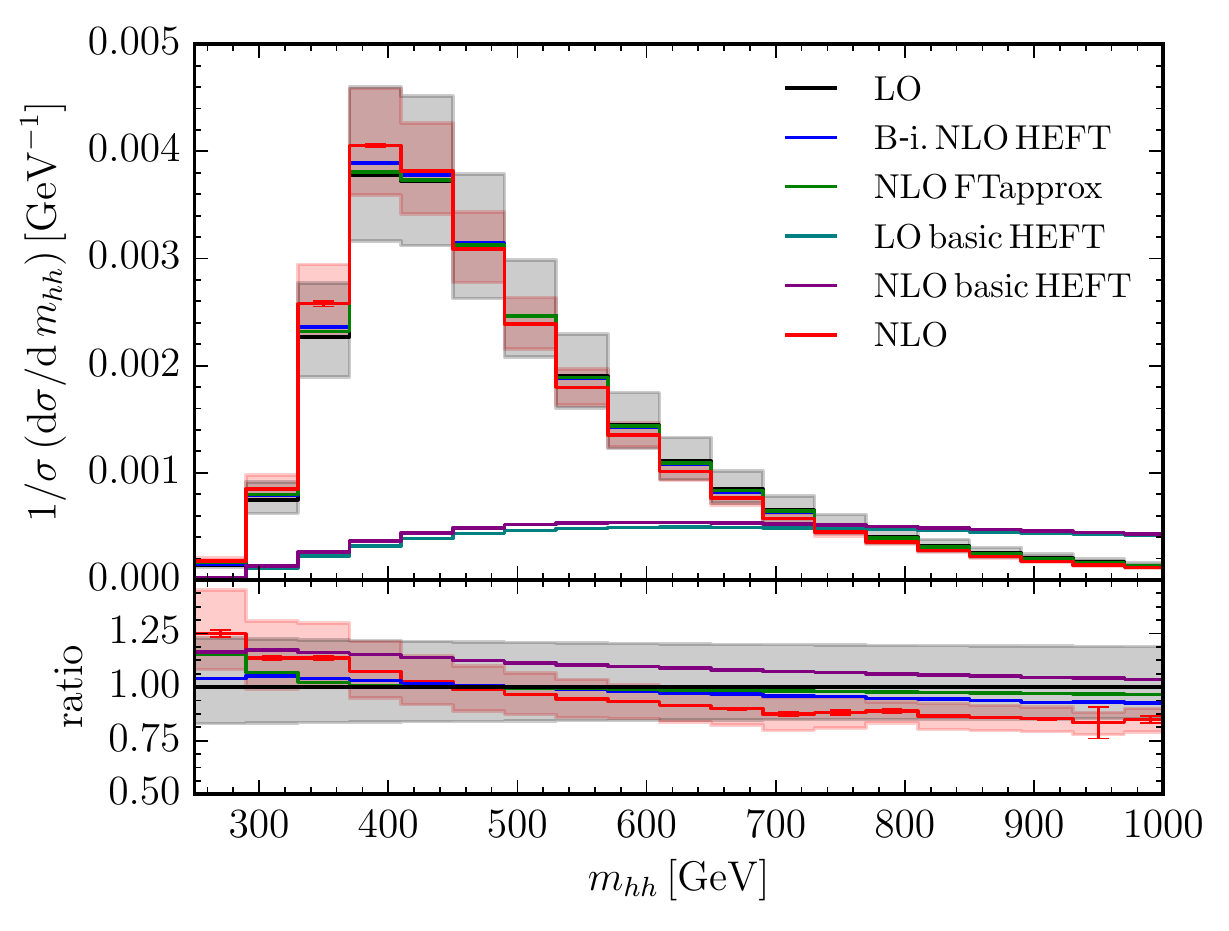}
\caption{100 TeV, normalised}
\end{subfigure}
\caption{Higgs boson pair invariant mass distribution $m_{hh}$  at 
  $\sqrt{s}=14$\,TeV and $\sqrt{s}=100$\,TeV  
  for absolute values (left panels) and  normalised to the corresponding total cross section (right panels).\label{fig:mhh}}
\end{figure}
In Fig.~\ref{fig:mhh} we show the Higgs boson pair invariant mass distribution $m_{hh}$ at 
$\sqrt{s}=14$\,TeV and $\sqrt{s}=100$\,TeV, comparing the full NLO result to various approximations. 
In particular, we compare to the ``basic HEFT" approximation at $\sqrt{s}=14$\,TeV, showing that it 
fails to describe the distribution.
Comparing the results at 14 TeV and 100 TeV, we observe that the differences of the full NLO result to the Born-improved HEFT
and also to the FT$_{approx}$ result are amplified at 100 TeV, as expected, as the HEFT approximation does not 
have the correct high energy behaviour. This scaling behaviour   will be discussed more in detail below.
We also see that the K-factor is far from being uniform for the $m_{hh}$ distribution, while the ``basic HEFT" results 
suggest a uniform K-factor.

\begin{figure}
\centering
\begin{subfigure}{0.49\textwidth}
\includegraphics[width=\textwidth]{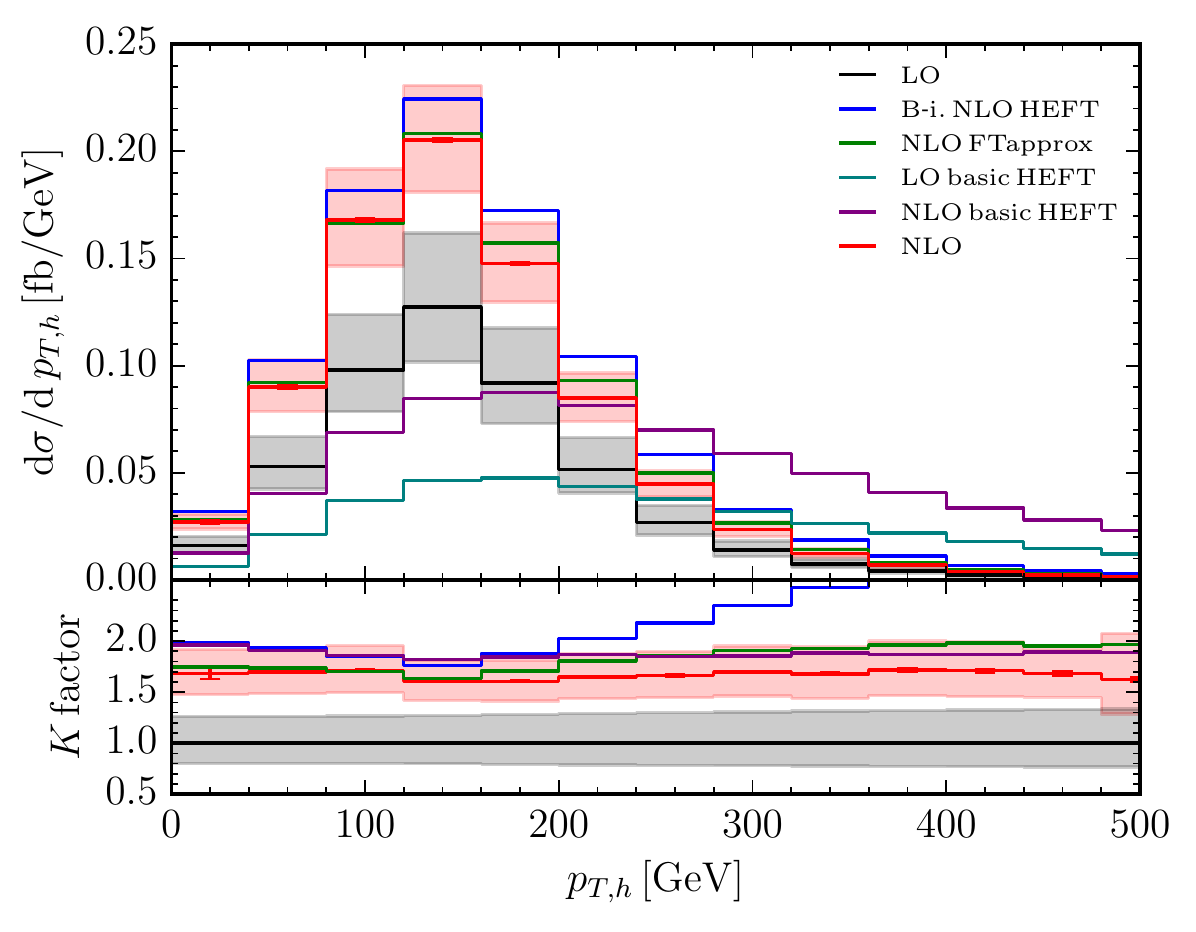}
\caption{14 TeV\label{subfig:pth14}}
\end{subfigure}
\begin{subfigure}{0.49\textwidth}
\includegraphics[width=\textwidth]{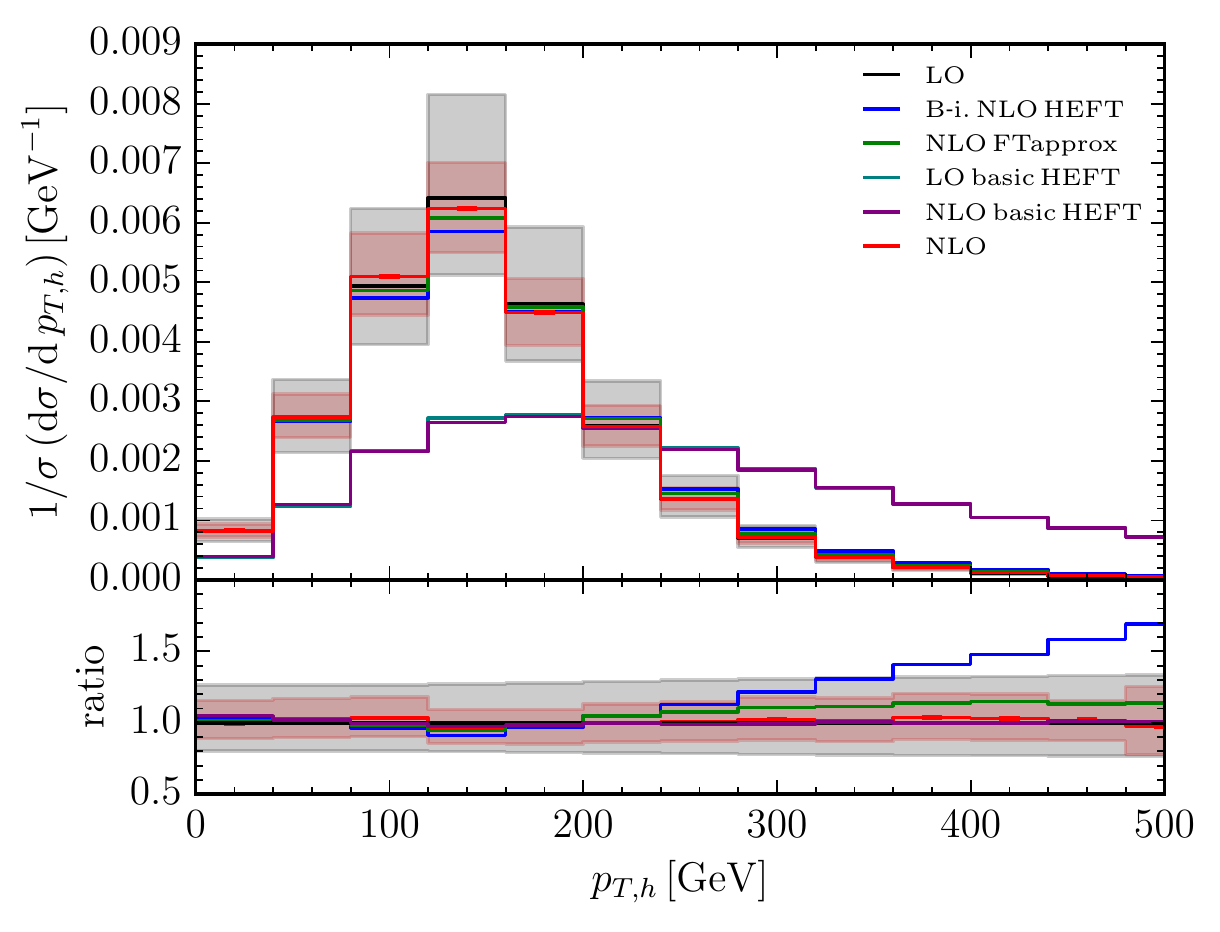}
\caption{14 TeV, normalised}
\end{subfigure}
\begin{subfigure}{0.49\textwidth}
\includegraphics[width=\textwidth]{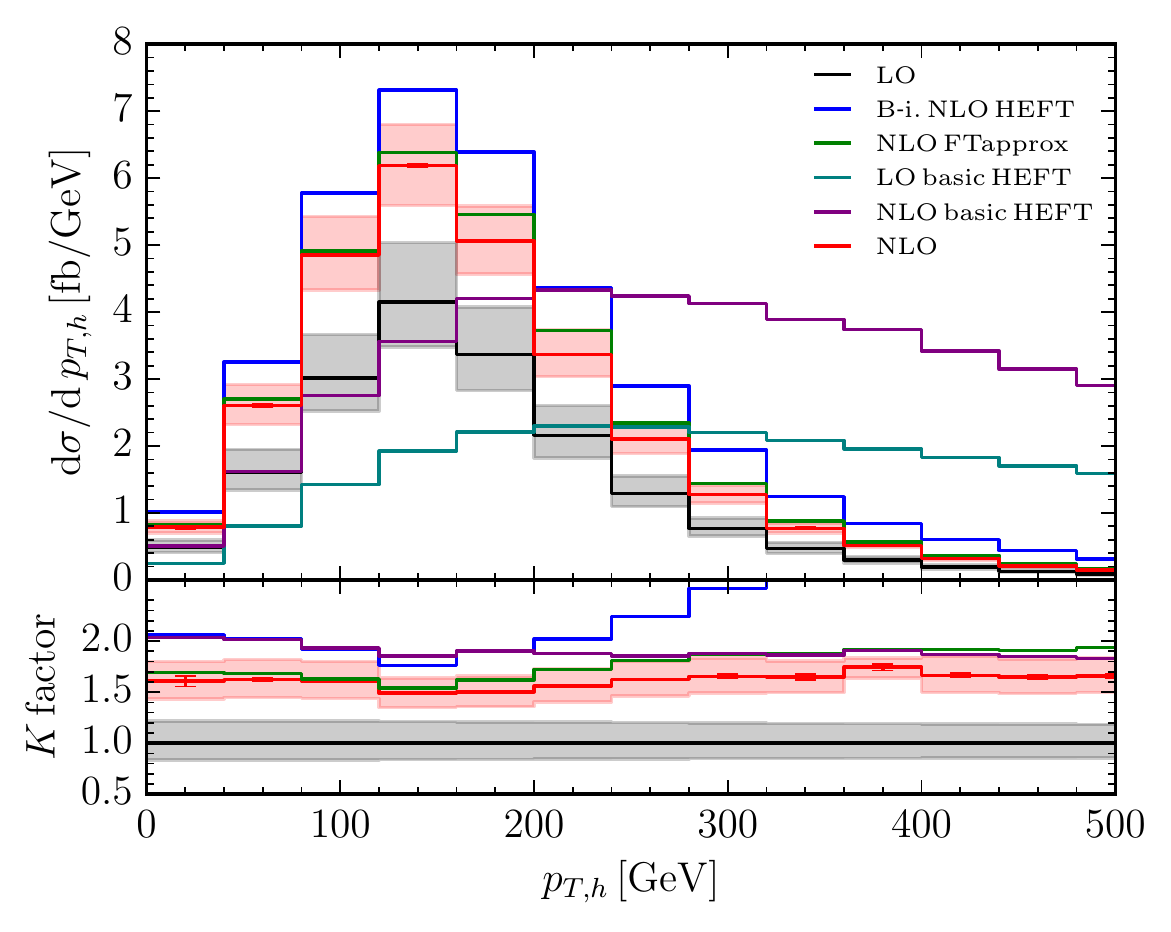}
\caption{100 TeV}
\end{subfigure}
\begin{subfigure}{0.49\textwidth}
\includegraphics[width=\textwidth]{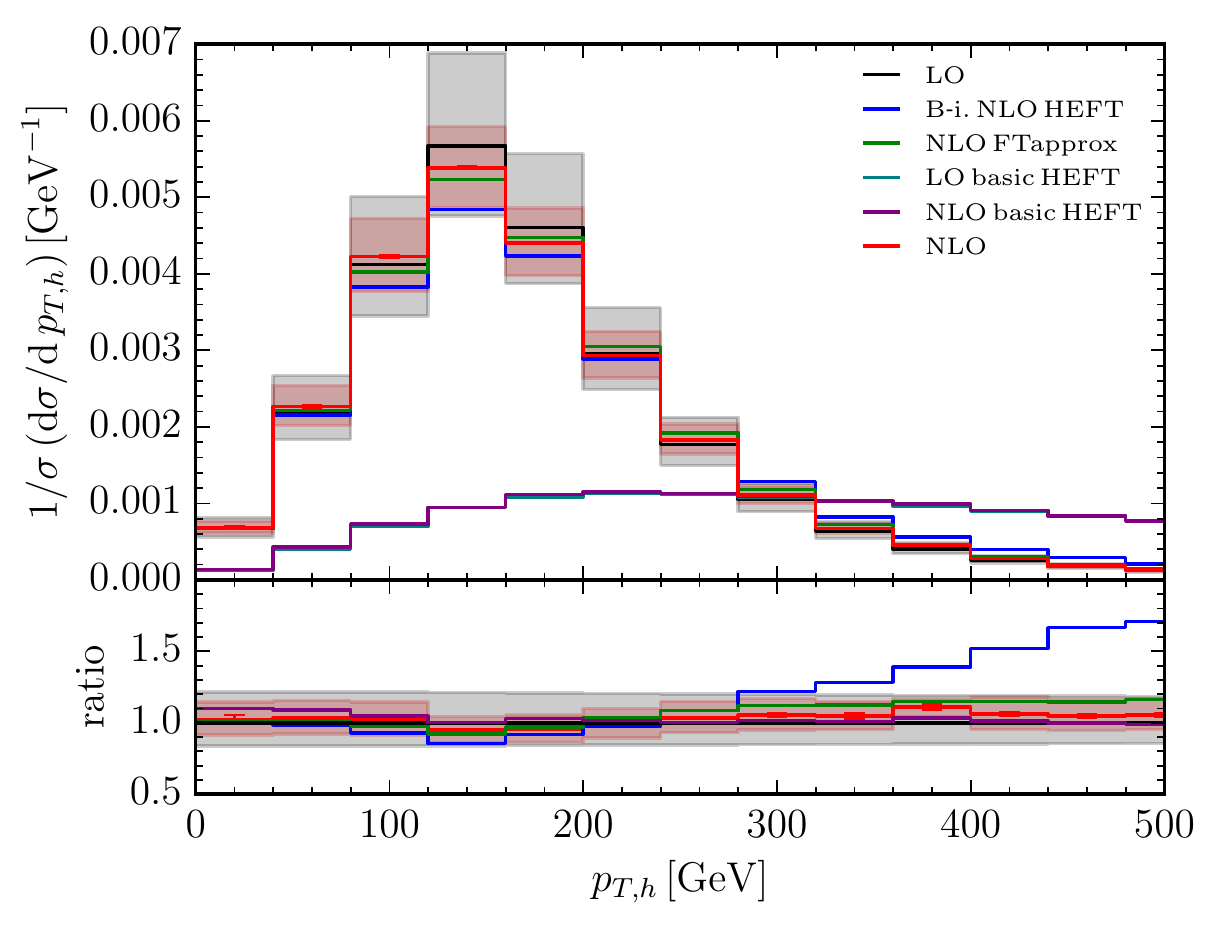}
\caption{100 TeV, normalised}
\end{subfigure}
\caption{Transverse momentum distribution of (any) Higgs boson 
at $\sqrt{s}=14$\,TeV and $\sqrt{s}=100$\,TeV.\label{fig:pth}}
\end{figure}
\begin{figure}
\centering
\begin{subfigure}{0.49\textwidth}
\includegraphics[width=\textwidth]{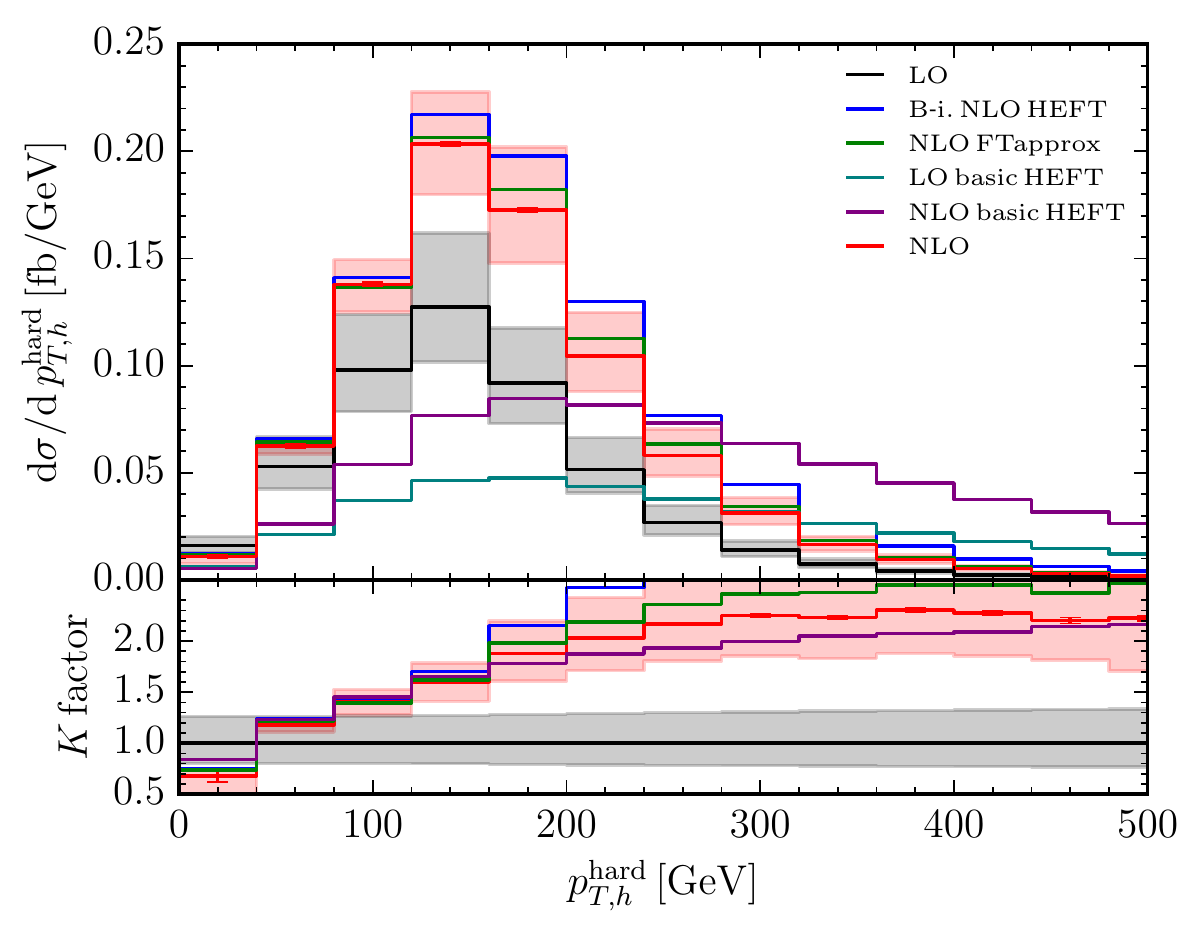}
\caption{14 TeV, leading $p_T$}
\end{subfigure}
\begin{subfigure}{0.49\textwidth}
\includegraphics[width=\textwidth]{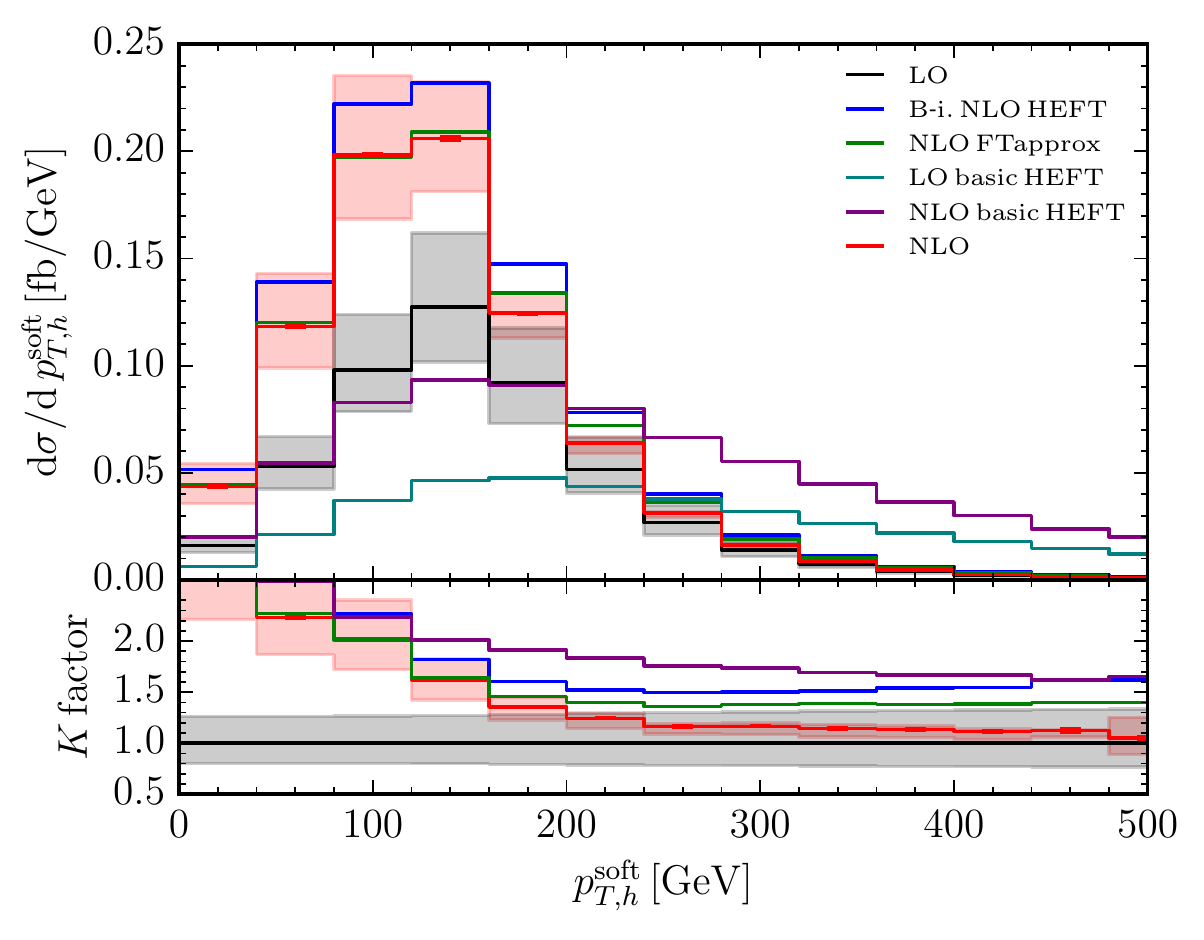}
\caption{14 TeV, subleading $p_T$}
\end{subfigure}
\begin{subfigure}{0.49\textwidth}
\includegraphics[width=\textwidth]{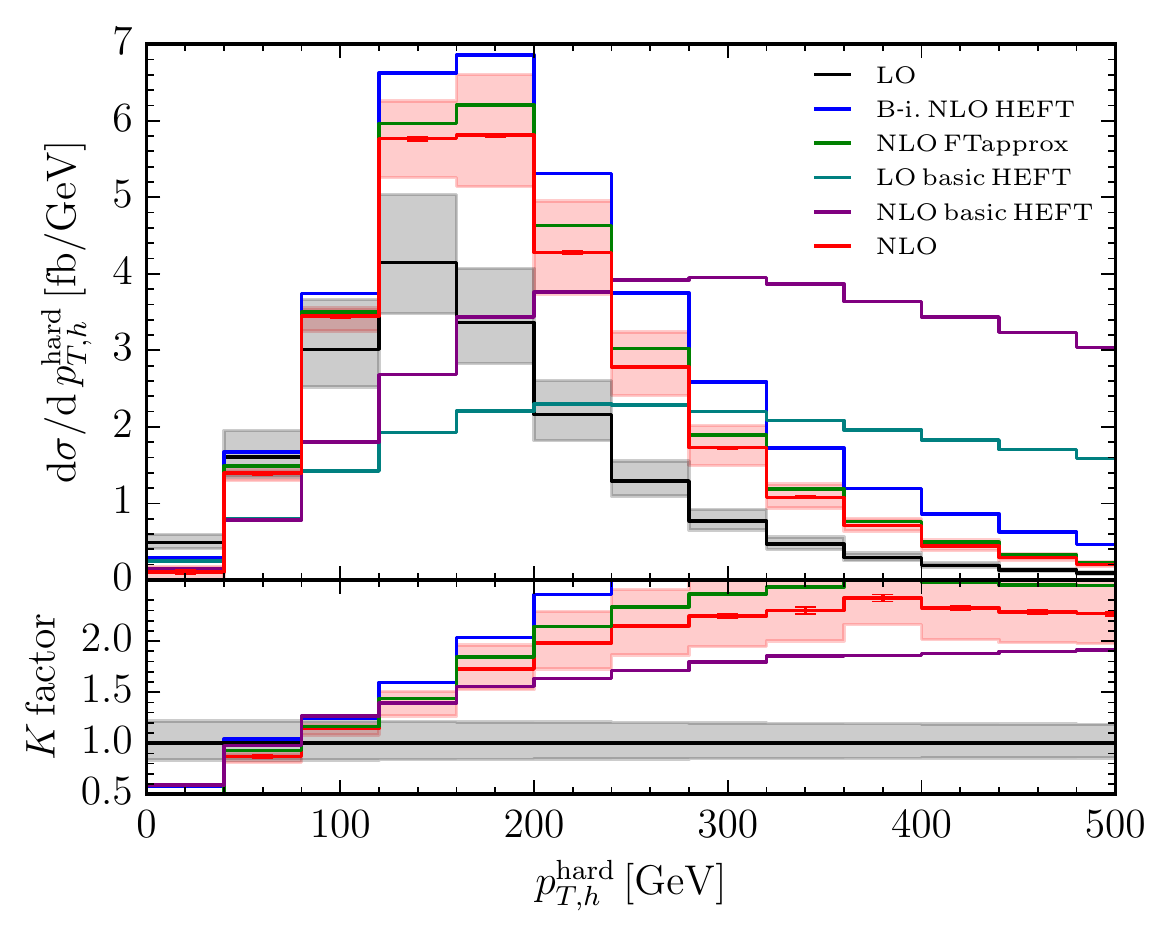}
\caption{100 TeV, leading $p_T$}
\end{subfigure}
\begin{subfigure}{0.49\textwidth}
\includegraphics[width=\textwidth]{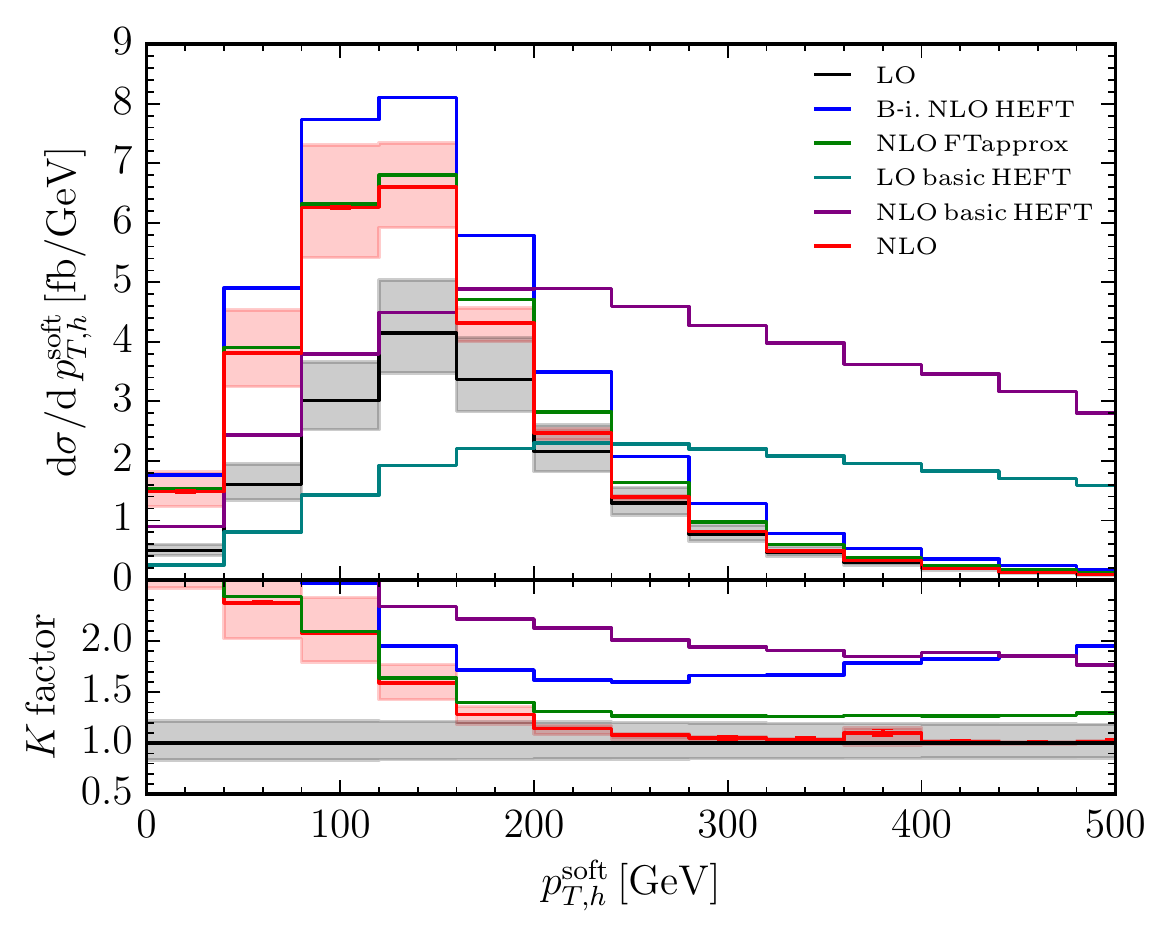}
\caption{100 TeV, subleading $p_T$}
\end{subfigure}
\caption{Transverse momentum distribution of the leading-$p_T$ Higgs
  boson (left panels) and the subleading-$p_T$ Higgs boson (right panels) at $\sqrt{s}=14$\,TeV and $\sqrt{s}=100$\,TeV.\label{fig:pth_hard_soft}}
\end{figure}

The  $p_{T,h}$ distribution  shown in Fig.~\ref{fig:pth} denotes the distribution 
of the ``single inclusive" Higgs boson transverse momentum, which denotes the transverse 
momentum distribution of any (randomly picked) Higgs boson. 
In contrast, Fig.~\ref{fig:pth_hard_soft} shows the transverse momentum distributions of 
the leading-$p_T$ (``harder") and subleading-$p_T$ (``softer") Higgs boson.
It again becomes very clear that reweighting the basic HEFT result is indispensable in order to get 
at least somewhat close to the shape of the full NLO result.
The $p_{T,h}$ distribution in Fig.~\ref{subfig:pth14} shows that, 
while the Born-improved NLO HEFT result starts moving out of
the scale variation band of the full NLO result at 14 TeV beyond $p_{T,h}\sim m_t$, 
the FT$_{approx}$ result stays within the scale uncertainty band of the full NLO result, 
(even though it is clear that it systematically overestimates the full result by about 20-30\%).
This is not surprising, as the tail of the $p_{T,h}$ distribution is to a large extent dominated by the real radiation 
contribution. At $\sqrt{s}=100$\,TeV, the FT$_{approx}$ result leaves the scale variation band of the full NLO result
beyond $p_{T,h}\sim 280$\,GeV, but still is much closer to the full result than the Born-improved NLO HEFT result. 
The differences of the latter to the full result are amplified at 100 TeV.\\
In any case, it is clear that the scale variation bands can only be indicative of missing higher order corrections 
in perturbation theory, while the top quark mass effects (or the omission of the exact top quark mass dependence) are in a different category.
Therefore one cannot expect that, for example, the NLO HEFT scale variation band would comprise the full NLO result.
It is also worth mentioning that the ``FT$_{approx}^{\prime}$"
approximation~\cite{Maltoni:2014eza}, where the partial two-loop results (known 
from  single Higgs production) were included, turned out to be a worse
approximation than ``FT$_{approx}$", where the virtual part is given
by the Born-improved NLO HEFT result, as it lead to a larger cross
section than the ``FT$_{approx}$" one, and the latter is still larger
than the full result.

 Note that for $2\to 2$ scattering the transverse momentum of the Higgs
 boson is given by $p_T^2=\frac{\hat{s}}{4}\,\beta_h^2\sin^2\theta$.
 Therefore, at leading order, the $p_{T,h}$ transverse momentum distribution directly
 reflects the angular dependence of the virtual amplitude. However, at NLO, 
 the angular dependence of the form factors is influenced to a large
 extent by the real radiation. 
This can be seen from the distributions of 
the leading-$p_T$ (``harder") and subleading-$p_T$ (``softer") Higgs bosons shown in 
Fig.~\ref{fig:pth_hard_soft}. The Higgs boson will pick up a 
large transverse momentum if it recoils against a hard jet, therefore the K-factor of the 
$p_{T,h}^{\mathrm{hard}}$ grows in the tail of the distribution, which is dominated by $2\to 3$ kinematics.

\begin{figure}
\centering
\begin{subfigure}{0.49\textwidth}
\includegraphics[width=\textwidth]{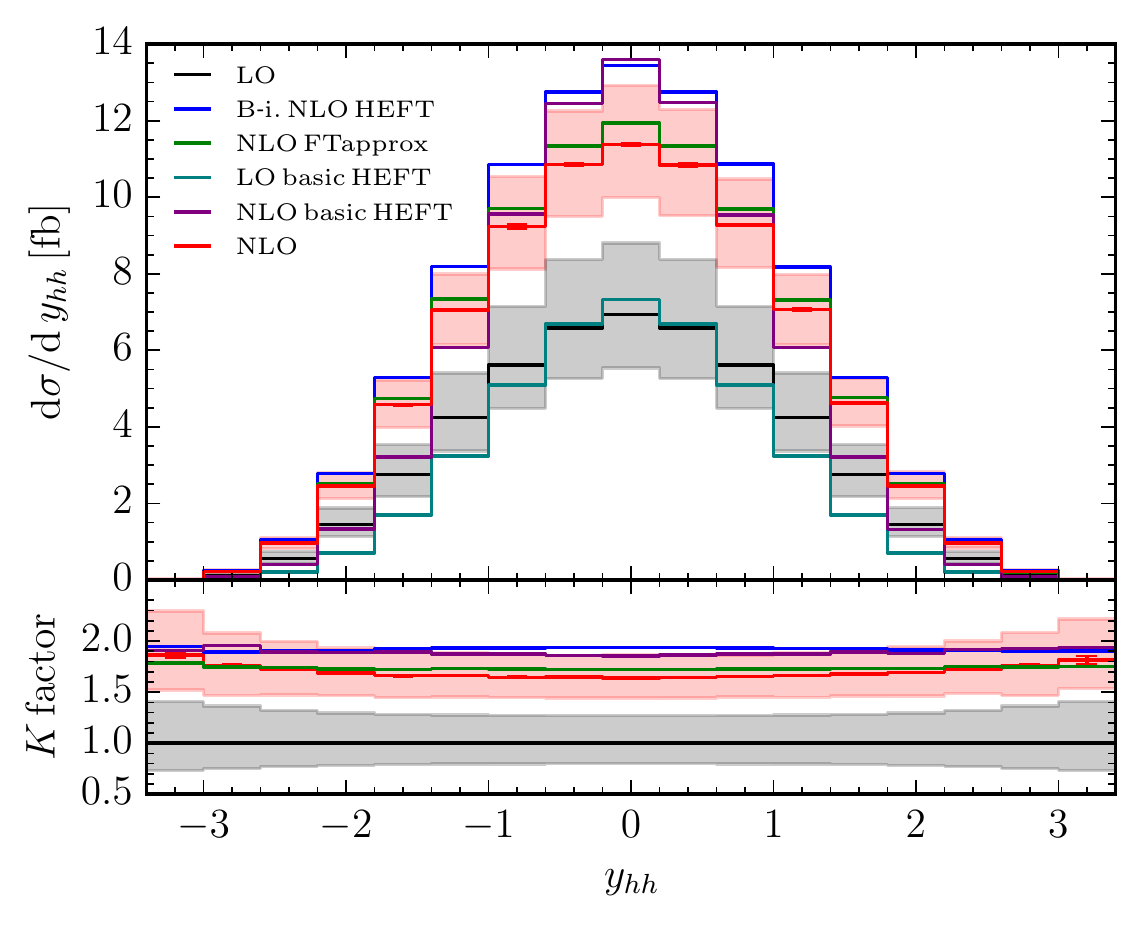}
\caption{14 TeV, rapidity of the pair}
\end{subfigure}
\begin{subfigure}{0.49\textwidth}
\includegraphics[width=\textwidth]{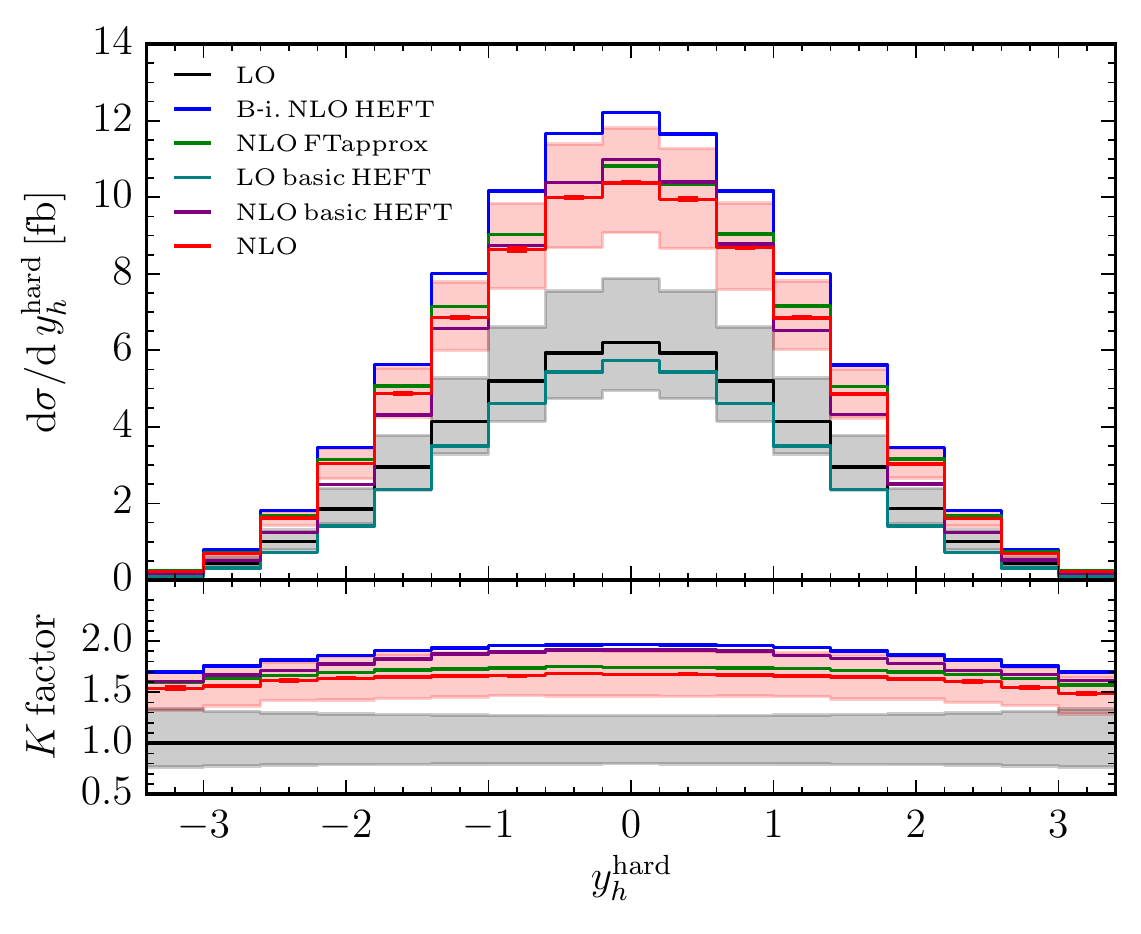}
\caption{14 TeV, rapidity of the leading-$p_T$ Higgs}
\end{subfigure}
\begin{subfigure}{0.49\textwidth}
\includegraphics[width=\textwidth]{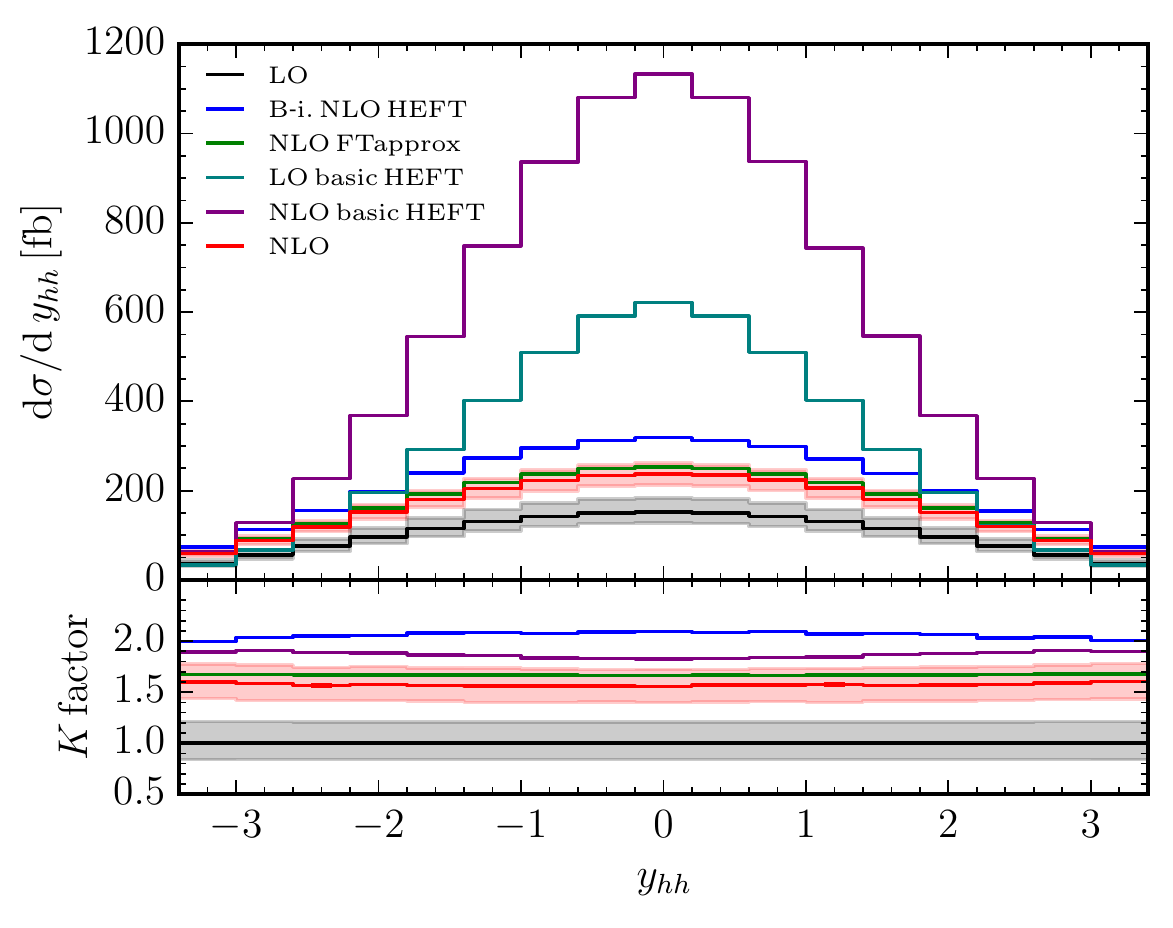}
\caption{100 TeV, rapidity of the pair}
\end{subfigure}
\begin{subfigure}{0.49\textwidth}
\includegraphics[width=\textwidth]{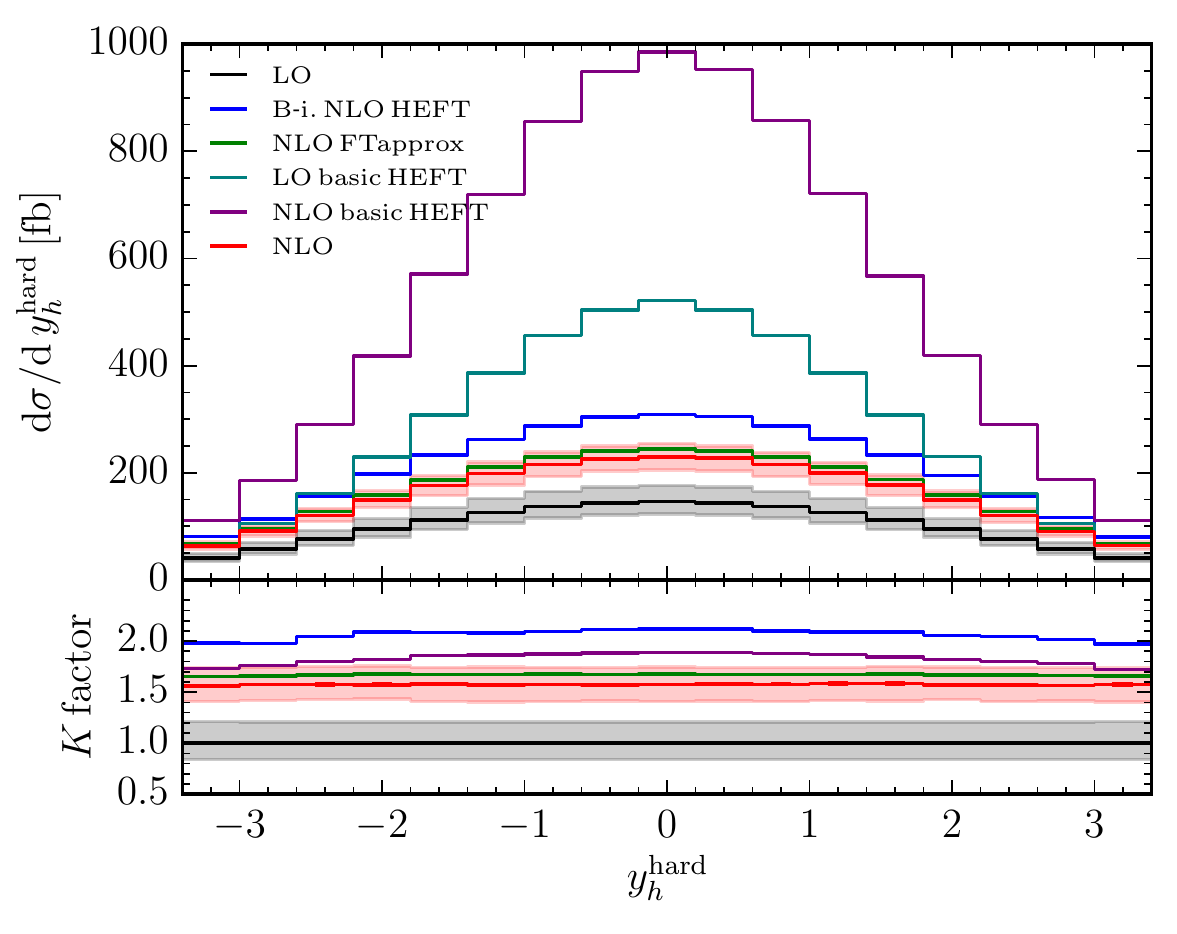}
\caption{100 TeV, rapidity of the leading-$p_T$ Higgs}
\end{subfigure}
\caption{Rapidity  distribution of  the  Higgs
  boson pair  and  the leading-$p_T$ Higgs
  boson  at $\sqrt{s}=14$\,TeV and $\sqrt{s}=100$\,TeV.\label{fig:yhh}}
\end{figure}
Fig.~\ref{fig:yhh} shows the rapidity distributions of both the Higgs boson pair and  the 
leading-$p_T$ Higgs boson. As the mass effects are uniformly distributed over the whole rapidity range, 
the K-factors are close to uniform for these distributions, and the FT$_{approx}$ result is within 10\% of the full result.
\begin{figure}
\centering
\begin{subfigure}{0.49\textwidth}
\includegraphics[width=\textwidth]{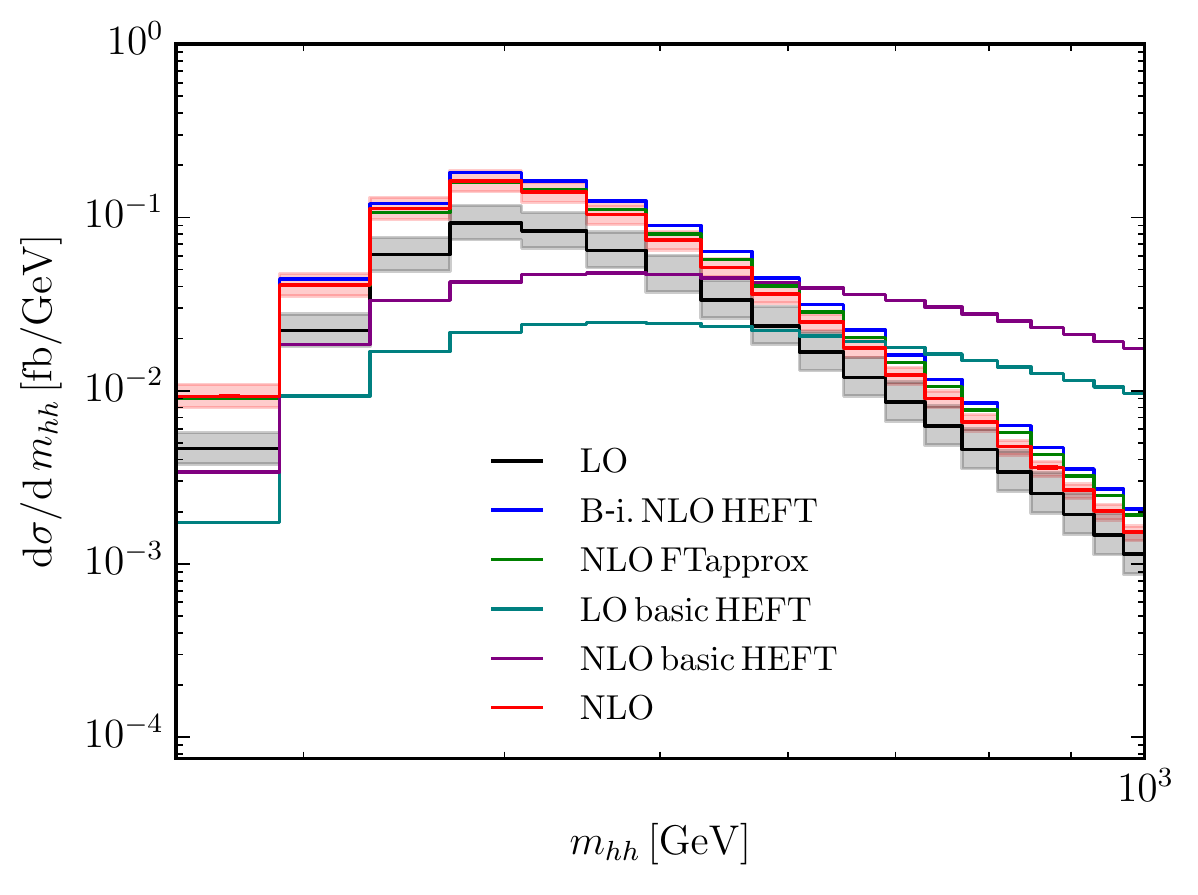}
\caption{14 TeV, scaling behaviour of $m_{hh}$}
\end{subfigure}
\begin{subfigure}{0.49\textwidth}
\includegraphics[width=\textwidth]{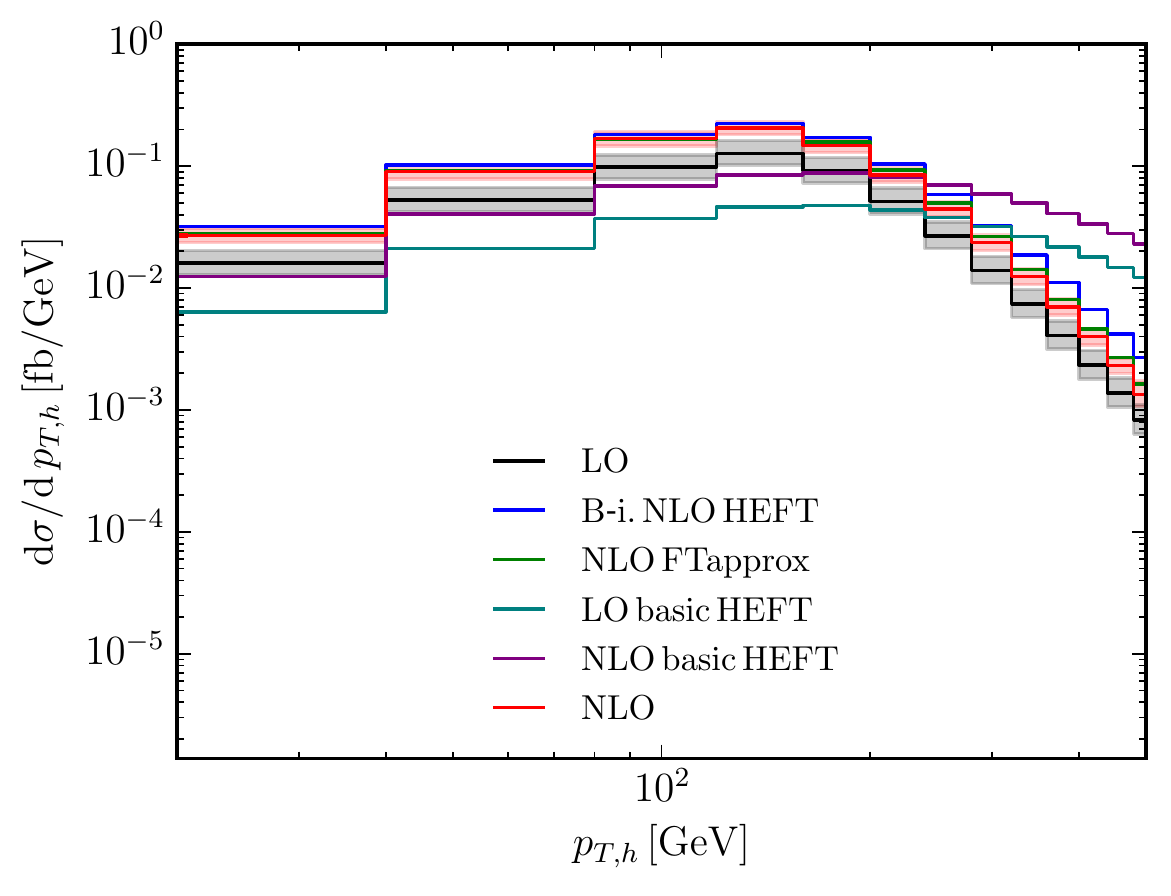}
\caption{14 TeV, scaling behaviour of $p_{T,h}$}
\end{subfigure}
\caption{Higgs boson pair invariant mass distribution (a) and transverse momentum distribution (b)  at 
  $\sqrt{s}=14$\,TeV on a logarithmic scale. The different high-energy
  scaling behaviour 
  of the amplitude in the full and the basic HEFT calculation can be clearly seen in the tails of the distributions.\label{fig:mhh14_logaxes}}
\end{figure}
In Fig.~\ref{fig:mhh14_logaxes} we display the tails of the $m_{hh}$ and 
 $p_{T,h}$ distributions on a logarithmic scale, in order to exhibit the scaling behaviour in the high energy limit.
Using leading-log high energy resummation techniques, it can be shown~\cite{Caola:2016upw} 
that at high transverse momentum, the differential partonic cross section for single Higgs (+jets) 
production $d\sigma/dp_{T,h}\sim 1/p_{T,h}^a$  scales with $a=2$ in the full theory, however with 
$a=1$ in the effective theory.
This behaviour also has been recently confirmed by a (leading order) calculation of 
Higgs + 1,2,3 jet production with full mass
dependence~\cite{Greiner:2016awe}.
\begin{figure}
\centering
\begin{subfigure}{0.49\textwidth}
\includegraphics[width=\textwidth]{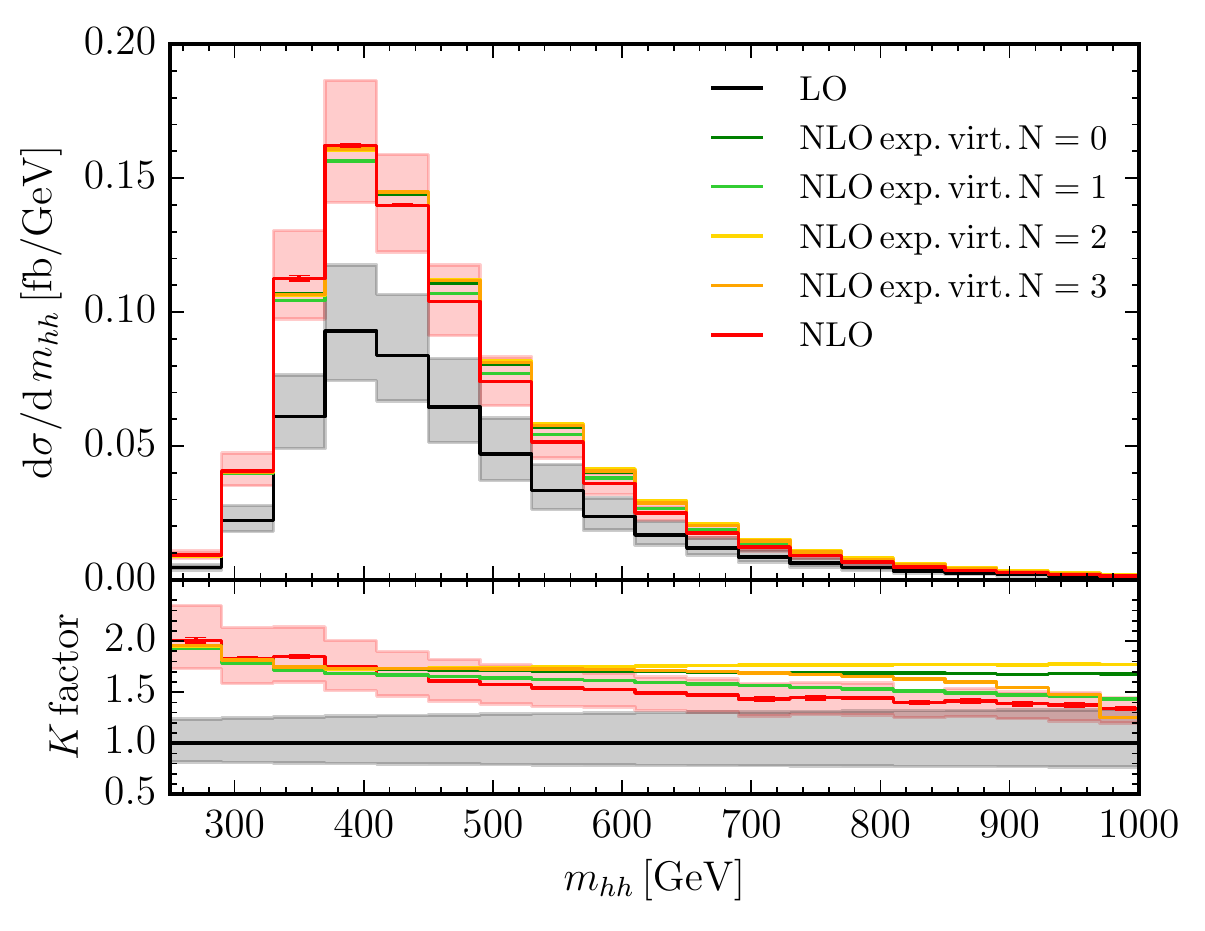}
\caption{14 TeV, $m_{hh}$}
\end{subfigure}
\begin{subfigure}{0.49\textwidth}
\includegraphics[width=\textwidth]{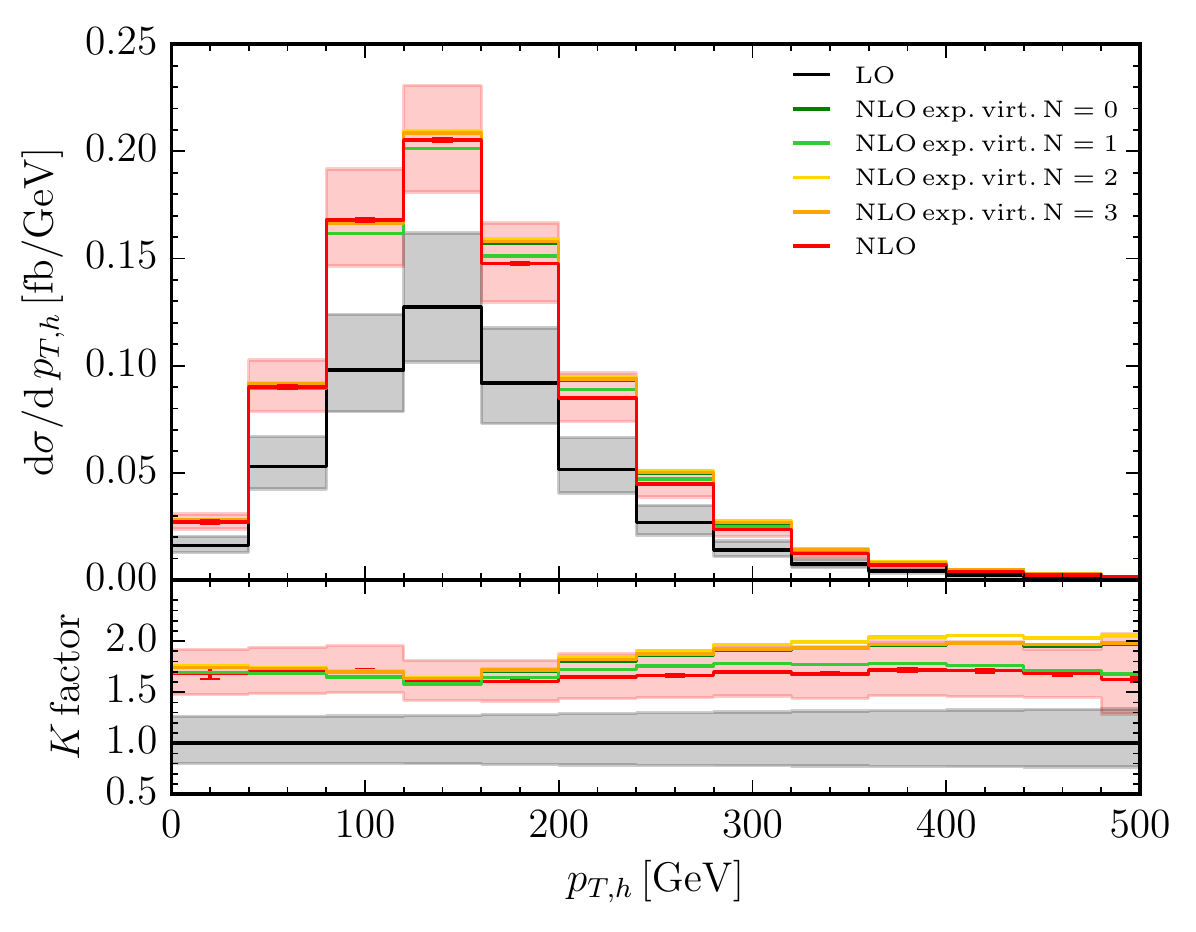}
\caption{14 TeV, $p_{T,h}$}
\end{subfigure}
\caption{Invariant mass distribution of the Higgs boson pair (a) and $p_{T}$ distribution of any Higgs (b) at $\sqrt{s}=14$\,TeV combining the full real emission with the virtual contribution expanded in $1/m_t^2$ up to order $N$. Note that $N=0$ corresponds to FT$_\text{approx}$.\label{fig:mtexp}}
\end{figure}
In order to investigate the high energy scaling behaviour we fitted a line to the tail of the leading order $m_{hh}$
distribution (with the luminosity factor set to one,  plotted 
logarithmically), and found the following
scaling behaviour: with full mass dependence, the scaling
is as $m_{hh}^{-3}$ for $d\hat\sigma/dm_{hh}$ i.e. the partonic cross
section scales as $\hat{s}^{-1}$, while in the basic HEFT approximation the scaling
is as $m_{hh}$ for $d\hat\sigma/dm_{hh}$ i.e. the partonic cross
section grows as $\hat{s}$. From Fig.~\ref{fig:mhh14_logaxes} one can
see that this relative difference in the high-energy scaling behaviour
between the full calculation and the basic HEFT approximation is
similar at NLO.

In Fig.~\ref{fig:mtexp} we show distributions for an improved
FT$_\text{approx}$, which is supplemented with higher order terms in
the expansion of the virtual amplitude in 
$1/m_t^2$ as given by Eq.~\eqref{eq:V+I}, dubbed ``exp. virt.'' for
``expanded virtuals''.
We see a trend similar to the one for the virtual (plus ${\bom  I}$-operator) part shown in Fig.~\ref{fig:ampexpand_TZJH}.

\begin{figure}
\centering
\begin{subfigure}{0.49\textwidth}
\includegraphics[width=\textwidth]{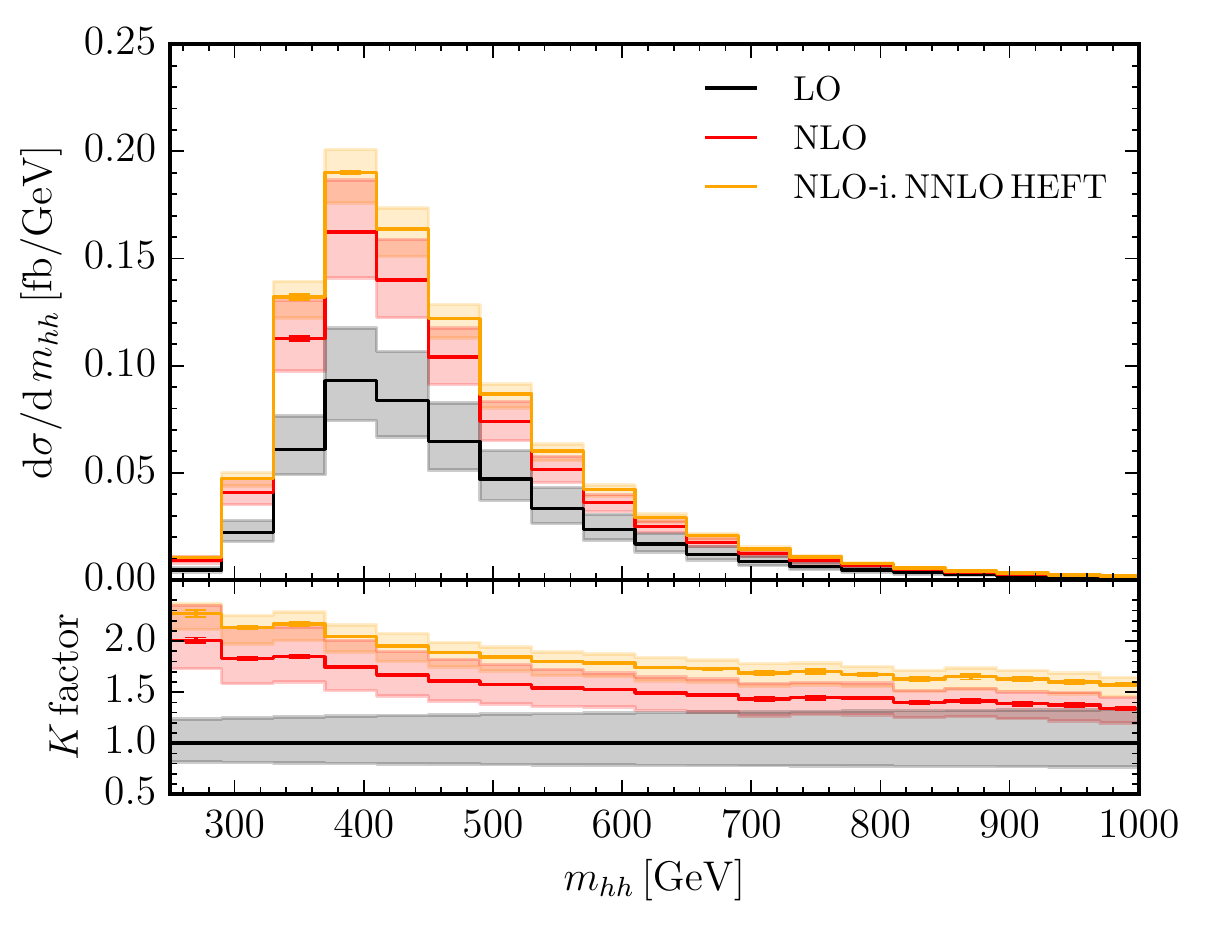}
\caption{14 TeV, $m_{hh}$}
\end{subfigure}
\begin{subfigure}{0.49\textwidth}
\includegraphics[width=\textwidth]{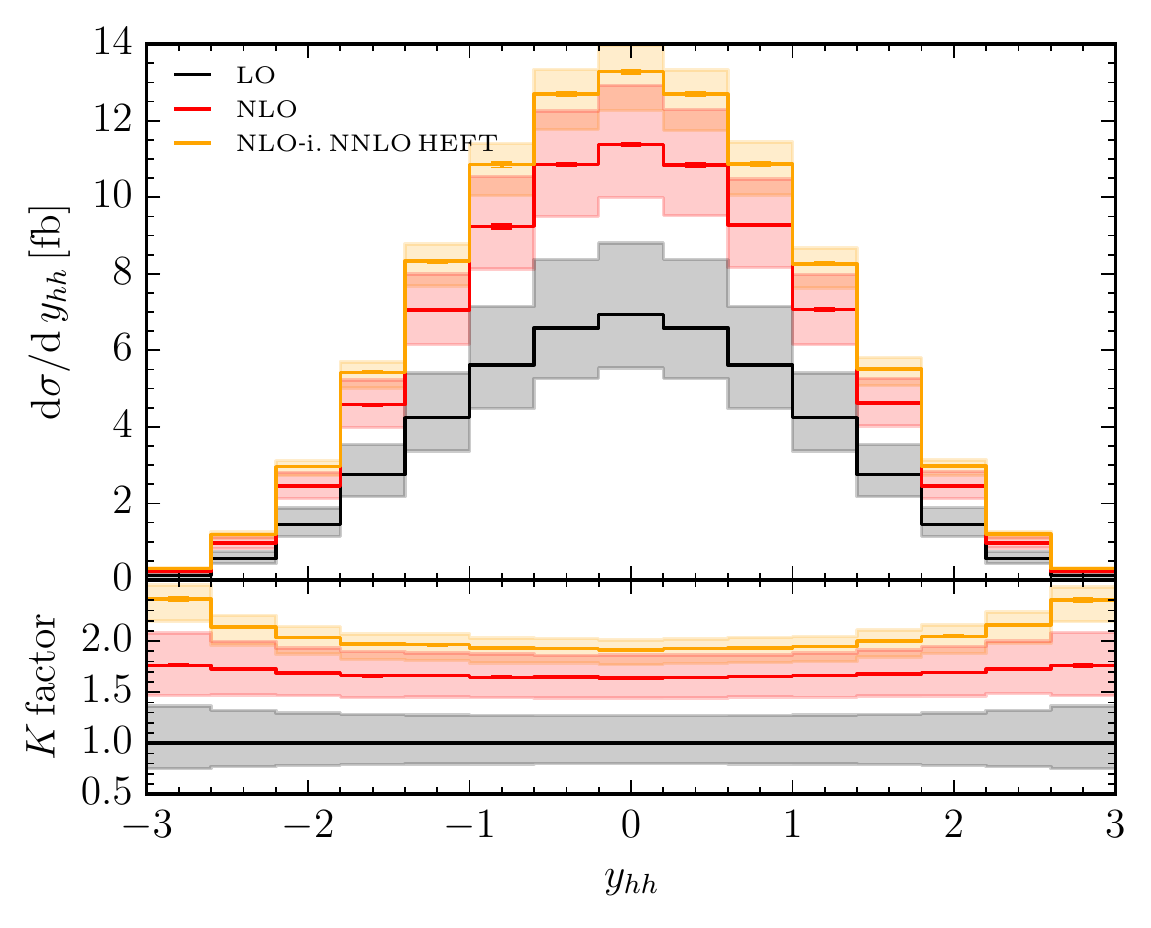}
\caption{14 TeV, $y_{hh}$}
\end{subfigure}
\begin{subfigure}{0.49\textwidth}
\includegraphics[width=\textwidth]{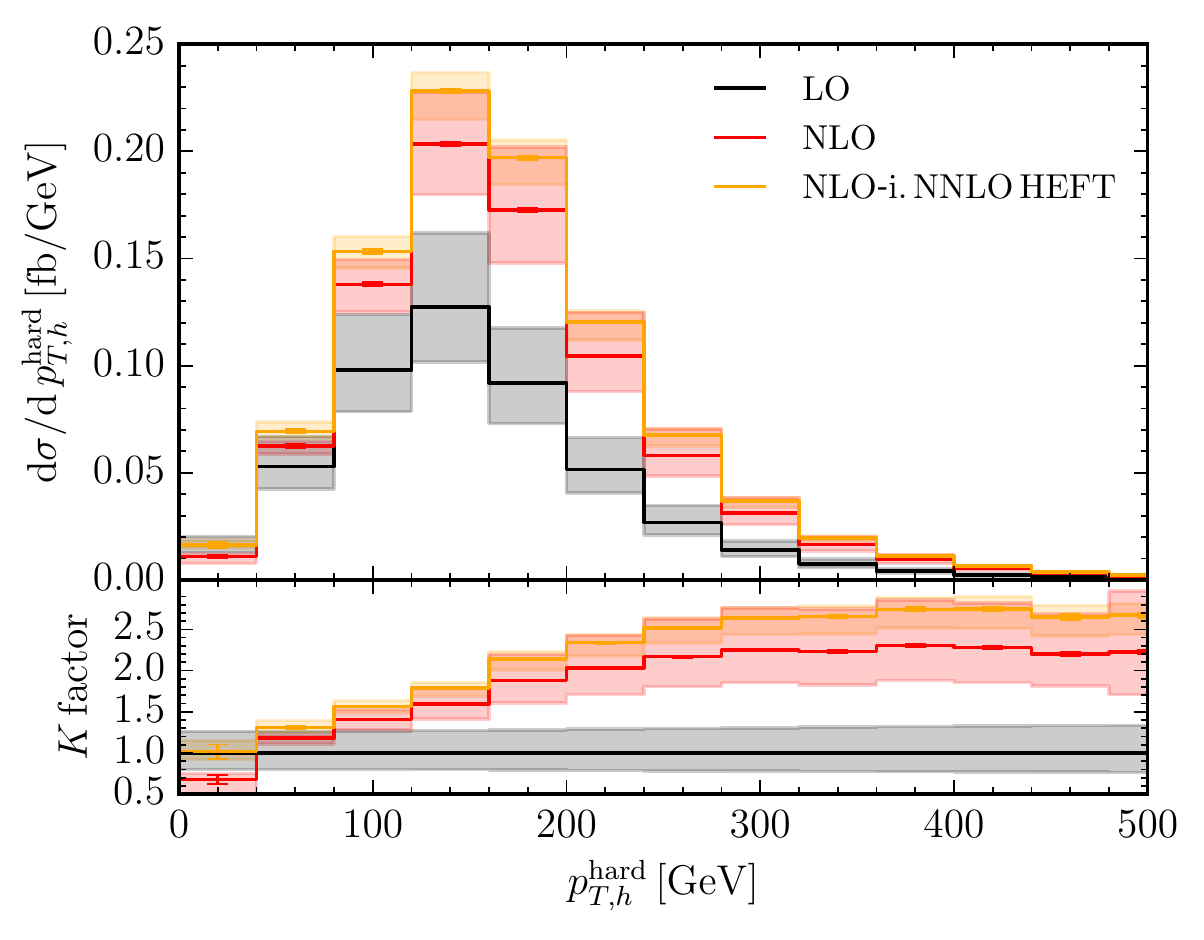}
\caption{14 TeV, leading $p_T$}
\end{subfigure}
\begin{subfigure}{0.49\textwidth}
\includegraphics[width=\textwidth]{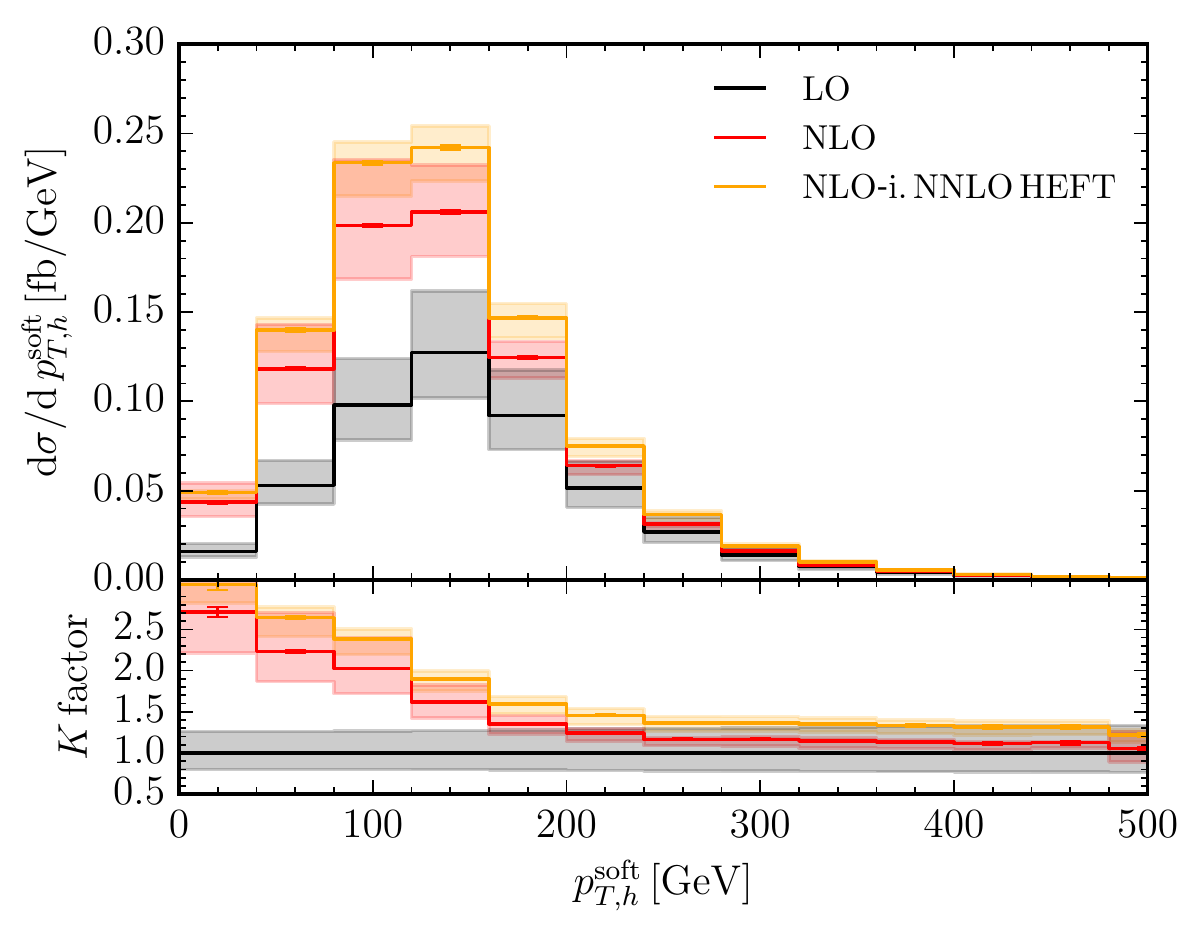}
\caption{14 TeV, subleading $p_T$}
\end{subfigure}
\caption{Invariant mass (a) and rapidity distribution (b) of the Higgs boson pair
and transverse momentum distribution of the leading-$p_T$ (c) and the subleading-$p_T$ Higgs boson (d)
at $\sqrt{s}=14$\,TeV including the combination with the NNLO HEFT results from Ref.~\cite{deFlorian:2016uhr} 
described in the main text.\label{fig:nnlo}}
\end{figure}
In order to better account for missing higher order corrections it is desirable to combine the full NLO with NNLO results obtained in the HEFT, ideally on a differential level. As a first attempt to achieve this, we take the NNLO to NLO ratio from Ref.~\cite{deFlorian:2016uhr} and calculate
\begin{align}
  \dd\sigma^{\text{NLO-i.\,NNLO HEFT} }= \dd\sigma^{\text{NLO}} \frac{\dd\sigma^{\text{NNLO basic HEFT}}} {\dd\sigma^{\text{NLO basic HEFT}}}
\end{align}
bin by bin, where ``NLO-i.\,NNLO HEFT'' stands for NLO-improved NNLO HEFT. Results for various distributions are shown in Fig.~\ref{fig:nnlo}. The error band is the NLO-rescaled scale uncertainty of the NNLO basic HEFT distributions, and the error on the central value is due to the error on the full NLO result.
Applying the same naive rescaling on the total cross section, one obtains
$\sigma^{\text{NLO-i.\,NNLO HEFT} } = 38.67^{+5.2\%}_{-7.6\%}$ for 14 TeV,
where we have neglected the numerical errors and simply quote the relative scale uncertainty given in Ref.~\cite{deFlorian:2016uhr} for the NNLO basic HEFT result.


\subsection{Sensitivity to the triple Higgs coupling}
\label{sec:tripleH}

As already mentioned in Section \ref{sec:amp}, the Higgs boson
self-coupling in the Standard Model is quite special. 
Not only that it is completely determined in terms of the Higgs boson mass and
VEV, but it also
leads to the fact that at the double Higgs production
threshold $\sqrt{\hat{s}}=2m_h^2$,  
the LO cross section is almost vanishing, due to destructive interference 
between box and triangle contributions.
Therefore a measurement of the Higgs boson self-coupling is a very
sensitive probe of New Physics effects. 

A more complete analysis of such effects would require an approach
where further operators are taken into account, for example
operators which mediate  direct $t\bar{t}HH$ couplings (and
Higgs-gluon couplings which can differ from the SM HEFT ones),
see e.g. \cite{Azatov:2015oxa,Grober:2015cwa,Ghezzi:2015vva}.
However, the conclusions drawn from the calculation of NLO corrections in the $m_t\to\infty$ limit 
to the extended set of EFT Wilson coefficients  have to be taken with a grain of salt,
as the full top quark mass dependence may affect them considerably. 

In this section we  would like to focus on just a single line in the
parameter space of possible non-SM Higgs couplings and investigate the
behaviour of the $m_{hh}$ distribution under variations of $\lambda$,
where we have defined $\lambda_{hhh} = 3 m_h^2\lambda$, see Eq.~(\ref{eq:lambda}).

\begin{figure}
\centering
\begin{subfigure}{0.49\textwidth}
\includegraphics[width=\textwidth]{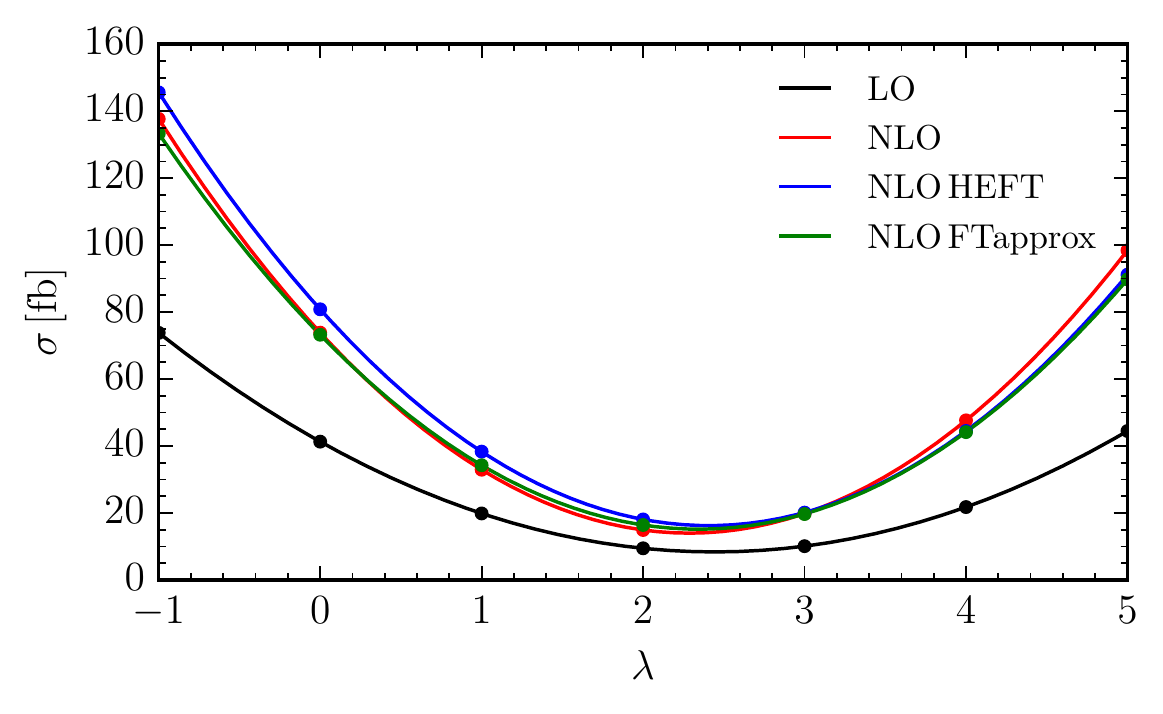}
\end{subfigure}
\begin{subfigure}{0.49\textwidth}
\includegraphics[width=\textwidth]{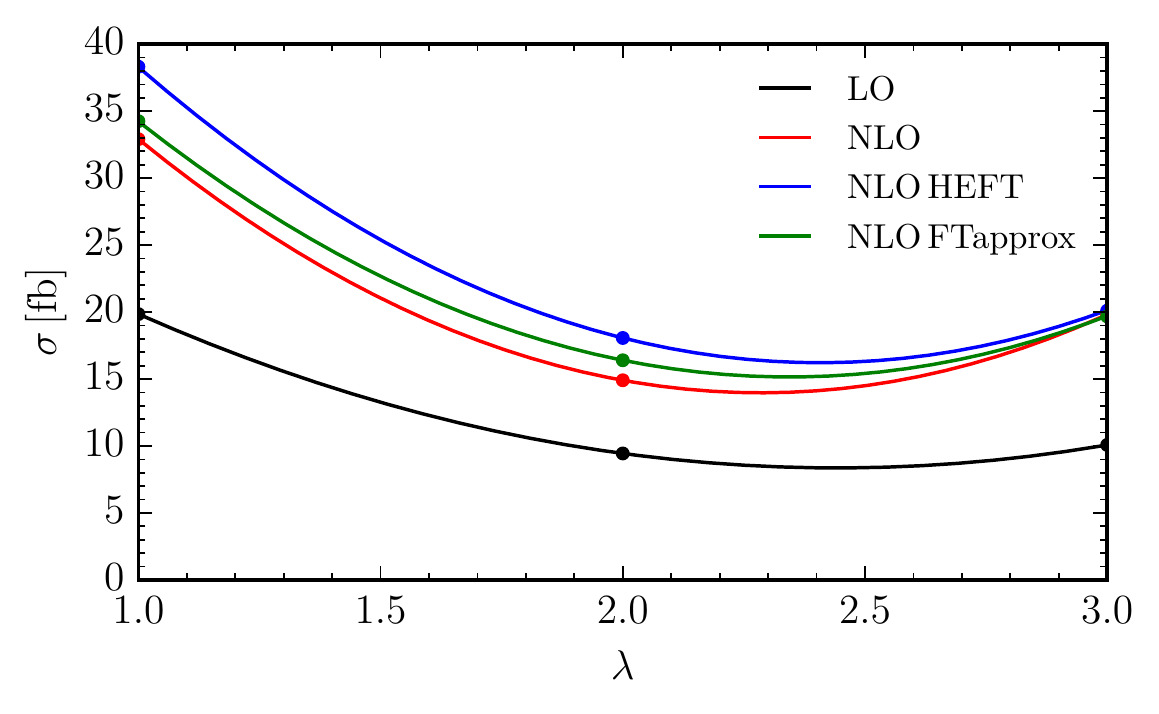}
\end{subfigure}
\caption{Total cross sections for various values of the triple Higgs coupling.
Panel (b) zooms into the region around the minimum. 
The curves are the result of an interpolation of integer values for $\lambda\in \{-1,\ldots, 5\}$. \label{fig:sigtotlambda}}
\end{figure}
 
\begin{figure}
\centering
\begin{subfigure}{0.49\textwidth}
\includegraphics[width=\textwidth]{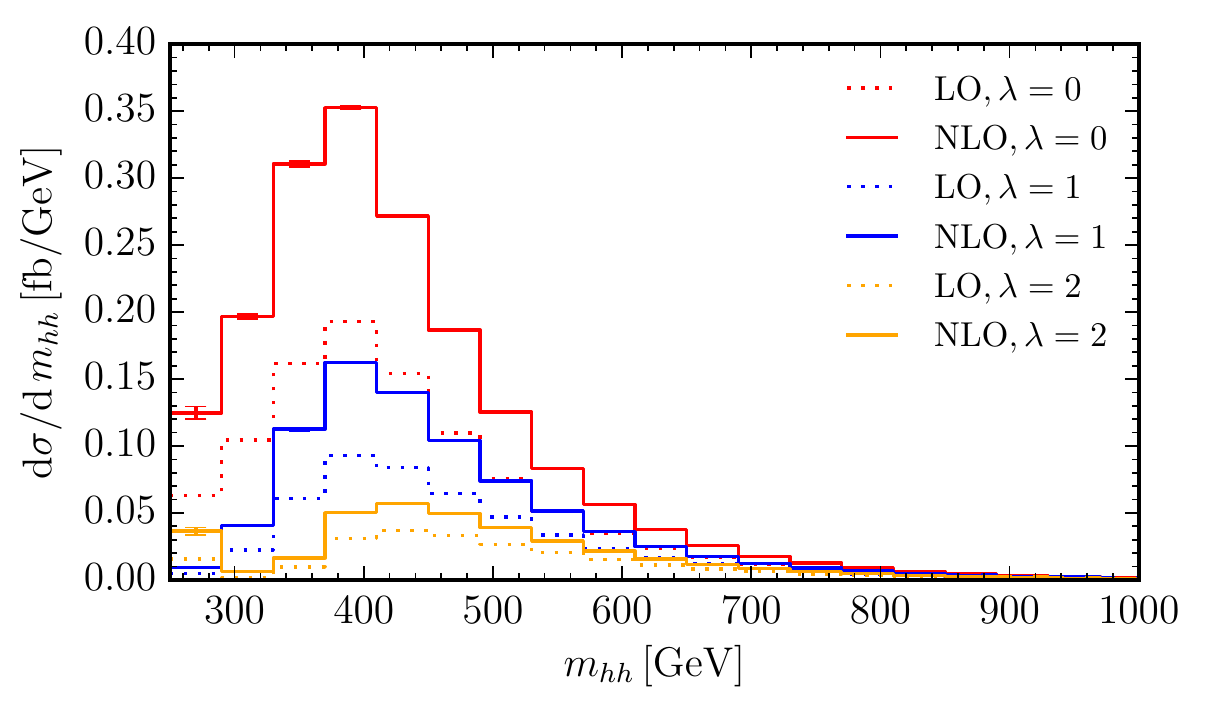}
\caption{14 TeV}
\end{subfigure}
\begin{subfigure}{0.49\textwidth}
\includegraphics[width=\textwidth]{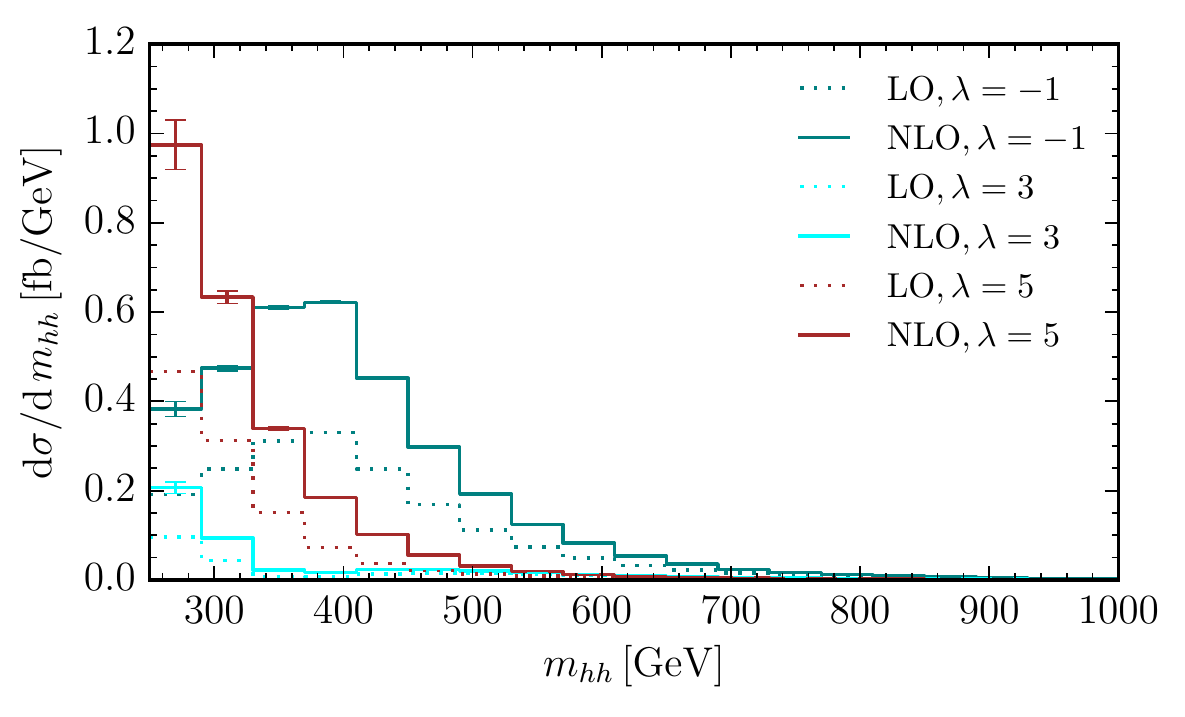}
\caption{14 TeV}
\end{subfigure}
\begin{subfigure}{0.49\textwidth}
\includegraphics[width=\textwidth]{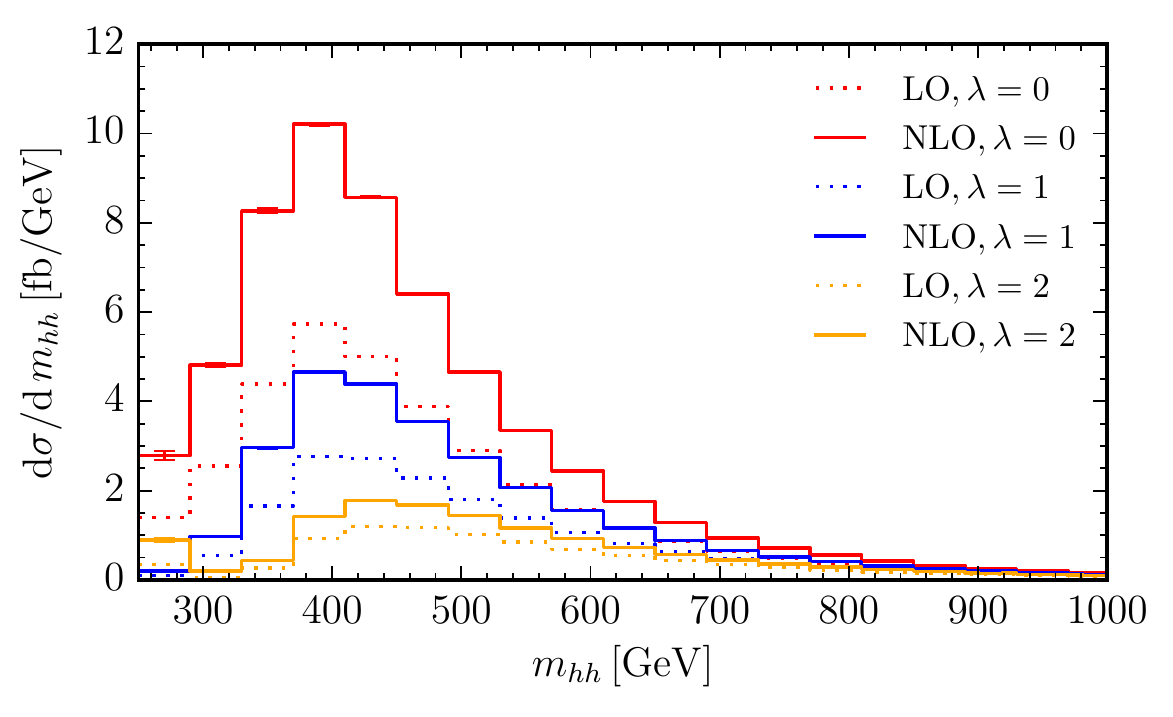}
\caption{100 TeV}
\end{subfigure}
\begin{subfigure}{0.49\textwidth}
\includegraphics[width=\textwidth]{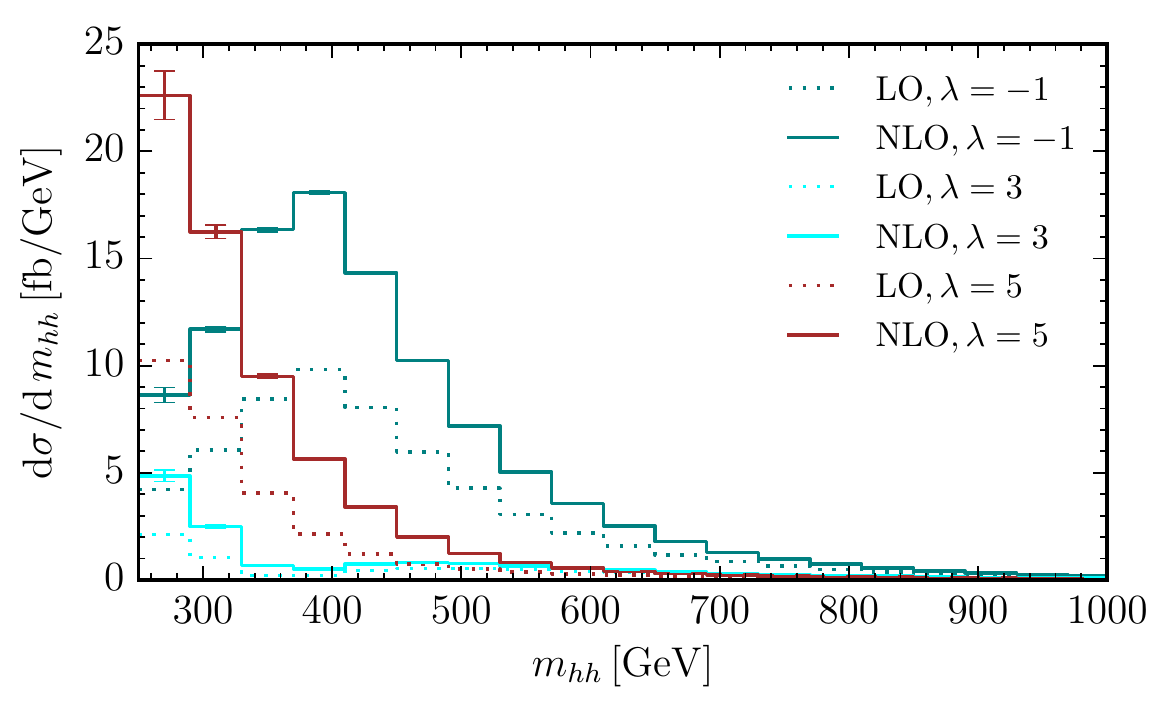}
\caption{100 TeV}
\end{subfigure}
\caption{NLO and LO results with full top quark mass dependence for the $m_{hh}$ distribution at 14 TeV and 100 TeV,
for various  values of the triple Higgs
coupling, where $\lambda=1$ corresponds to the Standard Model value.\label{fig:varylambda_small_large}}
\end{figure}

In  Fig.~\ref{fig:sigtotlambda} we show the total cross section as a
function of $\lambda$. As already observed for the LO cross
section~\cite{Baglio:2012np}, it has a minimum around $\lambda=2$.
Negative $\lambda$ values, which are not excluded neither theoretically nor experimentally 
(within certain broad limits given e.g. by vacuum stability), do not lead to destructive interference and therefore 
result in a much larger cross section.
For large positive values, $\lambda\sim 5$, the total cross section is
of comparable size to the one for 
$\lambda\simeq 0$, but the shape of the $m_{hh}$ distribution is completely different.
This can be seen in 
Fig.~\ref{fig:varylambda_small_large}, where we show the Higgs boson pair
invariant mass distribution for various values of the Higgs boson
self-coupling, at $\sqrt{s}=14$\,TeV and $\sqrt{s}=100$\,TeV. 
For $\lambda=5$, the
differential cross section is mainly dominated by contributions
containing the Higgs boson self coupling and peaks at low $m_{hh}$
values. In contrast,  the $\lambda=0$ case, which does not contain any
 triple Higgs coupling contribution, peaks shortly beyond the
$2m_t$ threshold at $m_{hh}\sim 400$\,GeV, as does the case
$\lambda=-1$. In the latter case, however, the
total cross section is much larger.
The case $\lambda=2$ shows a dip at $m_{hh}\sim 300$\,GeV, which is
due to destructive interference effects as mentioned above. 
At 100 TeV, the shape of the distributions is very similar.
However, the fact that the cross sections are much larger can be
exploited to place cuts which enlarge the sensitivity to the Higgs
boson self coupling. 
For example, one can try to enhance the self-coupling contribution 
by cuts favouring highly boosted virtual Higgs
bosons, decaying into a Higgs boson pair which 
could be detected in the $b\bar{b}\,b\bar{b}$ channel.
A highly boosted virtual Higgs boson must recoil against a high-$p_T$
jet. Therefore, an enhancement of the boosted component could be
achieved by imposing a $p_{T,jet}^{min}$ cut on the recoiling jet in
Higgs boson pair plus jet production~\cite{michelangelo,Mangano:2016jyj}. 
An additional advantage of boosted Higgs bosons is the fact that 
they lend themselves to the use of the $b\bar{b}b\bar{b}$ rather than the $b\bar{b}\gamma\gamma$ decay channel, 
as the decay channel into $b$-quarks is accessible through boosted
techniques. This leads to a gain in the rate which easily makes up for the loss in
statistics due to a high $p_{T,jet}^{min}$ cut.

\begin{figure}
\centering
\begin{subfigure}{0.49\textwidth}
\includegraphics[width=\textwidth]{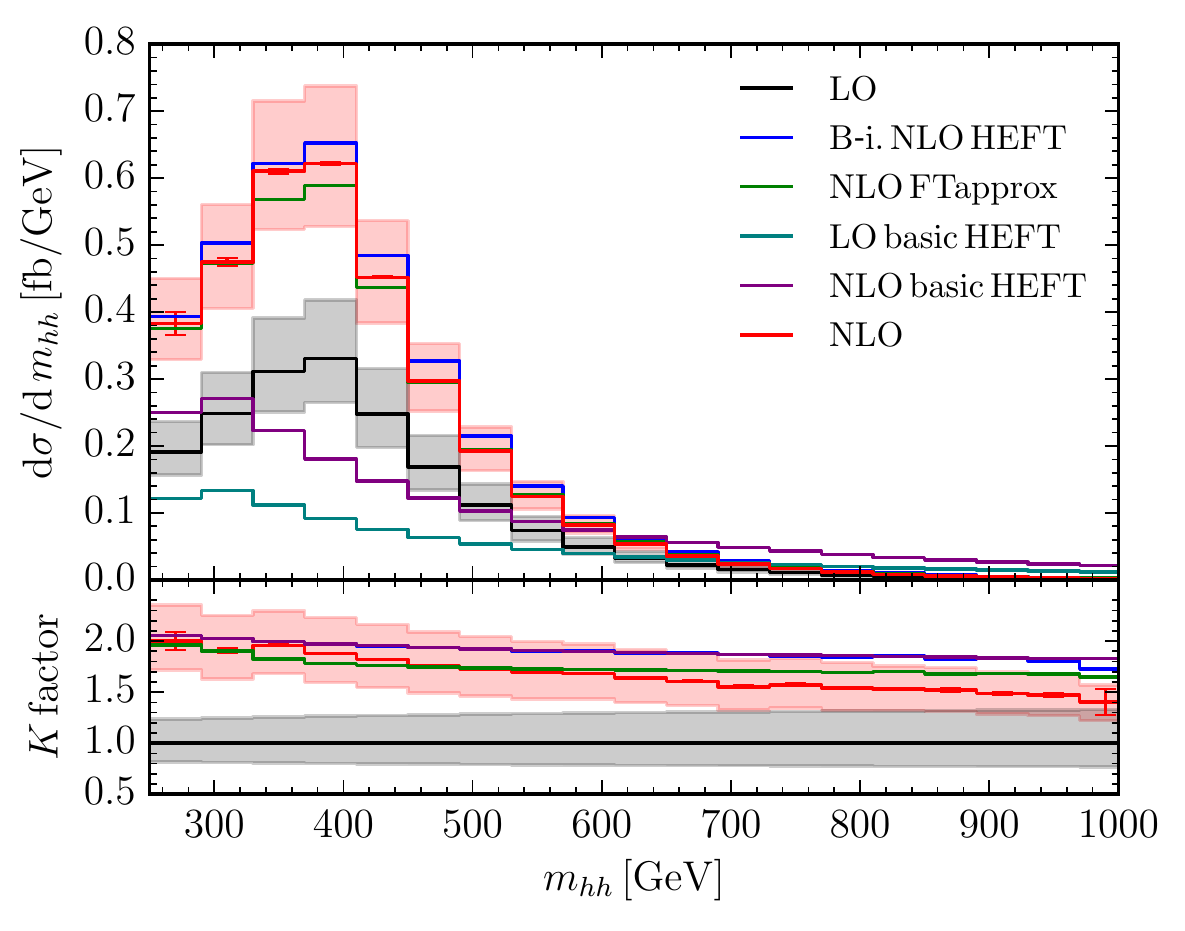}
\caption{14 TeV, $\lambda=-1$}
\end{subfigure}
\begin{subfigure}{0.49\textwidth}
\includegraphics[width=\textwidth]{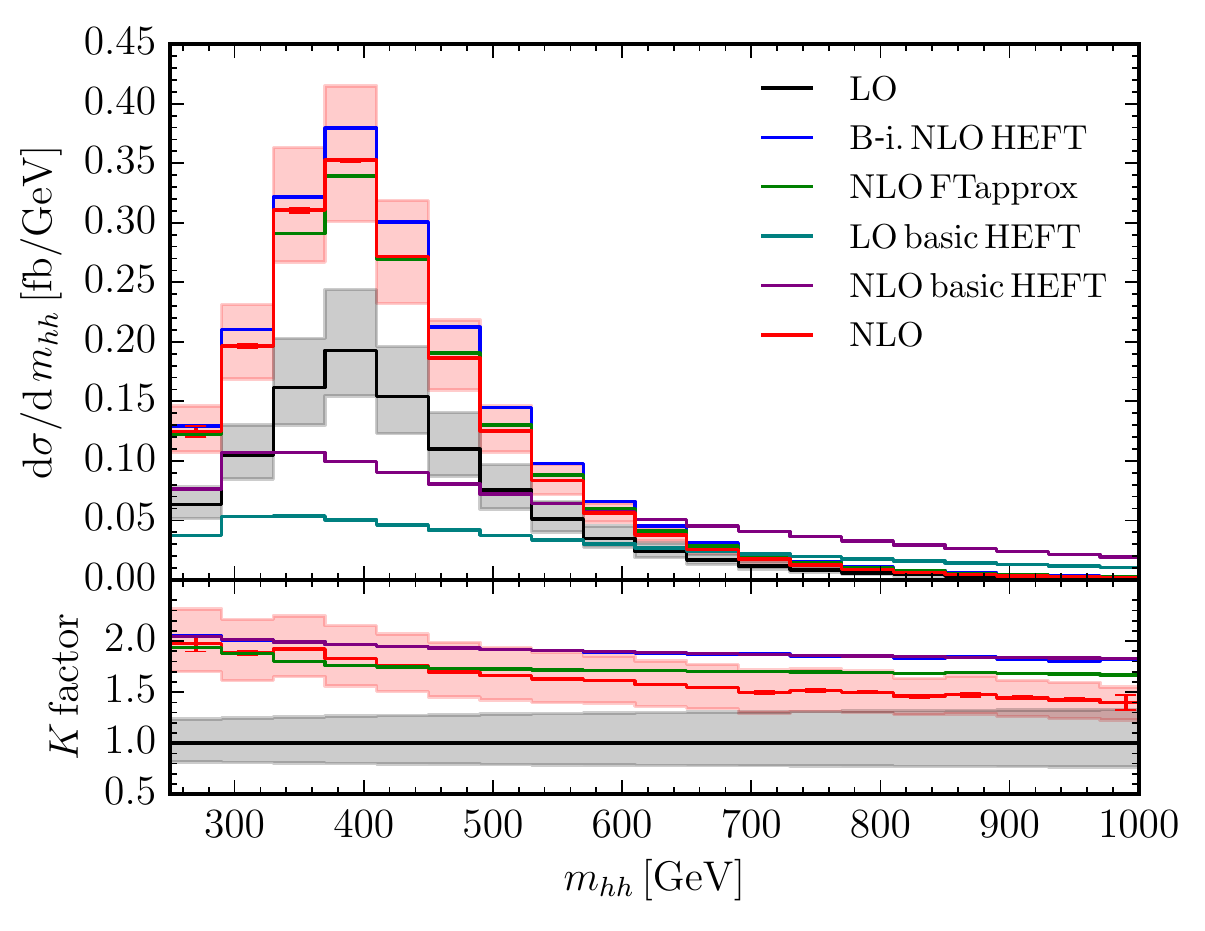}
\caption{14 TeV, $\lambda=0$\label{subfig:mhh14_lambda0}}
\end{subfigure}
\begin{subfigure}{0.49\textwidth}
\includegraphics[width=\textwidth]{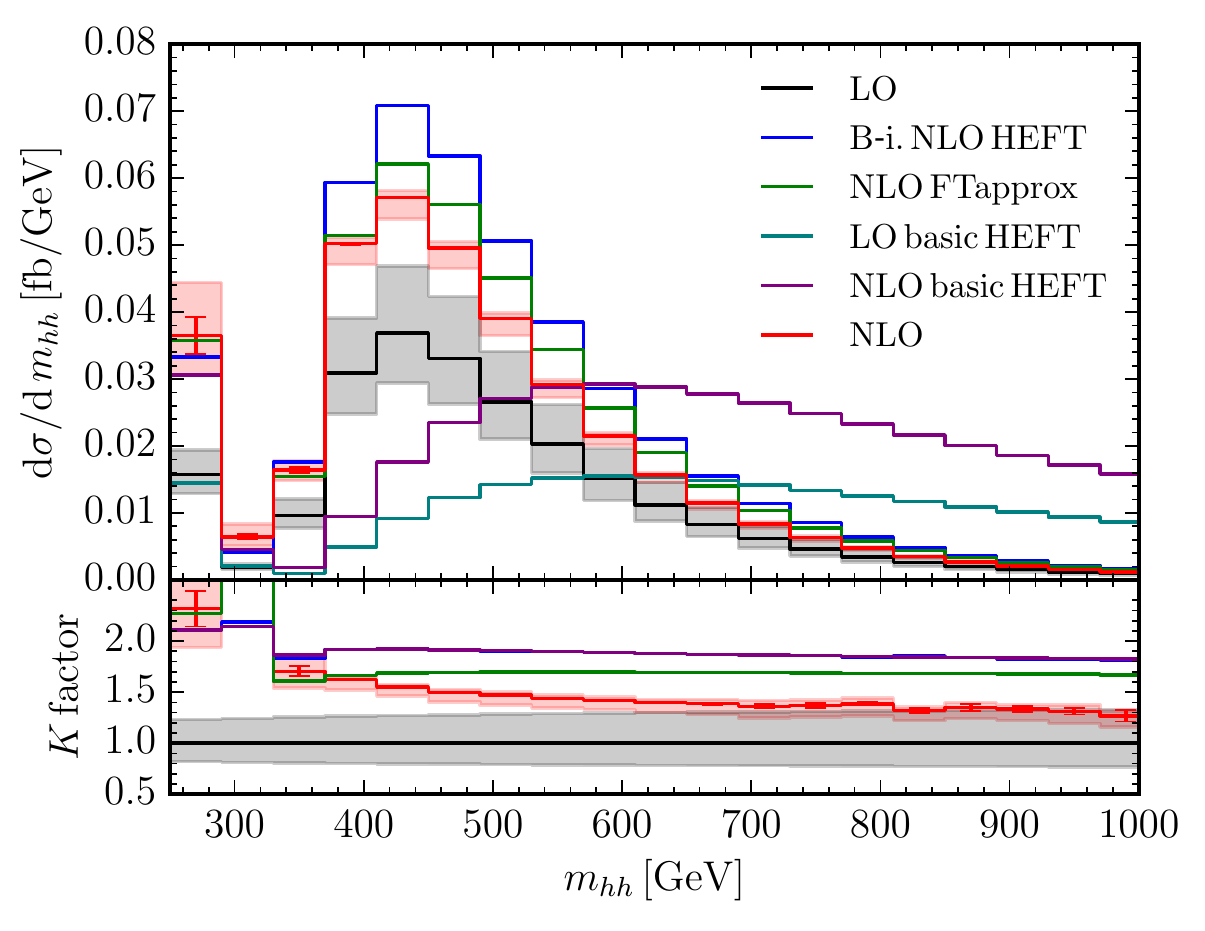}
\caption{14 TeV, $\lambda=2$}
\end{subfigure}
\centering
\begin{subfigure}{0.49\textwidth}
\includegraphics[width=\textwidth]{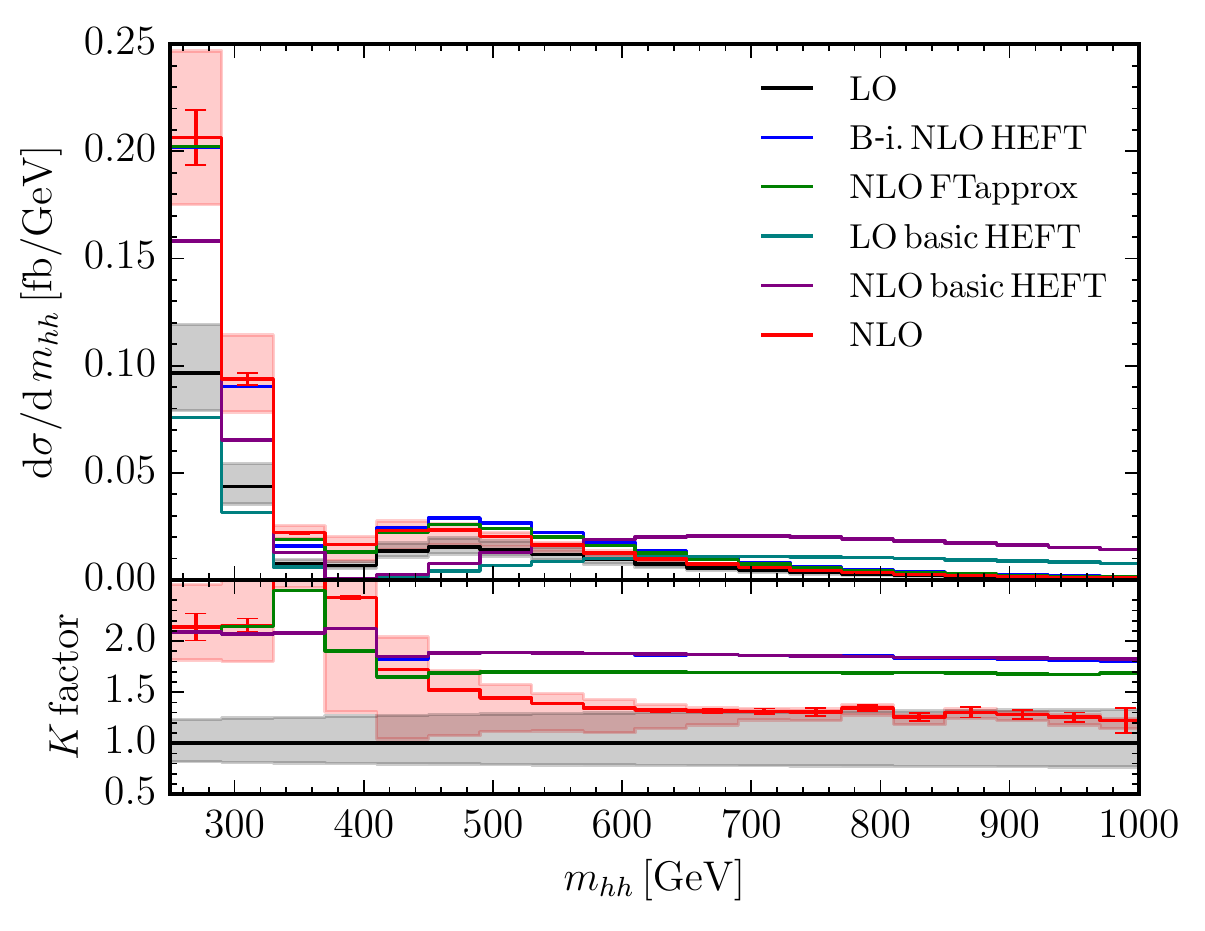}
\caption{14 TeV, $\lambda=3$}
\end{subfigure}
\begin{subfigure}{0.49\textwidth}
\includegraphics[width=\textwidth]{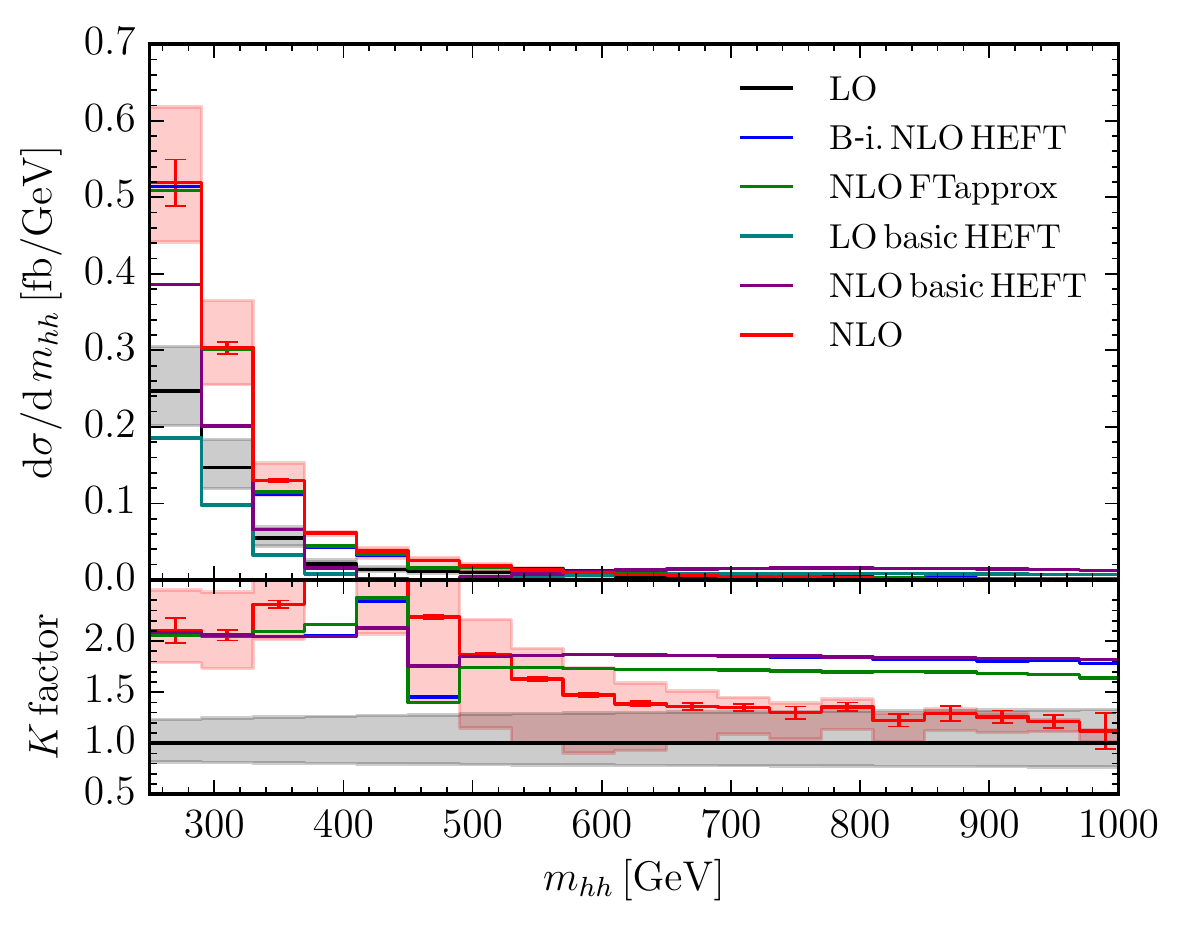}
\caption{14 TeV, $\lambda=4$}
\end{subfigure}
\begin{subfigure}{0.49\textwidth}
\includegraphics[width=\textwidth]{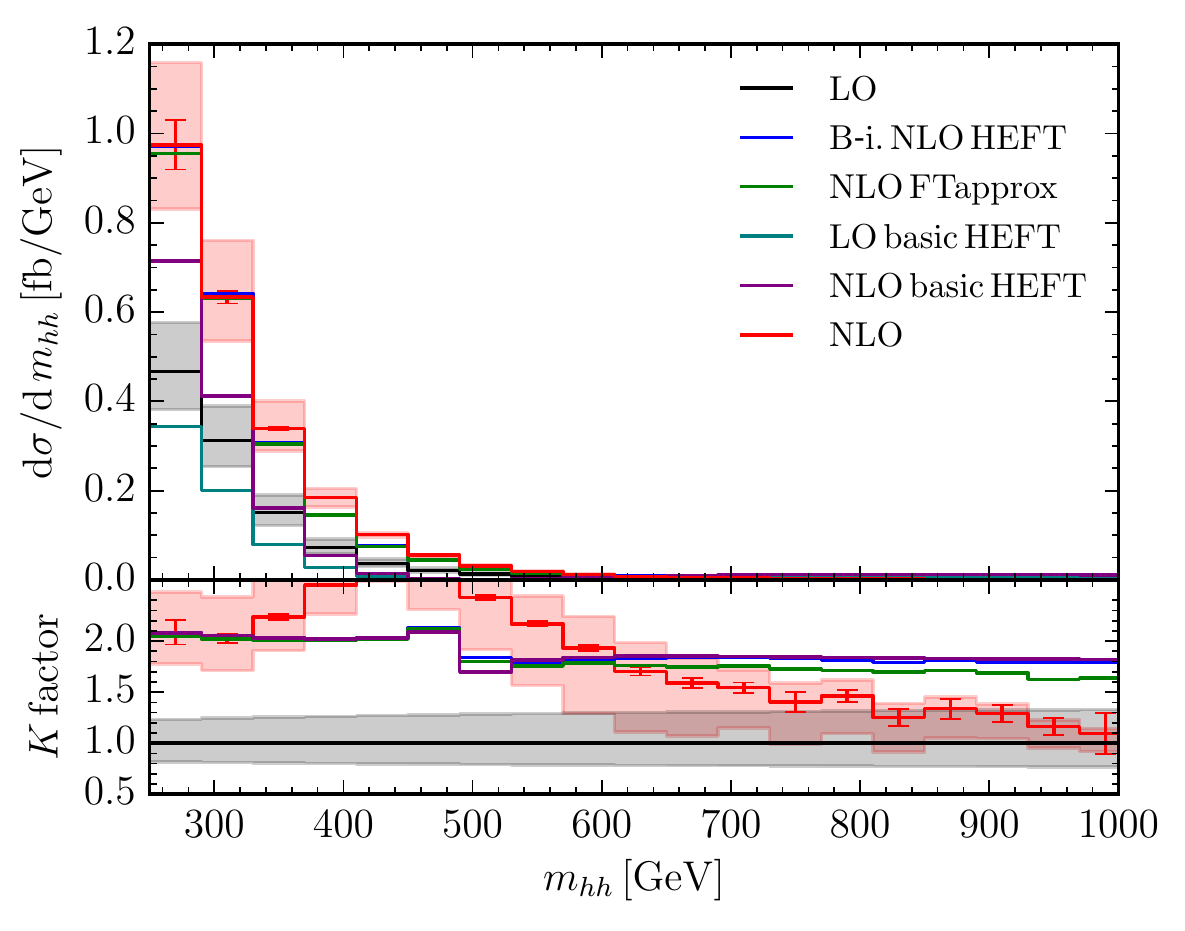}
\caption{14 TeV, $\lambda=5$}
\end{subfigure}
\caption{Higgs boson pair invariant mass distribution $m_{hh}$  at 
  $\sqrt{s}=14$\,TeV  for non-standard values of the triple Higgs coupling. 
\label{fig:mhh14_varylambda}}
\end{figure}

Fig.~\ref{fig:mhh14_varylambda} shows a comparison to the different
approximations for various values of $\lambda$, as well as the K-factors.
For all values of $\lambda$, the K-factors are far from being uniform,
while the HEFT approximation suggests almost uniform K-factors for
$\lambda\leq 1$.
For $\lambda=2$,  we  see a pronounced  ``interference dip'' at $m_{hh}\sim 330$\,GeV, 
which is present at LO already. 
We can get an idea about the destructive interference effect by observing the following:
In the basic HEFT approximation, the squared Born amplitude is
given by Eq.~(\ref{eq:loheft})
This expression has a double zero at
$\hat{s}=m_h^2(1+3\lambda)$. 
Therefore, the re-weighting factor $B_{FT}/B_{HEFT}$ can get large
when $B_{HEFT}$ approaches zero, i.e. at
$\sqrt{s}\simeq 330.72$\,GeV for $\lambda=2$, 
$\sqrt{s}\simeq 395.29$\,GeV
 for $\lambda=3$, $\sqrt{s}\simeq 450.7$\,GeV
 for $\lambda=4$ and  500\,GeV for $\lambda=5$.
In the full theory, the amplitude does not vanish completely at these points, 
but nonetheless also gets small, which should be the reason for the
dips in the $m_{hh}$ distributions
for $\lambda=2$ and 3.

\clearpage

\section{Conclusions}

We have presented results of a fully differential calculation of Higgs
boson pair production in gluon fusion at NLO retaining the exact top
quark mass dependence.
For the total cross section at $\sqrt{s}=14$\,TeV, we found a  reduction of 14\%
compared to the Born improved HEFT, 
and a 24\% reduction at $\sqrt{s}=100$\,TeV.
For differential distributions, the mass effects can be even larger.
In the tails
of the Higgs boson transverse momentum distributions, 
the differences to the Born improved NLO HEFT approximation amount to more
than 50\%, while the FT$_{approx}$ result, where the full top mass dependence
is included only in the real radiation part, stays within 20\% of the
full result.
The basic NLO HEFT approximation, where no reweighting by the
Born result in the full theory is performed, fails to properly
describe the shape of the $m_{hh}$ and $p_{Th}$ distributions, in
particular in the tails of the distributions, where 
we performed an analysis of the high-energy scaling behaviour.

We also studied the influence of non-standard values for the Higgs
boson self-coupling on the total cross sections and $m_{hh}$
distributions. 
As is known from leading order, there is destructive interference between various
contributions to the cross section, 
and this feature persists at NLO.
Varying $\lambda_{hhh}/\lambda_{SM}$ leads to a minimum in the value for the
total cross section around $\lambda_{hhh}/\lambda_{SM}\sim 2.3$.
The shape of the
$m_{hh}$ distribution is rather sensitive to variations of
$\lambda_{hhh}$, which alter the interference pattern.
For example, at 
$\lambda_{hhh}=0$, 
the total cross section is almost as large as for
$\lambda_{hhh}/\lambda_{SM}=5$,
but the shape of the distributions is very different.

Further, we made a first attempt to combine the full NLO results with
the NNLO results calculated in the basic HEFT
approximation~\cite{deFlorian:2016uhr}
at differential distribution level, which should lead to a
``NLO-improved NNLO HEFT'' result, which may still be 
improved in the near future in various directions, 
for example towards Higgs boson decays.

\section*{Acknowledgements}
We are grateful to Andreas von Manteuffel for his support
with the use of Reduze and to Jens Hoff for providing us results to compare to the $1/m_t$ expansion.
We also would like to thank Thomas Hahn, Stephan Jahn, Gionata
Luisoni, Fabio Maltoni, Michelangelo Mangano and Magdalena Slawinska for useful
discussions.
This research was supported in part by the 
Research Executive Agency (REA) of the European Union under the Grant Agreement
PITN-GA2012316704 (HiggsTools).
S. Borowka gratefully acknowledges financial support by the ERC
Advanced Grant MC@NNLO (340983).
NG was supported by the Swiss National Science Foundation under contract
PZ00P2\_154829.
GH would like to acknowledge the Kavli  Institute for Theoretical
Physics (KITP) for their hospitality. 
We gratefully acknowledge support and resources provided by the Max Planck Computing and Data Facility (MPCDF).




 

\providecommand{\href}[2]{#2}\begingroup\raggedright\endgroup

\end{document}